\documentclass[%
 reprint,
 amsmath,amssymb,
 aps,
]{revtex4-2}

\usepackage[english]{babel}
\usepackage[utf8]{inputenc}
\usepackage{dsfont}

\usepackage[shortlabels]{enumitem}

\usepackage{float}
\makeatletter
\let\newfloat\newfloat@ltx
\makeatother
\usepackage{algorithm}
\usepackage{algpseudocode}
\usepackage{algorithmicx}
\usepackage{algorithm}

\usepackage{graphicx}
\usepackage{dcolumn}
\usepackage{bm}

\usepackage{hyperref}
\hypersetup{
    colorlinks=true,
    linkcolor=blue,
    filecolor=magenta,      
    urlcolor=cyan,
    pdftitle={Overleaf Example},
    pdfpagemode=FullScreen,
    }

\begin{document}



\title{Optimization complexity and resource minimization of emitter-based photonic graph state generation protocols}

\author{Evangelia Takou}
\email{etakou@vt.edu}
\affiliation{Department of Physics, Virginia Tech, Blacksburg, VA 24061, USA}
\affiliation{Virginia Tech Center for Quantum Information Science and Engineering, Blacksburg, VA 24061, USA} 
 \author{Edwin Barnes}%
 \email{efbarnes@vt.edu}
 \affiliation{Department of Physics, Virginia Tech, Blacksburg, VA 24061, USA}
\affiliation{Virginia Tech Center for Quantum Information Science and Engineering, Blacksburg, VA 24061, USA}
\author{Sophia E. Economou}%
 \email{economou@vt.edu}
\affiliation{Department of Physics, Virginia Tech, Blacksburg, VA 24061, USA}
\affiliation{Virginia Tech Center for Quantum Information Science and Engineering, Blacksburg, VA 24061, USA}

\date{\today}

\begin{abstract}
Photonic graph states are important for 
measurement- and fusion-based quantum computing, quantum networks, and sensing. They can in principle be generated deterministically by using emitters to create the requisite entanglement. Finding ways to minimize the number of entangling gates between emitters and understanding the overall optimization complexity of such protocols is crucial for practical implementations. Here, we address these issues using graph theory concepts. We develop optimizers that minimize the number of entangling gates, reducing them by up to 75$\%$ compared to naive schemes for moderately sized random graphs. While the complexity of optimizing emitter-emitter CNOT counts is likely NP-hard,  we are able to develop heuristics based on strong connections between graph transformations and the optimization of stabilizer circuits. These patterns allow us to process large graphs and still achieve a reduction of up to $66\%$ in emitter CNOTs, without relying on subtle metrics such as edge density. We find the optimal emission orderings and circuits to prepare unencoded and encoded repeater graph states of any size, achieving global minimization of emitter and CNOT resources despite the average NP-hardness of both optimization problems. We further study the locally equivalent orbit of graphs. Although enumerating orbits is $\#$P complete for arbitrary graphs, we analytically calculate the size of the orbit of repeater graphs and find a procedure to generate the orbit for any repeater size. Finally, we inspect the entangling gate cost of preparing any graph from a given orbit and show that we can achieve the same optimal CNOT count across the orbit.
\end{abstract}

\maketitle


\section{Introduction}

Graph states are garnering an increasing amount of interest for quantum networks~\cite{AzumaNatCommun2015,PantPRA2017,BorregaardPRX2020,HilaireQuantum2021,HilairePRA2021,ZhanQuantum2023,HilairePRXQ2023,AzumeRevModPhys2023,ShapourianNPJ2023,SaikatarXiv2024}, error correction~\cite{LooiPRA2008,BellNatCommun2014,LiaoPRA2022}, measurement-based or fusion-based quantum computing~\cite{FBQCSparrow2021}, and enhanced sensing~\cite{ShettellPRL2020,WangPRA2020}. They are stabilizer states that can be represented by graphs, providing a graphical method to understand their entanglement structure~\cite{HeinPRA2004,fattal2004arxiv,TothPRA2005,hein2006arxiv,PlenioNewJPhys2005,MuraoNewJPhys2013} and utility for error correction codes~\cite{WenerPRA2001,BellPRXQ2023}. Stabilizer and graph states are also crucial for understanding the complexity of quantum circuits and the cost of the Clifford part of universal quantum circuits~\cite{MaslovQInfComput2016,BravyiIEEE2021}. 

In all-photonic architectures, photonic graph states can be prepared using fusion gates and multiplexing to increase the success probability of probabilistic fusions~\cite{FBQCSparrow2021}. An alternative way to prepare photonic graph states is by utilizing quantum emitters such as color centers~\cite{WrachtrupPRB2015,TrupkeNPJQInf2020,DorianQuantum2021}, quantum dots~\cite{EconomouPRL2010,GershoniScience2016,VezvaeePRA2022}, or atoms~\cite{ThomasNature2022,RempeNature2024}. In this approach, the emitters mediate the entanglement between the photons, and the preparation is deterministic. There are several proposals for how to create graph states from quantum emitters, with well known examples being cluster states~\cite{LindnerPRL2009,EconomouPRL2010,Russo2019,VezvaeePRA2022} and repeater graph states (RGSs)~\cite{AzumaNatCommun2015,ButerakosPRX2017,HilaireQuantum2021}. For arbitrary graphs, the generation schemes typically assume unrestricted connectivity~\cite{CabelloPRA2011} or constrained connectivity, which might require a large number of SWAP gates. A recent  algorithm  that incorporates connectivity constraints was developed in Ref.~\cite{BikunnpjQI2022}; this algorithm takes an arbitrary photonic graph state as input and outputs the minimal number of emitters needed to produce the state and an explicit state generation circuit. The runtime of the algorithm and the depth of the circuit it produces are both polynomial in the size of the target graph state. 

Although there has been a lot of progress in developing algorithms for generating photonic graph states, the question of how to prepare them optimally using minimal resources is still a line of ongoing research. Ref.~\cite{BikunnpjQI2022} showed that the minimal number of emitters depends on the emission ordering. Moreover, for a given target graph, the complexity of finding the photon emission ordering that requires the fewest emitters is generically NP-hard. Despite this, there exist heuristics that can be used to identify optimal orderings. Even if one fixes the emission ordering, there remain multiple options for what gates to perform at each step of the algorithm of Ref.~\cite{BikunnpjQI2022}, and each choice can lead to a very different state generation circuit. In addition to the number of emitters, another very costly resource is the number of emitter-emitter CNOT gates appearing in the generation circuit. Ref.~\cite{BikunnpjQI2022} did not address the question of how to minimize the number of these CNOTs. On the other hand, several other works have examined the question of how to reduce CNOT gate counts more generally in stabilizer circuits. The Gaussian elimination algorithm of Ref.~\cite{GottesmanPRA2004} corresponds to unrestricted connectivity and requires at most $\mathcal{O}(n^2)$ CNOT gates. The algorithm developed in Ref.~\cite{Patel2008} for unrestricted connectivity achieves at most $\mathcal{O}(n^2/\log2(n))$ CNOT gates, using parallel row-elimination matrices. Ref.~\cite{BujiaoPRR2023} developed an algorithm for restricted connectivity which achieves at most $\mathcal{O}(n^2/\log2(\delta))$ circuit size, where $\delta$ is the minimum node degree of the graph. These works have shown progress in reducing CNOT counts, but the complexity of minimizing the number CNOTs in stabilizer circuits for constrained architectures, as well as the complexity of the photonic graph preparation based on Ref.~\cite{BikunnpjQI2022}, are still open problems. Very recently, Ref.~\cite{Ghanbari2024arXiv} showed that the number of CNOTs needed to produce an RGS of $2n$ photons is $n-2$. This was done by searching through the space of graphs locally equivalent to the target state (i.e., the local complementation (LC) orbit~\cite{hein2006arxiv}) and applying the algorithm of Ref.~\cite{BikunnpjQI2022} to each to find the graph with the minimal CNOT count. 

In this paper, we develop several different optimization algorithms to reduce the emitter CNOTs required to prepare arbitrary photonic graph states. Using these algorithms, we obtain the following results:
\begin{itemize}[leftmargin=*]
\item
Using our most computationally expensive algorithm, we show a reduction of emitter CNOT gates up to $75\%$ for modest-sized graphs compared to a naive implementation of the algorithm developed in Ref.~\cite{BikunnpjQI2022}. Although there are strong indications that this CNOT minimization problem is NP-hard, we identify and incorporate into our algorithms heuristics that enable them to reduce the CNOT counts up to $66\%$ even for large graph states containing hundreds of photons. 
\item
We find the optimal emission ordering that minimizes the number of required emitters for both unencoded and logically encoded RGSs of any size, although the general problem is NP-hard. 
\item 
We find that any LC-equivalent RGS can be generated with $n-2$ CNOTs, where $2n$ is the number of photons in the state, in agreement with Ref.~\cite{Ghanbari2024arXiv}. Unlike in Ref.~\cite{Ghanbari2024arXiv}, we do not need to search through LC-equivalent graphs to find such minimal CNOT circuits. Instead, we can start from any graph in the LC orbit and directly construct a circuit that generates it using only $n-2$ CNOTs. Our results imply that there is not an intrinsic correlation between the number of edges in the target graph state and the minimal number of CNOTs needed to produce it, as was suggested in Ref.~\cite{Ghanbari2024arXiv}. This correlation was likely due to specific algorithmic settings used in Ref.~\cite{BikunnpjQI2022} rather than to fundamental features of the algorithm. We also show that RGSs with logical encodings that protect against photon loss exhibit the same CNOT scaling.
\item 
We find analytical expressions for the size of the LC orbits of complete graphs and of RGSs. This is despite the fact that this counting problem is $\#$P-complete for arbitrary graphs. We further show that if we exclude graphs that differ from others in the orbit by only a relabeling of vertices, then the resulting orbit is linear in the size of the RGS. We show this for RGSs without and with extra leaf qubits.
\item 
RGSs with different photon orderings generally require different numbers of emitters and different numbers of emitter-emitter CNOT gates to produce, as was shown in Ref.~\cite{BikunnpjQI2022}. However, here we find numerical evidence that there may exist a simple correlation between these two resources: Photon orderings that require the same number of emitters also require the same number of CNOTs, even if the graphs with these different orderings are not LC-equivalent.
\item 
The fact that LC-equivalent graphs can be generated with the same number of CNOTs follows from the fact that such states are the same up to single-photon gates. However, this does not mean that a given graph state generation algorithm will necessarily find circuits with the same minimal number of CNOTs for two LC-equivalent graphs. We perform an exhaustive search across all graph states of six photons and find that, with our optimized algorithms, LC-equivalent graphs always require the same number of emitter-emitter CNOTs to produce. We conjecture that, unlike the algorithms of Refs.~\cite{BikunnpjQI2022} and \cite{Ghanbari2024arXiv}, our optimized graph state generation algorithms are capable of constructing minimal-CNOT circuits for any target graph state.
\end{itemize}
The code for our algorithms can be found in Ref.~\cite{ETgithubCode2024}.

Our paper is organized as follows. In Sec.~\ref{Sec:Optimization_CNOTcost}, we explain the algorithm of Ref.~\cite{BikunnpjQI2022} in terms of graph modifications, discuss the complexity of the optimization problem, and test the performance of our optimizers for random graphs. In Sec.~\ref{Sec:CircleGraphsandLC}, we discuss graph theory concepts related to circle graphs and local complementations. In Sec.~\ref{Sec:LCOrbitSize}, we find analytically the size of LC orbits of well-known graphs. In Sec.~\ref{Sec:OptPrepRGS}, we study the optimal preparation of RGSs. Finally, in Sec.~\ref{Sec:RandomLCOrbits}, we focus on the optimal preparation of LC orbits of random graphs.

\section{Minimizing the number of entangling gates in photonic graph state generation \label{Sec:Optimization_CNOTcost}}

\subsection{Graph visualization of the algorithm}

\begin{figure*}[!htbp]
    \centering
    \includegraphics[scale=0.64]{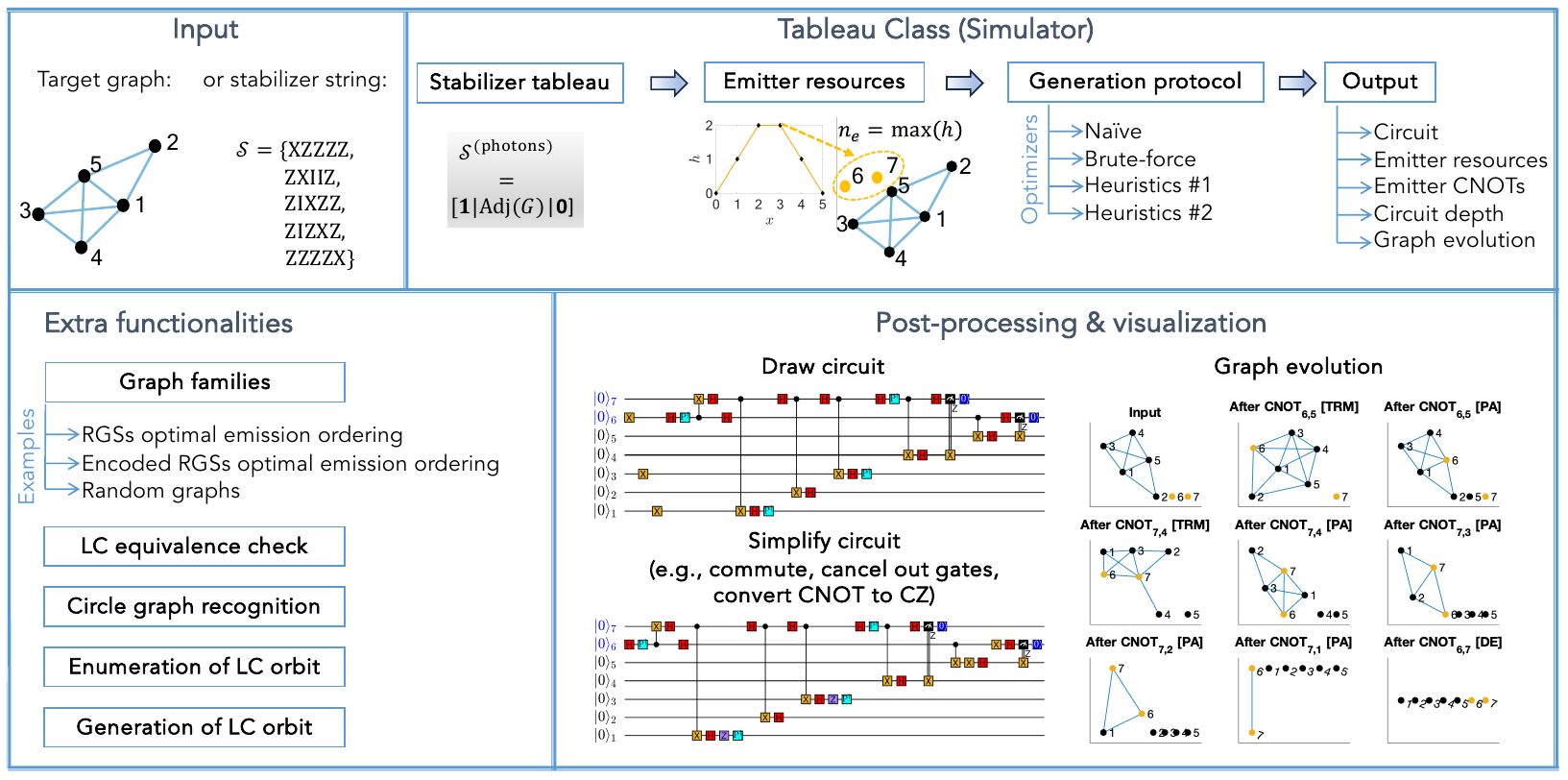}
    \caption{Overview of optimization of photonic graph state generation. The input photonic graph is fed into the tableau class simulator, from which we can choose different optimizers to minimize its preparation cost. The outputs of the simulator can be used for visualization of the results (e.g., circuit, or graph state evolution). Extra external functionalities further allow us to characterize the graphs, such as enumerating or generating their LC orbits. }
    \label{fig:Overview}
\end{figure*}

The photonic graph state generation algorithm introduced in Ref.~\cite{BikunnpjQI2022} was developed using the stabilizer formalism. Here we develop various optimizers with improvements and additional features compared to this algorithm. A summary of our toolbox is shown in Fig.~\ref{fig:Overview}, which we will analyze throughout the paper. First, we briefly review how the algorithm of Ref.~\cite{BikunnpjQI2022} works and also show how we can visualize the main steps of the algorithm in terms of graph rather than stabilizer operations.

The algorithm starts from a target stabilizer state of entangled photons with a specified photon emission ordering. The algorithm first calculates the number of emitters required to generate the target state by computing the maximum bipartite entanglement entropy for the given emission ordering. This is facilitated by putting the tableau of stabilizers in the row-reduced echelon form (RREF)~\cite{PlenioNewJPhys2005}. Once the minimal number of emitters is obtained, the generation circuit is then constructed step-by-step by working backwards in time. That is, the algorithm starts with the stabilizer tableau of the target photonic graph state and decoupled emitters and successively modifies it through a series of Gaussian eliminations (each time restoring it to RREF) until the tableau is converted into that of a product state for all photons and emitters. This product state is the actual initial state corresponding to decoupled emitters and photons that have not yet been emitted, and the actual generation circuit is obtained by inverting the time-reversed circuit produced by the algorithm. The resulting circuit must obey the constraints that photon-photon gates are forbidden, and that photons are emitted from the quantum emitters according to the specified emission ordering. At intermediate stages of the algorithm, emitters and photons are entangled with one another, but there is no emitter-photon entanglement at the beginning or end of the generation circuit. All the entanglement generated during the algorithm (and hence all the entanglement in the final photonic graph state) originates either from the photon emission process or from emitter-emitter entangling gates.

The time-reversed circuit that the algorithm builds up from the sequence of Gaussian eliminations consists of two main primitives: photon absorption and time-reversed measurement (TRM). The overall goal of the time-reversed circuit is to absorb all of the photons into the emitters in the reverse of the chosen emission ordering. However, it is not always possible to absorb a photon at a given step of the algorithm since the ability to do so depends on the structure of the photon-emitter state at that stage. Photon absorption is the time-reverse of photon emission. Since photon emission creates photon-emitter entanglement, photon absorption removes it, but this can only happen if the photon in question is already entangled with emitters. When a photon absorption is not possible, we instead perform a TRM to prepare the state for the next absorption. TRMs are operations that act on emitters that are not entangled with photons or other emitters. This is because TRMs are time-reversed versions of ordinary projective measurements on emitters, and since the measured emitter is completely decoupled in the post-measurement state, TRMs can only act on decoupled emitters in the time-reversed circuit. TRMs generate photon-emitter entanglement, thus enabling a subsequent photon absorption. In particular, the result of a TRM is to connect the decoupled emitter with the neighborhood of the photon to be subsequently absorbed (see, for example,  2nd step in Fig.~\ref{fig:6NodeGraph}, and proof in Appendix~\ref{App:Graphical_Proofs}). The first step of the graph state generation algorithm always involves TRMs, because a photon absorption is not possible when all photons are decoupled from emitters.  TRMs can also happen mid-circuit, and they might require emitter-emitter CNOTs beforehand. The emitter CNOTs can disentangle a chosen emitter from the rest, allowing it to then undergo a TRM to create more emitter-photon entanglement and prepare the state for the next photon absorption. 

\begin{figure}[!htbp]
    \centering
    \includegraphics[scale=0.675]{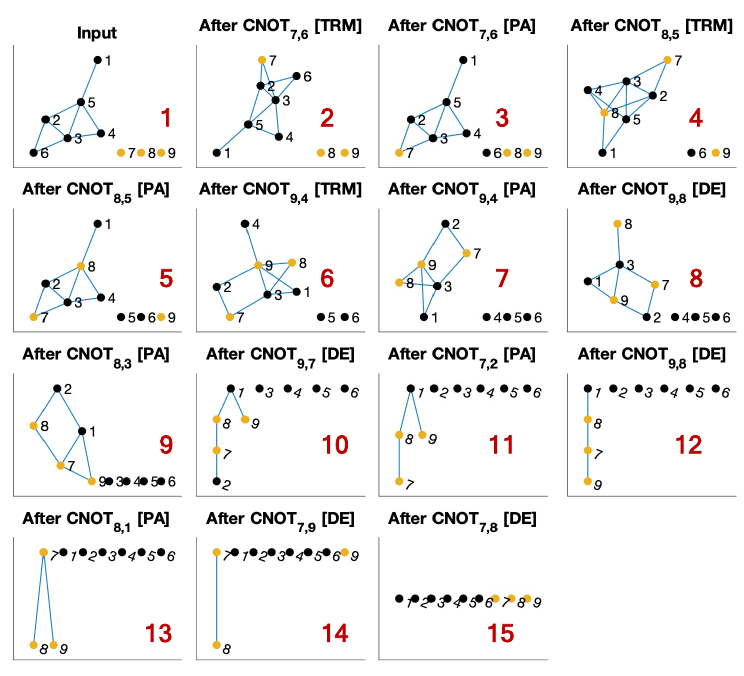}
    \caption{Time-reversed evolution of the photon-emitter graph state during the graph state generation algorithm. The input target state shown in panel 1 contains of 6 entangled photons (black nodes) and 3 decoupled emitter qubits (orange nodes). The algorithm converts the input state into the disentangled state shown in panel 15 through a sequence of time-reversed measurements (TRMs) and photon absorptions (PAs), interspersed with single-qubit gates and emitter-emitter CNOT gates (DE). We omit showing steps that involve single-qubit gates.}
    \label{fig:6NodeGraph}
\end{figure}

Let us take a closer look at photon absorption. When photon absorption is possible, the tableau has at most two stabilizer rows (due to RREF) in which the first non-identity Pauli starts from the photon to be absorbed. In some cases, photon absorption happens \textit{freely}, meaning that there is an available emitter we can use, and no emitter-emitter CNOTs are required upfront. In the tableau formalism, this means that a stabilizer of the form $I^{\text{photon} ~(1)}\dots I^{\text{photon} ~(j-1)}\sigma_j^{\text{photon}~(j)}\sigma_{k}^{\text{emitter}~(l)}$ exists, such that the Pauli weight on emitter qubits is one. The absorption is performed by transforming both the photon and the emitter Paulis into $Z$ via local gates, and then applying CNOT$_{\text{em},\text{ph}}$. In other cases, the photonic stabilizer row(s) could have more than one non-identity emitter Pauli, in which case we need to apply emitter-emitter CNOTs first to reduce the Pauli weight on emitters to one, which then enables a single emitter to absorb the photon. (The emitter choice is not unique, as we discuss further later on.) Graphically, free absorption happens in any of the following three scenarios:
\begin{itemize}
    \item An emitter is a leaf of the photon to be absorbed. See steps 8-9 of Fig.~\ref{fig:6NodeGraph}.
    \item The photon to be absorbed is a leaf of the emitter. See steps 10-11 of Fig.~\ref{fig:6NodeGraph}.
    \item The emitter is connected to the neighborhood of the photon to be absorbed. See steps 2-3 of Fig.~\ref{fig:6NodeGraph}.
\end{itemize}
By ``leaf" of a qubit, we mean a neighboring qubit that itself has no other neighbors.
Note that steps 8, 10, 12, 14, and 15 of Fig.~\ref{fig:6NodeGraph} correspond to emitter CNOTs. The decision of how we apply these  CNOTs, meaning which emitter we free up for the next photon absorption, is not unique, and our decision affects the total CNOT count. For example, the generation path of Fig.~\ref{fig:6NodeGraph} is not optimal, and in Appendix~\ref{App:Opt_Circuit}, we show that the graph can be prepared with only 3 emitter CNOTs.

\subsection{Optimization of CNOT cost and complexity of minimizing CNOTs in stabilizer circuits}

Minimizing the number of emitter-emitter entangling gates used to prepare an arbitrary photonic graph state is, in general, a non-trivial task. In principle, there are several choices we can make in the construction of the circuit to minimize this metric, and it is not always clear a-priori which decision path we should follow. When we can absorb a photon using an available emitter (without needing emitter CNOTs), we can follow this path because it does not increase the number of CNOTs.  It is important that we maximize the number of free photon absorptions, but it is not always obvious how to do this from the RREF gauge. This is because the RREF gauge on its own does not always ensure that the stabilizers have minimal Pauli weight. For this reason, we might also need to perform ``back-substitution" while preserving the RREF gauge, by which we mean that we try multiplying each stabilizer with ``shorter" stabilizers further down in the RREF tableau---if a product has lower weight, then we replace the stabilizer with this product. In Appendix~\ref{App:Algorithms}, we  provide Algorithm~\ref{alg:BackSubs}, which performs the back-substitution. That appendix also includes
Algorithm~\ref{alg:Free_Photon_Absorption}, which searches through all possible conditions for free absorption.

When the absorption cannot be performed ``freely'' (i.e., without needing emitter CNOTs), we can choose to free up any of the available emitters with a non-identity Pauli in the photonic stabilizer row(s). For example, consider the simplest case where we have the stabilizer $I\dots I Z^{\text{photon}~(j)}Z^{\text{em}_1}Z^{\text{em}_2}$. (More generally, such a stabilizer can have Pauli's different than $Z$'s, and we can transform them with local gates into $Z$'s.) In this case, we can choose to do CNOT$_{\text{em}_1,\text{em}_2}$ or CNOT$_{\text{em}_2,\text{em}_1}$. Because CNOT$_{ij}$ transforms $Z^{(i)}Z^{(j)}$ into $I^{(i)} Z^{(j)}$, the Pauli of the target qubit is preserved in the stabilizer row, and this qubit can then absorb the photon. Although at the current step we cannot avoid a CNOT gate between the emitters, which emitter we choose to absorb the photon could affect the number of CNOTs required in later stages of the algorithm.

Because the stabilizers that represent the state are not unique, another freedom in the optimization comes from multiplying stabilizer rows. We can explore this freedom when emitter CNOTs are necessary before photon absorption. At every step, there are at most two photonic stabilizer rows that start from the photon that we want to absorb (due to the RREF gauge). Multiplying these rows together changes the Pauli operators of emitters and can also change which emitters remain in the photonic stabilizer row, and thus which are available for absorbing the photon. Consequently, we end up with new local gates that we can perform before deciding how to apply the emitter CNOTs that leave one emitter to absorb the photon. Additionally, at some time step, we could also have a stabilizer (or stabilizers) that has support only on emitter sites. We could multiply such a stabilizer with the photonic stabilizers (of the photon to be absorbed), which once again changes the local gates and emitter CNOTs we apply. The same effect of requiring different local gates comes also from performing back-substitution. Instead of only putting the tableau in the RREF gauge before every photon absorption step (which is done irrespective of whether free absorption is possible), we can go one step further and do the back-substitution. We find that this also affects the total gate count and, in most cases, leads to reduced circuit size.

Finally, we can also perform local complementations (LCs) before making an emitter choice. Complementing locally the graph $G$ about a vertex $v$, which we denote as $G*v$, inverts the neighborhood of the node $v$, $N_v=\{w\in V(G)|(w,v)\in E(G)\}$, where $V(G)$ is the vertex set and $E(G)$ is the edge set. That is, for every pair of vertices $(w,p)\in N_v$, we either remove an existing edge between them or add one if there wasn't one already. An LC translates to the following Clifford operation~\cite{ElliottPRA2008,BorisPRA2022,AdcockQuantum2020}:
\begin{equation}\label{Eq:LC_Gate}
    U_{v}^{\text{LC}}= \sqrt{-iX_a}\prod_{j\in N_v}\sqrt{iZ_j}\sim H_aP_aH_a \prod_{j\in N_v}P_j^\dagger,
\end{equation}
where $H_a$ ($P_a$) is a Hadamard (phase) gate acting on qubit $a$.

Multiple LC rounds form an LC sequence about vertices $vw\dots p$ such that $G'=G*vw\dots p$. We will explain in detail the role of LCs in the optimization of stabilizer circuits in Sec.~\ref{Sec:RandomLCOrbits}. For now, we stress that if a graph is the local complement of another graph, then the graphs are related by local graph transformations, and the stabilizer states differ up to local Clifford gates. Similar to the new choices over local gates generated by row multiplications of the stabilizers, the extra degree of freedom of inspecting local complements can be incorporated into the optimization of generating photonic graph states.

Overall, we need to inspect several degrees of freedom to ensure optimal preparation of any generic graph, and the number of choices we can make grows quickly with the number of emitters and graph size. This indicates that the problem of 
minimizing the number of CNOTs to generate arbitrary photonic graph states utilizing the algorithm of Ref.~\cite{BikunnpjQI2022} is, on average, a hard problem. There exist related works in the literature that all involve mapping the problem of the optimization of CNOT circuits (known as linear reversible circuits) to NP-complete (for the decision variant), or NP-hard (for the optimization variant) problems. We summarize the conclusions of these other works here:

\begin{itemize}
    \item In Ref.~\cite{ZhangSocietyforAppliedMathematics2020}, the authors mention that optimizing the size or depth of an $n$-qubit CNOT circuit, using $m$ ancillas and with topological constraints on the connectivity, is NP-hard. They prove that it is also NP-hard to optimize a corresponding sub-circuit.
    \item Ref.~\cite{CA0InfandComp2020} discusses the minimum fill-in problem. To obtain circuits of minimum size, we need to minimize the steps in the Gaussian elimination where zero entries might become non-zero. The question of whether it is possible to do the Gaussian elimination with at most $k$ fill-ins is NP-complete as was proven in Ref.~\cite{Yannakakis1981}.
    \item In Ref.~\cite{Allouche2020}, the problem of synthesizing   linear reversible circuits is translated into the syndrome decoding problem, which is known to be an NP-hard optimization problem. The authors develop heuristics to minimize the CNOTs for restricted or unrestricted architecture connectivity.
    \item In Ref.~\cite{JialinPRR2023}, the authors use Steiner trees to optimize the size of the Clifford circuit. The decision variant of the minimum Steiner tree is NP-complete, and its optimization variant is NP-hard. Thus, the authors construct 2-approximations to Steiner trees to achieve $\mathcal{O}(n^2/\log(\delta))$ circuit size, where $\delta$ is the minimum degree of the graph. 
\end{itemize}

Since from the perspective of our optimization problem, we also find that minimizing the number of CNOTs requires that we inspect exponentially many decisions, it is unlikely that a polynomial time algorithm can be developed for the general case. However, later on, we will show how to use different optimization methods to obtain nearly optimal CNOT counts and show under which cases we can make more optimal choices for the construction of the circuit, without exploring the entire search space.

\subsection{Description of Brute-force algorithm}
\begin{figure*}[!htbp]
    \centering
    \includegraphics[scale=0.65]{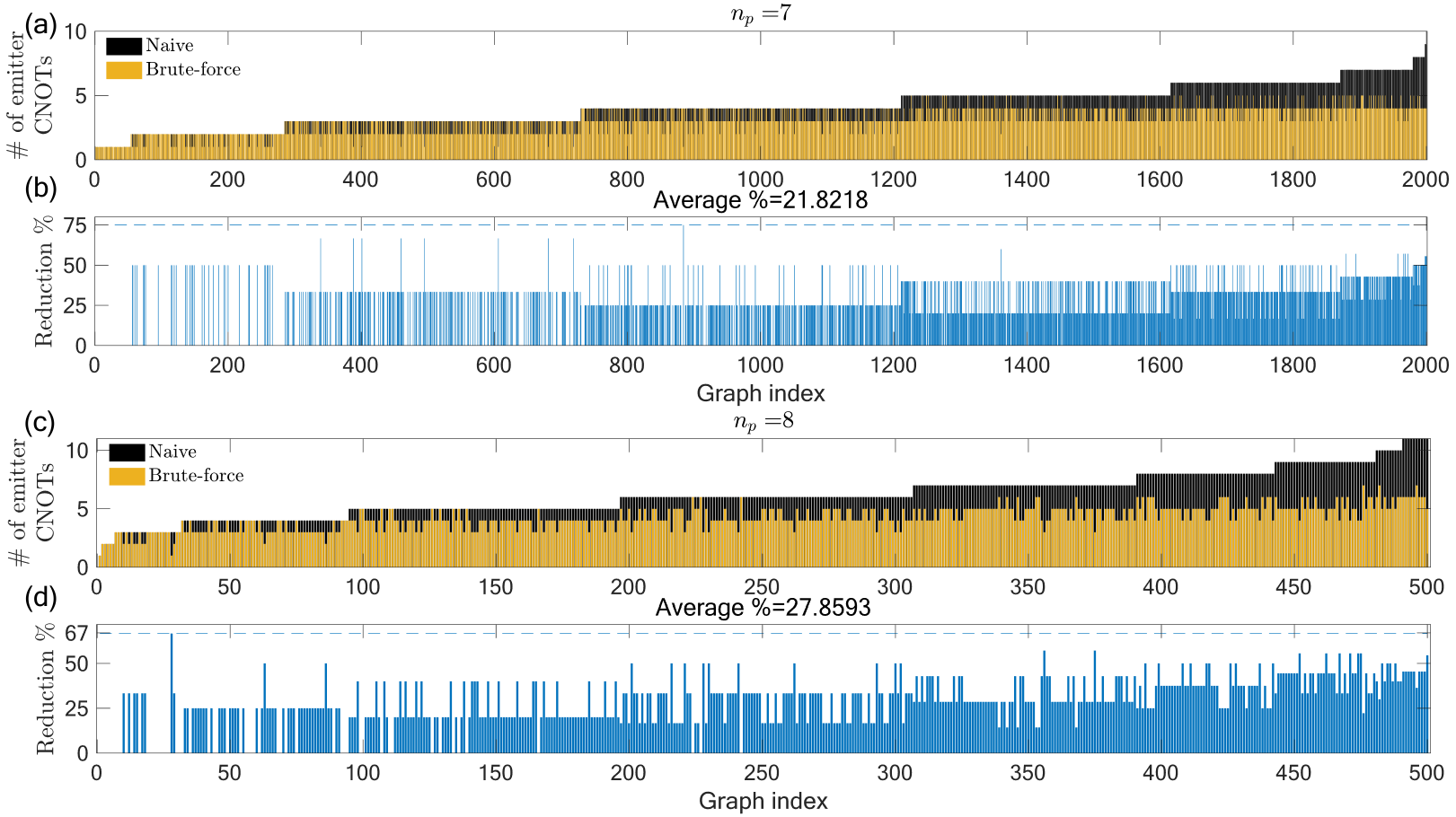}
    \caption{(a) Emitter CNOT counts obtained from the Naive (black), and Brute-force optimizer (yellow), for 2000 random graphs of order $n_p=7$. (b) Percentage reduction of CNOT counts for each graph of (a). (c) Same as in (a), for 500 random graphs of $n_p=8$. (d) Percentage reduction in emitter CNOT counts for the graphs of (c). }
    \label{fig:BruteForce}
\end{figure*}
We now describe our Brute-force algorithm, which we use to address many degrees of freedom, while targeting choices that matter the most. It is a tree decision algorithm that inspects multiple paths we can follow on the most crucial optimization step, the step that requires emitter CNOTs before absorption. When absorption of photon $j$ is possible but we first need to apply emitter CNOTs, the algorithm explores the following paths:
\begin{itemize}
    \item It picks any of the available emitters of the two photonic stabilizer rows to absorb the photon. 
    \item It multiplies the photonic rows (if two of them exist) and then picks, again, any of the available emitters to absorb the photon.
    \item It multiplies stabilizers that have support on emitters (if such stabilizers exist) on photonic rows and then inspects emitter choices again.
    \item It performs LCs, based on the cutoff of LC rounds and nodes we choose to complement about, and for each such realization tests, again, photonic absorption utilizing any of the available emitters.
\end{itemize}
All the above choices are explored for the current photon absorption level, and each tableau is fed to the next algorithmic steps of time-reversed measurements and absorptions. In those next levels, new tableaux are generated and processed next. Once all tableaux are processed (we have a product state of all qubits), the circuit with minimal CNOT counts across all realizations is selected as optimal. The algorithm has also a pruning option of how many realizations to keep per level. This option allows us to prune the search since the choices we explore increase quickly with increasing graph size or emitter resources.

In Fig.~\ref{fig:BruteForce}, we compare the performance of the Naive approach against the Brute-force method for random graphs of $n_p=7$ and $n_p=8$ photons. The Naive approach does not perform back-substitution and does not exhaust all checks for free photon absorption (PA). For the absorption step, it first attempts to perform free PA. It finds the photonic stabilizer row(s) which starts with a non-trivial Pauli on the photon to be absorbed, and tests if there is only one emitter site with a non-trivial Pauli in these rows. If this is true, the emitter is used for absorption. If this is not true and there are two photonic rows, it multiplies them together and tests again if there is only one emitter site in the new photonic row that we obtain. If this is true, the emitter is used for free PA. Otherwise, we search for stabilizers with support only on emitter sites. If such a stabilizer exists, we test if free PA is possible after multiplying the stabilizer with support on emitters with the photonic stabilizer row. (If there are two photonic stabilizer rows, we multiply first on the first photonic stabilizer row, and test for free PA, and repeat the same for the second photonic row. If there are multiple stabilizers with support on emitters, we multiply one of them at a time at a given photonic row, and re-check the free PA, till we have looped over all emitter stabilizers and photonic stabilizers.) If this condition also fails, then we return with a flag that free PA was not successful, and resort to a PA that requires that we first perform emitter-emitter CNOTs. Of the two photonic rows that could exist, we pick the one with minimal Pauli weight on emitter sites, and we free-up the first emitter we encounter in this row via emitter-emitter CNOTs. Thus, the Naive approach picks an emitter randomly in this final decision step.  In Figs.~\ref{fig:BruteForce}(a),(b), we search over all degrees of freedom we mentioned for the Brute-force method, and enable one round of LCs. We note a substantial reduction in emitter CNOTs, with a maximum of up to 75$\%$. For several graphs, we see a reduction of $50-67\%$ in emitter CNOT counts, and the average reduction across the 2000 graphs of order $n_p=7$ is $\sim 22\%$. In Figs.~\ref{fig:BruteForce}(c),(d), we repeat the same calculation, but for graphs of $n_p=8$ photons. For these random graphs, we again enable the final optimization step of LC operations before the photon absorption, but we prune some of the choices (i.e., potential tableaux that we will explore per step). Although we prune away choices, we still note that the reduction in emitter CNOTs can reach up to $\sim 67\%$. 

\begin{figure*}[!htbp]
    \centering
    \includegraphics[scale=0.81]{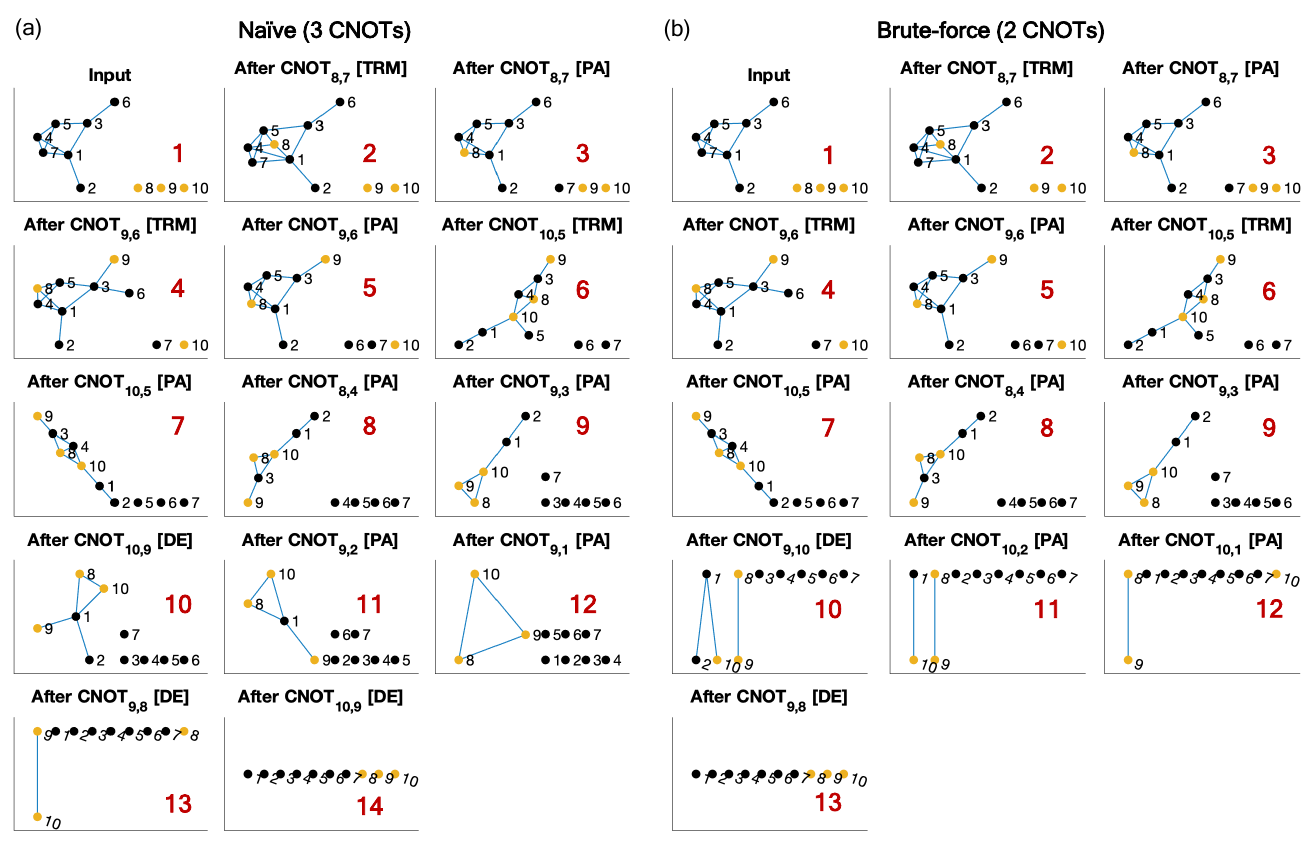}
    \caption{Evolution of a 7-node graph prepared with the Naive, (a), or Brute-force method, (b). The top-left panel is the input graph and the bottom-right panel the final state. All steps are identical between the two methods till the decision of which emitter absorbs photon 2 (step 10). The Naive (Brute-force) method picks emitter 9 (10), and leads to 3 (2) emitter CNOTs in total. TRM is time-reversed measurement, DE  disentanglement, and PA photon absorption. }
    \label{fig:Graph_Evol}
\end{figure*}

\subsection{Description of Heuristics \texorpdfstring{$\#$}{a}1 optimizer}

Although our Brute-force algorithm obtains close to optimal CNOT counts for small-sized graphs, it is inefficient to inspect all degrees of freedom  we mentioned for large graphs. Even if we were to prune choices, we would end up chopping our search space severely and in a blind way. However, in several cases, we observed that the Brute-force optimizer uncovers a particular pattern associated with the optimal decision. Here, we will explain and use this pattern to develop efficient heuristic methods that improve substantially the CNOT counts.

As we mentioned previously, back-substitution on the RREF tableau tends to lead to shorter circuits and fewer CNOTs. Thus, in our heuristics method we incorporate the option to perform back-substitution right before every photon absorption, and we also enable the option to exhaust all possibilities for free photon absorption. Additionally, as the size of the graph and emitter resources increase, we end up with a large number of entangled emitters at the final step where we disentangle only emitter qubits. If we perform back-substitution again in this last step, we find that in several cases we reduce the CNOT counts, because back-substitution tends to reduce the weight of the stabilizers. All these extra improvements do not increase substantially the runtime (see Appendix~\ref{App:Runtime}). 

An important decision step is which emitter to ``free up'' for absorption via local gates and CNOTs. The emitter with a non-trivial Pauli on the photonic stabilizer row can then absorb the photon. In several cases, the Brute-force optimizer selects local gates and an emitter for absorption, such that the application of emitter CNOTs increases the number of disconnected subgraphs. In other words, if we track the graph evolution, we find that the graph splits into smaller ones, and subsets of emitters belong to different subgraphs. This means that certain emitters will never interact again before future absorptions, and they will consume photons of disjoined subgraphs.

We illustrate this feature in Fig.~\ref{fig:Graph_Evol}, which depicts the evolution of a 7-node graph prepared by three emitters, labeled as 8-10. We display the updated graph after each CNOT gate (corresponding to time-reversed measurement, photon absorption, or an emitter-emitter gate); we do not explicitly show the local gates that are also applied between graph updates. All decisions are identical between the Naive and Brute-force methods until the step of free-ing up an emitter to absorb photon 2 (step 10 in the figure). The Brute-force method picks emitter 10, which creates two disconnected subgraphs, and ensures that emitter 10 never interacts again with emitters 8 and 9. Consequently, in the end, we need to disentangle only emitters 8 and 9, whereas in the Naive method, we need to disentangle all of them. 

This intuitive pattern also arises in the so-called nested dissection technique of Gaussian elimination on sparse matrices~\cite{Gilbert1986}. In this context, a non-zero entry of a sparse matrix represents an edge between graph nodes (note that this graph is not the same as a graph state, rather it is an artificial graph that represents the matrix). In nested dissection techniques, one tries to identify separators, i.e., nodes whose removal will create subgraphs, so that then Gaussian elimination can be performed on each piece separately.

The search for an increase in the number of disconnected subgraphs can be used as a heuristic inspection on top of the Naive circuit generator. Our first level of heuristics explores if we can achieve this pattern during the generation. The increase in the number of disconnected subgraphs is accomplished by ensuring that the application of local gates followed by a single CNOT gate either removes an emitter from the remaining graph or breaks the graph into separate subgraphs that each have more than one node (e.g, as in Fig.~\ref{fig:Graph_Evol}). We first prioritize removing single emitters whenever this is possible. We find that we can deterministically choose local gates and the emitter to be removed from the graph at a particular time-step by performing back-substitution. Back-substitution is important, because it reveals whether or not stabilizers of weight 2 with support only on emitters exist. If such stabilizers exist, we use a look-up table related to what local operations we apply on the qubits before the emitter CNOT (see transformation rules in Appendix~\ref{App:CNOT_Rules}).

Once no more weight-2 stabilizers with support on emitter sites exist, we then resort to a more extensive search for potentially creating disconnected components. We select the photonic stabilizer row of the photon to be absorbed (if two exist, we pick the one with minimal weight on emitter sites), and identify the emitters with non-identity Paulis. We take any emitter pair $\{i,j\}$ and apply one of 36 possible combinations of local gates from the set $\{I,H,P,HP,PH,HPH\}$ on each emitter. After that, we apply CNOT$_{ij}$ or CNOT$_{ji}$ and check whether a combination of these gates leads to a greater number of disconnected components.
Note that $PHP$ gives the same tableau transformation as $HPH$ (up to different phase updates, since $HPHPHP=e^{i\pi/4}I$), so we ignore the former. 
\begin{figure*}[!htbp]
    \centering
    \includegraphics[scale=0.66]{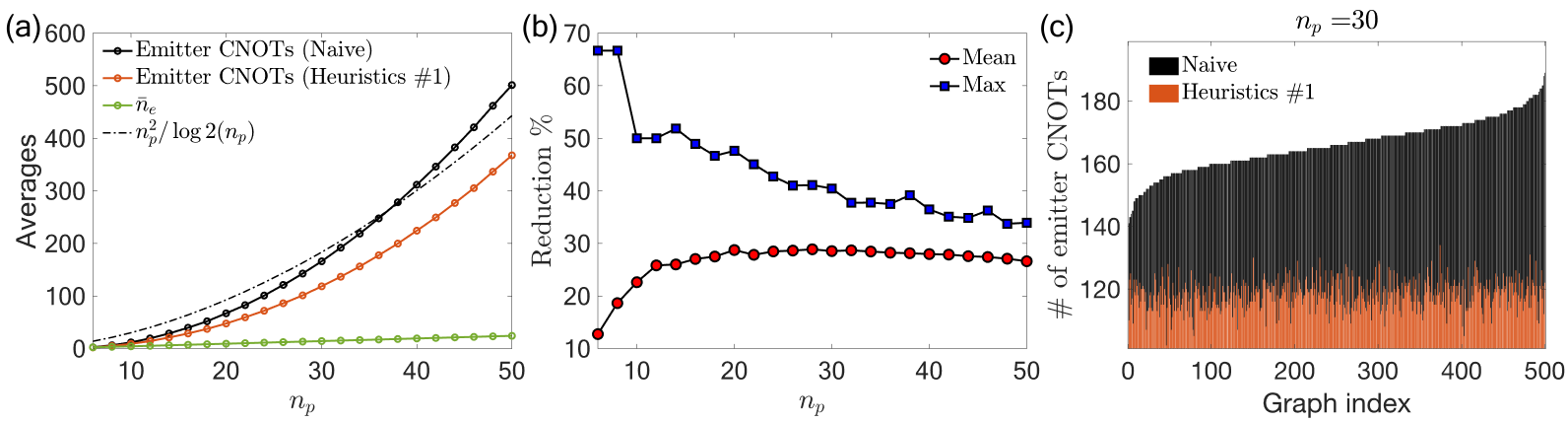}
    \caption{(a) Average number of emitter CNOTs needed to prepare photonic graphs from $n_p=6$ till $n_p=50$, using the Naive (black), or Heuristics $\#1$ optimizer (orange). We sample 500 random graphs per $n_p$. The dashed line indicates the bound $n_p^2/\log2(n_p)$. The green line shows the average number of emitters. (b) Mean and maximum reduction of emitter CNOTs per $n_p$. (c) Number of emitter CNOTs required by the Naive or Heuristics $\# 1$ method for 500 random graphs of $n_p=30$ photons. }
    \label{fig:Naive_vs_Heur1}
\end{figure*}

The above heuristic method is implemented only before an absorption and only when no emitter is freely available. We terminate the search once we find a choice that increases the number of disconnected components, and then we re-attempt the free photon absorption. If the free absorption is still not possible, we transform all $\tilde{n}_e$ non-identity Paulis acting on emitters to $Z$ via local gates, for the photonic row we are inspecting. We then scan through these $\tilde{n}_e$ emitters, which are all potential absorption sites, and for each we apply a series of $\tilde{n}_e-1$ CNOT gates, all with the chosen emitter as the target, so that the Paulis on the remaining $\tilde{n}_e-1$ emitters are transformed to identities (as needed for photon absorption). The emitter that yields a graph with the fewest edges following these CNOTs is chosen as the photon absorption site. We find that this procedure typically leads to a greater CNOT reduction compared to selecting an emitter randomly. 

Let us also examine the cost in runtime for searching for an increase in disconnected components. In the worst-case scenario and for a particular time step, the complexity of the search scales as follows:

\begin{itemize}
    \item For a fixed emitter pair, we apply any of the 36 combinations of local gates from the set $\{I,P,H,HP,PH,HPH\}$ before the CNOT.
    \item We inspect any emitter pair that appears in the photonic stabilizer row at a given time step: $36\times \tilde{n}_e(\tilde{n}_e-1)/2$. $\tilde{n}_e$ is the number of available emitters for photon absorption for the current photon (emitter sites with non-identity Paulis). We inspect only the photonic stabilizer row with minimal weight on emitter sites. Due to the RREF gauge, there can be at most two stabilizer rows we can use.
    \item We can also exchange the role of control/target qubits, which increases the number of choices by a factor of 2, giving rise to $2\times 36 \times \tilde{n}_e(\tilde{n}_e-1)/2$.
    \item We need to extract the adjacency  from the tableau to count the number of connected components for each of the above choices: $n^3 \times 2 \times 36 \times \tilde{n}_e(\tilde{n}_e-1)/2$. The factor of $n^3$ arises from Gaussian elimination.
\end{itemize}

Note that the above list does not include the final step of scanning through the potential absorption sites and applying CNOTs to find the site for which the updated graph has the fewest edges, because this step makes a sub-leading contribution to the runtime cost. [This is because we only need to apply $\tilde{n}_e-1$ emitter CNOTs $\tilde{n}_e$ times to free up any of the available emitters.] Also, we resort to this final step only if none of the above patterns succeed in increasing the number of disconnected components and we still fail to absorb the photon. By inspecting several graphs, we find that from the set of local gates $\{I,H,P,HP,PH,HPH\}$, we can discard $HP$ and $HPH$ because these are not usually selected by the optimizer. Thus, we reduce the search of local gates to 4 per emitter and 16 combinations per emitter pair. Further, the above worst-case scaling is for only one step of the algorithm, and so the complexity of the algorithm increases for every step in which it is not immediately clear which emitter to use for the next photon absorption. However, this algorithm is much more efficient than the Brute-force one because it does not open recursive search levels. 
Further analysis of this algorithm, which we refer to as the Heuristics \#1 optimizer, is given in Appendix~\ref{App:Flowcharts}. We should also comment that it is not always guaranteed that the choice revealed by the increase in disconnected subgraphs is globally optimal. This is because the minimum fill-in problem is NP-complete, meaning we do not know if the Gaussian elimination ordering we follow (which is also subject to constraints on which vertices can be removed) will be the most optimal one. Nevertheless, as we will show shortly, the patterns we describe lead to substantial reductions in emitter CNOTs compared to the Naive optimizer.

\begin{figure*}[!htbp]
    \centering
    \includegraphics[scale=0.68]{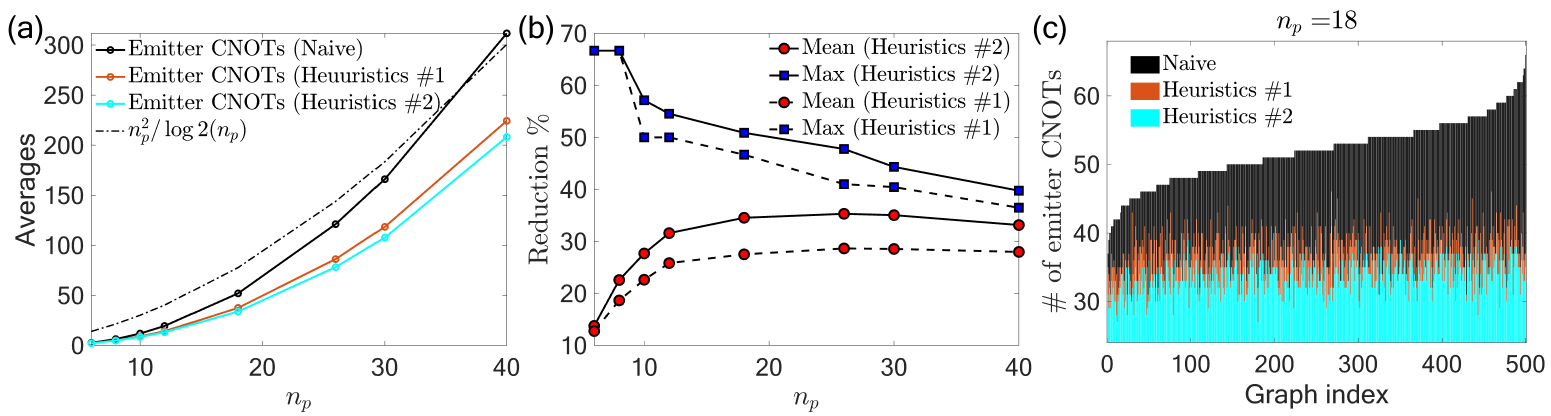}
    \caption{(a) Average number of emitter CNOTs obtained by the Naive (black), Heuristics $\# 1$ (orange), and Heuristics $\# 2$ (cyan) method as a function of $n_p$. We sample over the same 500 random graphs per $n_p$ from Fig.~\ref{fig:Naive_vs_Heur1}. (b) Reduction of emitter CNOTs obtained by the two heuristics methods. With solid (dashed) lines we show the mean and max reductions obtained by the Heuristics $\# 2$ (Heuristics $\#1$) method. (c) Emitter CNOTs obtained by the three optimizers for the graphs with $n_p=18$.}
    \label{fig:Comp_2_heuristics_w_Naive}
\end{figure*}

This summarizes the subroutines of our Heuristics $\#1$ optimizer.  In Fig.~\ref{fig:Naive_vs_Heur1}(a), we compare the average number of emitter CNOTs obtained by the two methods for 500 random graphs of size $n_p=6$ up to $n_p=50$. We explore three variants of the Heuristics $\# 1$ method, which have to do with whether or not we allow back-substitution globally, and whether or not we exhaust all checks for free photon absorption. From these three methods, we keep the one that gives the lowest CNOT counts. For $n_p+n_e\geq 30$, we truncate the emitters we inspect as possible photon absorption sites to up to 1/3 of the available emitters in a given photonic stabilizer row. For $n_p+n_e<30$, we explore all emitter pairs and local gate combinations. The black curve in Fig.~\ref{fig:Naive_vs_Heur1}(a) shows the average number of emitter CNOTs the Naive method utilizes, and the orange curve shows the number of emitter CNOTs used by the Heuristics $\#1$ method. The dashed line shows the bound $n_p^2/\log2(n_p)$. Ref.~\cite{Patel2008} developed a Gaussian elimination algorithm for unrestricted connectivity which achieves at most $\mathcal{O}(n^2/\log2(n))$ CNOT gates. Here we  show instead the bound $n_p^2/\log2(n_p)$, since the emitter resources per $n_p$ are not fixed, and depend on the input graphs.
In Fig.~\ref{fig:Naive_vs_Heur1}(b), we further show the average and maximum reduction in emitter CNOTs for each value of $n_p$. We note that the maximum reduction across $n_p$ is $\geq 35-40\%$. The average reduction across all $n_p$ is $\sim 30\%$. Figure~\ref{fig:Naive_vs_Heur1}(c) shows a slice of the data of Fig.~\ref{fig:Naive_vs_Heur1}(a) for $n_p=30$. The black (orange) bars are the emitter CNOTs obtained by the Naive (Heuristics $\#1$) method. For illustration purposes, we sort the data in increasing number of CNOTs required by the Naive method. We see that the percentage reduction in  CNOT counts for the graphs with $n_p=30$ we sampled over can reach up to $\sim 40 \%$.

\subsection{Description of Heuristics \texorpdfstring{$\#$}{a}2 optimizer}

The Heuristics $\# 1$ optimizer sets a new upper bound on emitter CNOTs compared to the Naive scheme. To reduce the emitter CNOT counts further, we can add more optimization levels. We developed another algorithm, Heuristics $\#2$ (see Appendix~\ref{App:Flowcharts}), which follows most of the steps of Heuristics $\#1$ discussed above, but differs from it in the last step. Instead of picking the emitter that leads to the fewest edges in the graph (right before the photon absorption), Heuristics $\#2$ looks further ahead and examines CNOT counts that would be accrued in subsequent iterations of the algorithm if any of the $\tilde{n}_e$ available emitters were to be chosen to absorb the next photon. To ``free up'' one of them, we apply $\tilde{n}_e-1$ CNOTs between it and each of the remaining available emitters as before. Heuristics $\#2$ continues to examine potential future steps until all qubits are disentangled or until a cutoff on the number of future steps to be examined is reached. The algorithm explores these potential next steps by repeatedly calling the same subroutine, so that options are explored recursively following all steps of the Heuristics $\# 1$ optimizer, and the additional steps of Heuristics $\#2$. The algorithm keeps track of the number of emitter CNOTs for each potential future trajectory, and after it has explored all the possible paths made available to it, it returns to the present iteration of the algorithm and chooses as the next absorption site the emitter that it projects will lead to the fewest CNOTs overall. Note that this is not an exhaustive search since, although we open recursions, we pick emitters greedily. Moreover, we monitor the CNOT counts constantly, making informed choices about which emitter to choose at each level.

The Heuristics $\# 2$ optimizer can also be controlled by setting a cutoff on the number of emitters we consider as potential absorption sites, and we can choose when to perform a recursive search or, when to proceed with only the steps of Heuristics $\# 1$. Because we interleave recursions with Heuristics $\# 1$, we save up computational time, since for certain graphs the Heuristics $\# 1$ patterns will be identified and the optimizer will exit before searching exhaustively through the decision space. Additionally as mentioned above, we can set a ``future cutoff'', i.e., monitor the CNOT counts up to a future reference point, and then return to make a decision for a previous emitter choice. This optimizer is more efficient than the Brute-force one, and it allows us to push the computation to larger-sized graphs by pruning choices in a controlled way.  

To test the performance of the Heuristics $\# 2$ optimizer, we consider the same 500 random graphs that we used in Fig.~\ref{fig:Naive_vs_Heur1} and inspect if the new optimization levels reduce the CNOT counts further. (Again, we test three variants of Heuristics $\# 2$ which have to do with whether we perform back-substitution globally and whether we exhaust the conditions for free photon absorption. From the three methods, we select the one that gives fewer CNOTs.) We set a future cutoff of 2 photon absorptions and constrain the available emitters we inspect as potential absorbers to at most 5. We also allow recursions to enter the new optimization steps of Heuristics $\# 2$ if no more than half of the graph has been consumed yet. We set this last condition so that we explore choices early on in the backwards generation.  For $n_p> 20$, we further turn off the optimization part that involves a search over local gates and emitter pairs for disconnected components, because this is a more computationally intensive step, and because we already inspect emitter choices recursively. In Fig.~\ref{fig:Comp_2_heuristics_w_Naive}(a), we show the mean number of emitter CNOTs obtained by the two Heuristic methods and compare them against the Naive optimizer. Based on Fig.~\ref{fig:Comp_2_heuristics_w_Naive}(b), the Heuristics $\# 2$ optimizer gives an extra boost to the mean and maximum reduction of up to $\sim 6\%$ compared to Heuristics $\# 1$. It is noteworthy that despite the small cutoff on emitters we inspect per photon absorption, and the small future cutoff, we still obtain further reductions in emitter CNOTs. (Recall that the Heuristics $\#1$ method inspects all available emitters up to $n_p+n_e<30$.)  In Fig.~\ref{fig:Comp_2_heuristics_w_Naive}(c), we show the number of emitter CNOTs obtained by the three methods for graphs of size $n_p=18$. Compared to the Naive scheme, the Heuristics $\# 2$ method requires even $\sim  50\%$ fewer emitter CNOTs, and the reduction is more pronounced for graphs that scored a higher CNOT count in the Naive method.

\subsection{Computational runtime}

\begin{figure}[!htbp]
    \centering
    \includegraphics[scale=0.74]{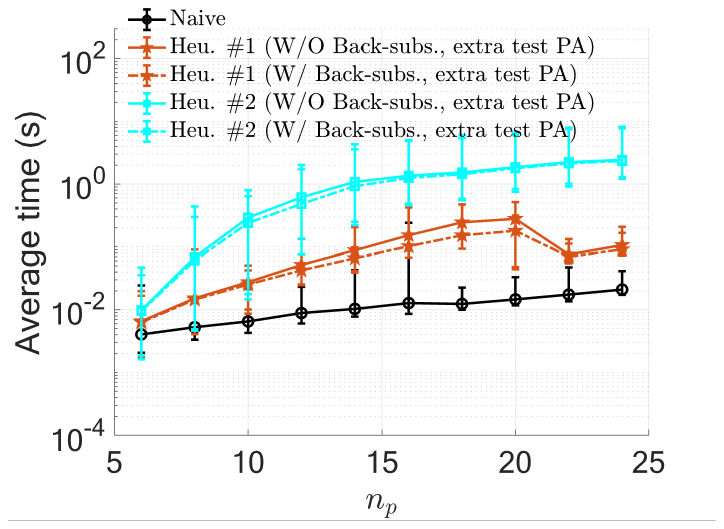}
    \caption{Average runtime of Naive (black), Heuristics $\#1$ (orange), and Heuristics $\# 2$ (cyan) optimizers for finding the generation circuit of a photonic graph of size $n_p\in[6,24]$. We sample 200 graphs per $n_p$. The error bars mark the minimum and maximum computational time for each $n_p$. }
    \label{fig:Runtime}
\end{figure}

Let us briefly discuss the computational time of the different schemes we introduced. All our results were simulated with MATLAB 2021a on an 8-core Apple M1 Chip with 8 GB of RAM. In Fig.~\ref{fig:Runtime}, we show the average runtime for extracting the CNOT counts for graphs of up to $n_p=24$, and for 200 samples per $n_p$. The fastest method is the Naive approach (black line), which involves no substantial optimization. In particular, we obtain the generation circuit of a 100-node photonic graph state within 0.3-0.4~s on average (see Appendix~\ref{App:Runtime}). 
The overhead of the Heuristics $\#1$ method (orange line) leads to an increase in the runtime, which is still well below 1~s up to $n_p=24$.  The Heuristics $\# 2$ method (cyan line) is more computationally intensive but can still run fast, depending on the options we set. Here, we set a cutoff on emitters to be inspected for a given photonic absorption of up to 3, and a future cutoff of 2 photon absorptions. We also enable the optimizer to recursively call itself if we have not yet consumed more than half of the total number of photons. This method also runs approximately within the range of a few seconds for up to $n_p=24$. The solid lines for the Heuristic methods correspond to no back-substitution and do not exhaust all tests for free photon absorption. The dashed lines include these extra steps. We see that, on average, including these refinements globally does not affect the runtime substantially, and this is owed to how fast the optimizer will detect a heuristic pattern and make a decision to exit before further inspection steps. Also, note that the runtime of the Heuristics $\# 2$ optimizer approximately saturates for particular ranges of $n_p$. This is not a coincidence, and is owed to the cutoffs we set. This is another advantage of our optimizers, because based on the optimization levels we include and the cutoffs we set, we can have very good control on the runtime. In other words, our algorithms can be made fixed-parameter tractable with respect to the number of emitters (which typically increases with increasing graph size) since the choices that arise are mainly owned to the increase in emitter resources. Additionally, we have optimized the runtime of several subroutines so that we achieve fast processing of the graphs. For example, in Appendix~\ref{App:Algorithms} we provide Algorithm~\ref{alg:rowsum} to perform the phase update of the rowsum faster than in Ref.~\cite{GottesmanPRA2004}. Throughout the search for disconnected subgraphs, we avoid performing the phase updates of stabilizers, because those are unnecessary (we only care about extracting the adjacency matrix after the gates). Once we find the successful pattern (if it exists), we then perform the phase updates to keep the entire procedure consistent. Finally, let us comment that our optimizers achieve a similar performance irrespective of how dense the graph is (see further Appendix~\ref{App:CNOT_cnts_Vs_Edge_Prob}). It is very important to keep our optimization unbiased, and not rely on subtle metrics such as graph density, since as we will see later on, two locally equivalent graphs can differ in their edge density, but incur exactly the same entangling gate cost in their preparation.

\section{Properties of LC graphs and summary of graph theory tools \label{Sec:CircleGraphsandLC}}
In this section, we will inspect in detail the role of local gates in the circuit optimization. As we already mentioned, LCs on a graph  translate into local Clifford gates on the respective stabilizer state. To proceed with our discussion, we will first mention useful graph theory concepts from the literature and use them to understand the structure and properties of particular LC families.

\subsection{Local complementation and LC equivalence}

Local complementations were initially introduced by Kotzig~\cite{Kotzig1968,KotzigAnnalsOfDiscreteMathematics1980} and further studied later on by Bouchet~\cite{BouchetCombinatorica1987,BouchetDiscreteMathematics1993,BOUCHETJournalofCombTheory1994}, and Van den Nest~\cite{VanDenNestPRA2004,VanDenNestPRA2004Second}. The LC about vertex $v$ has the effect of inverting the edges in the neighborhood of $v$, denoted as $N_v$ [see Fig.~\ref{fig:CircleG}(a)]. The local operation $U_v^{\text{LC}}$ of Eq.~(\ref{Eq:LC_Gate}) provides a one-to-one correspondence between stabilizer and graph states as long as the tableau is already in canonical form (i.e., the $\mathcal{S}_Z$ part is the adjacency, and the $\mathcal{S}_X$ part is the identity). In other words, we can write more accurately:
\begin{equation}
    |G*v\rangle=U_v^{\text{LC}}|G\rangle.
\end{equation}
If one performs all possible LCs on a given graph, this creates a family of graphs often termed the LC orbit. In Appendix~\ref{App:Algorithms}, we provide  Algorithm~\ref{alg:MapOutOrbit} that creates the LC orbit of a given graph by either discarding or not the isomorphs (i.e., graphs that look the same but have different node labeling). Bouchet developed an algorithm for testing if two graphs are equivalent under any series of LCs~\cite{BouchetDiscreteMathematics1993}. This equivalence can be tested in polynomial time by solving a linear system of equations; the complexity of this algorithm is $\mathcal{O}(|V(G)|^4)$ due to the need to perform Gaussian elimination to find a basis for the solution space.

\subsection{Circle and split-free graphs}
\begin{figure}[!htbp]
    \centering
    \includegraphics[scale=0.66]{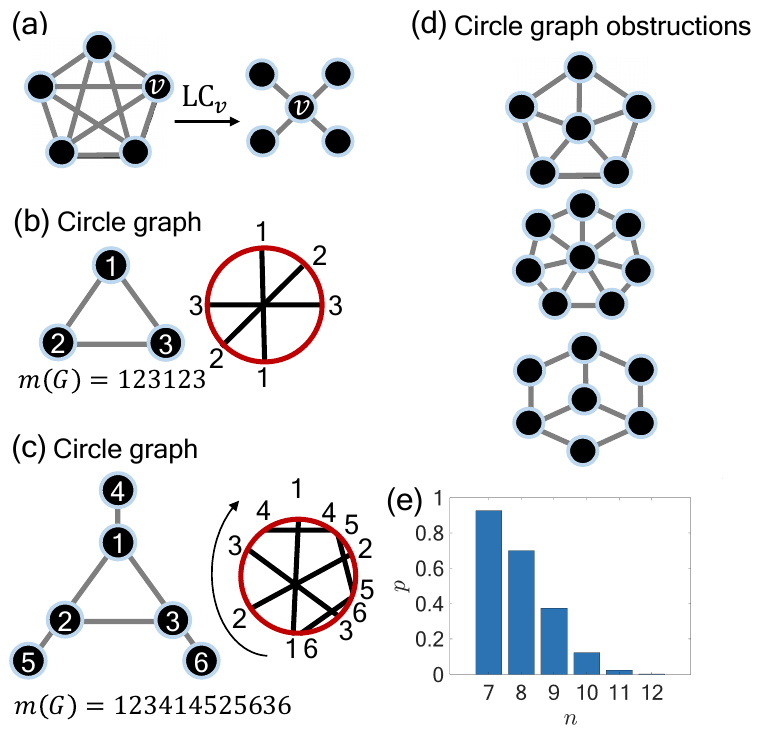}
    \caption{(a) LC operation about $v$ and resulting graph. (b) Circle graph representation of $K_3$. (c) Circle graph representation of RGS of three core nodes. (d) Circle graph obstructions of Ref.~\cite{BOUCHETJournalofCombTheory1994}. (e) Probability to find a circle graph as a function of its size. We sample 5000 graphs per $n$.}
    \label{fig:CircleG}
\end{figure}

\begin{table}[!htbp]
    \centering
    \begin{tabular}{|c|c|}
    \hline
         $P_n$& Path graph  \\
         \hline
         $K_n$& Complete graph \\
         \hline
         $K_n^n$& Unencoded repeater graph \\
         \hline
         $C_n$ & Cycle graph \\
         \hline
         $S_n$ & Star graph\\
         \hline
    \end{tabular}
    \caption{Notations of graph families and the family each one represents. These families are examples of circle graphs.}
    \label{tab:GraphFam}
\end{table}
Several properties of LC graphs have been formulated from the perspective of a unique family of graphs called circle graphs. Formally, they are defined as graphs whose nodes are represented as the chords of a circle, and whose edge set is represented as intersections of chords. Two chords intersect if there is an edge between the nodes they represent in $E(G)$. A circle graph is also described via an alternance (or double occurrence) word, such that if $(v,w)\in E(G)$, then the pattern $v\dots w \dots v \dots w$, or $w\dots v \dots w \dots v$ exists in the word. For example, the path graph $P_2$ is represented by the word $m(P_2)=v_1v_2v_1v_2$. We also show the circle graph representation of the complete graph $K_3$ in Fig.~\ref{fig:CircleG}(b). The alternance word is formed by reading the labels on the circle clockwise. An LC about a vertex $v$ reverses the subword between the two occurrences of $v$. In other words, if $m(G)=A v B v C$, then $m(G*v)=A v\bar{B} v C $, where $\bar{B}$ is a subword that contains the nodes of $B$ in reverse order.

Circle graphs have been studied from both a graph theory and quantum information theory perspective~\cite{AxelJofMathPhysics2020,DahlbergQuantum2020,DahlbergQuantumScieAndTech2020,DAHLBERGInfProcessLett2022}. Among others~\cite{Gabor1985,Naji1985,SPINRAD1994,GASSE1997,Gioan2014}, Bouchet introduced an algorithm to recognize circle graphs in polynomial time~\cite{BouchetCombinatorica1987}. For completeness, we describe this algorithm in Appendix~\ref{App:Circle_Graphs}. Here, we give a brief overview of how the algorithm works. To understand the circle graph recognition algorithm of Bouchet, we need to first review some graph theory concepts, namely splits, and split-free graphs.

Let us begin with the definition of a split. A split in a graph is a partition of the nodes into two sets $V_1,V_2$ with $|V_1|,|V_2|\geq 2$, such that $A\subseteq V_1$ and $B\subseteq V_2$ together form a complete bipartite graph [see also Appendix~\ref{App:Circle_Graphs}]. A split-free graph is also known as prime (note that primality can have a different definition, but we use it here to refer to graphs that have no splits). For a prime graph of order $n\geq 6$ there exists a node $v$ such that at least one out of the following subgraphs (known as vertex minors) is always prime~\cite{BouchetDiscreteMathematics1993}:
\begin{equation}\label{Eq:VertexMinors}
    G\backslash v, G*v \backslash v, G*vwv\backslash v,~~ w\in N_v.
\end{equation}
In other words, a prime graph necessarily contains a vertex minor which is also a prime graph~\cite{AllysCombinatorica1994,OumDiscreteAppliedMathematics2024}.
In the quantum information community, the above three graphs are known as the graphs obtained after a $Z$-, or $Y$-, or $X$-measurement of $v$. The operation corresponding to an $X$-measurement is not unique, which is consistent with the fact that one can choose any $w\in N_v$~\cite{HeinPRA2004}. The implication of the prime reduction is that there exists a closure condition, such that the vertex minors of split-free graphs remain split-free. Splits are also invariant under LCs. These important features form the basis of recognizing circle graphs~\cite{BouchetCombinatorica1987}, or enumerating their LC orbit~\cite{BouchetDiscreteMathematics1993}.

The circle graph recognition algorithm first tests if the graph is prime. If it is not prime, the graph is partitioned into prime components and the alternance word is built for each piece. 
The construction of the word uses the reduction principle of Eq.~(\ref{Eq:VertexMinors}). In the end, the total  word is formed by piecing together the smaller words. If, at any step, the alternances cannot be satisfied, then some of the prime subgraphs is not a circle graph, and hence the input graph is also not a circle graph~\cite{OumDiscreteAppliedMathematics2024}. Bouchet also proved that non-circle graphs have as a vertex minor a graph which is isomorphic to at least one graph that belongs in a finite set of graphs~\cite{BOUCHETJournalofCombTheory1994}. The members in this family of vertex minors are often referred to as obstructions. For circle graphs, these vertex minors are called circle graph obstructions~\cite{BOUCHETJournalofCombTheory1994}, shown in Fig.~\ref{fig:CircleG}(d). 

A unique property of circle graphs is that one can count the size of their orbit, i.e., the number of graphs, which together with $G$, are LC equivalent~\cite{BouchetDiscreteMathematics1993}. Ref.~\cite{AxelJofMathPhysics2020} showed that this counting task maps to the enumeration of Euler Tours in 4-regular multigraphs, which is known to be a $\#$P-complete problem. While a single Euler Tour can be found in poly-time (via a depth-first search), counting all of them is hard. Although the problem is, on average, $\#$P-complete, it is possible to count the LC-orbit size efficiently for certain circle graphs. Bouchet~\cite{BouchetDiscreteMathematics1993} showed how to count analytically the orbit of the cycle, $C_n$, the path, $P_n$, and the 5-wheel, $W_5$, graph. Here, we will show how to count the orbit of complete and repeater graphs, which also belong in the family of circle graphs.

\subsection{Circle graph statistics}

It is interesting to study how many graphs of order $n$ are circle graphs. For $n\leq 5$, all graphs are circle graphs as the first obstruction is of order $n=6$. For $n=6$, all graphs are circle graphs besides those belonging to a single LC class. Although circle graphs are recognized in poly-time, it is hard to find the exact number of them for a large order $n$ due to the exponential increase in the number of connected graphs. However, we can still sample random graphs and count how many of them are circle graphs. We generate 5000 random graphs per $n$, for $n\in[7,12]$, and apply Bouchet's recognition algorithm. For every $n$, we count the occurrence of circle graphs, and to get the probability, we divide by the number of samples. The results are plotted in Fig.~\ref{fig:CircleG}(e). The probability of encountering a circle graph drops quickly with $n$. This is expected because, for $n\geq 7$, the probability to encounter an obstruction becomes large. For $n=12$, the probability that we find a circle graph is vanishingly small.

Since most graphs are non-circle graphs, it is interesting how to formulate the problem of counting the size of their orbit and what is the complexity of this problem. To the best of our knowledge, the complexity of counting the size of, or generating the LC orbit for non-circle graphs is an open problem.

In the next section, we will study particular families of circle graphs, such as the complete, star graph, and repeater graph. These are examples of graphs that remain circle graphs irrespective of their size $n$, and we can find the scaling of their LC orbit analytically.

\section{Size of LC orbit of known families of graphs\label{Sec:LCOrbitSize}}

\subsection{Summary of the LC orbit counting algorithm}
\begin{figure*}[!htbp]
    \centering
    \includegraphics[scale=0.5]{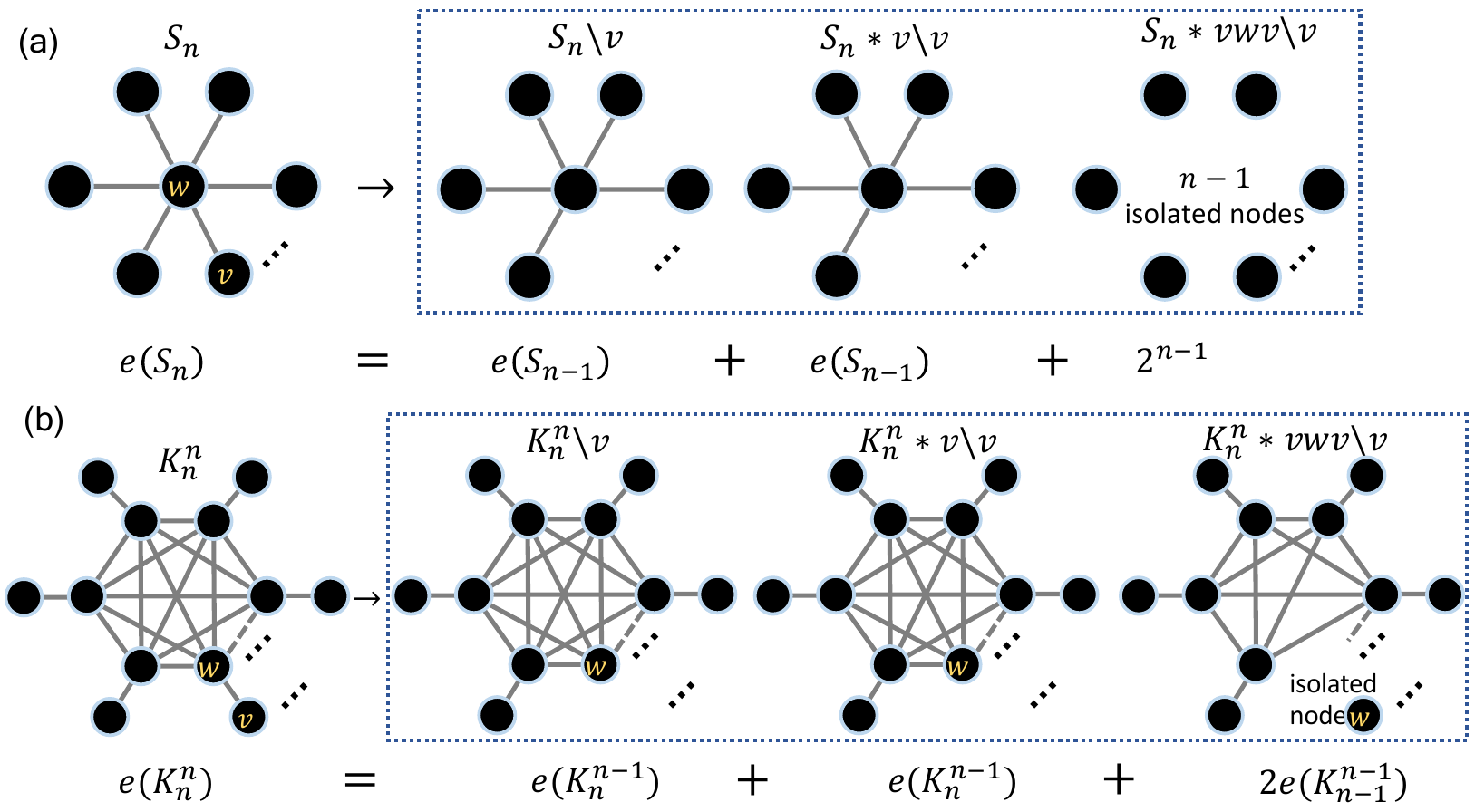}
    \caption{(a) Star graph, $S_n$, and its decomposition. (b) Repeater graph, $K_n^n$, and its decomposition.}
    \label{fig:ReductionStarRGS}
\end{figure*}
The LC orbit counting algorithm for circle graphs was developed by Bouchet in Ref.~\cite{BouchetDiscreteMathematics1993} and counts the number of local complements of a graph $G$ (including $G$). The size of the LC orbit includes isomorphs and is defined as the ratio:
\begin{equation}
    l(G):=\frac{e(G)}{k(G)},
\end{equation}
where $e(G)$ can be found by the recursive relation:
\begin{equation}\label{Eq:e(G)}
    e(G)=\begin{cases}
    2e(G\backslash v), \text{if } v \text{ is isolated}\\
    e(G\backslash v)+e(G*v\backslash v)+e(G*vwv\backslash v), \text{else}.
    \end{cases}
\end{equation}
A node is isolated if it has no incident edges. The index $e(G)$ is recursively calculated based on this reduction formula [see relation with Eq.~(\ref{Eq:VertexMinors})], and it corresponds to the number of Euler Tours in 4-regular multigraphs. For $|V|=1$, the index $e(G)$ is initialized with $e(G)=2$~\cite{BouchetDiscreteMathematics1993}. The index $k(G)$ relates to the dimension of the bineighborhood space (see also Ref.~\cite{AxelJofMathPhysics2020}). Because the meaning of this index and how to calculate it is more involved, we provide the details in Appendix~\ref{App:index_k(G)}. In our following results, we will report the value of $k(G)$ for certain graph families and refer the reader to our proofs in Appendices~\ref{App:k(G)_of_Kn},~\ref{App:k(G)_of_RGS}.

\subsection{Complete and star graphs}

Let us begin with the well-known families of complete, $K_n$, and star, $S_n$, graphs. To prove that a complete graph is a circle graph, it suffices to find a valid alternance word. It is easy to verify that the word for $K_n$ is $m(K_n)=v_1v_2\dots v_n v_1v_2\dots v_n$. For example, it contains the alternance $v_1 v_2 \dots  v_1 v_2$ up to $v_1 \dots v_n v_1 \dots  v_n$, and similarly, all other expected alternances exist in this word [see also Fig.~\ref{fig:CircleG}(a)]. The star graph, obtained by complementing about any node of $K_n$, is also a circle graph. This is because circle graphs are closed under the LC operation. The LC operation on the word reverses the subword between the two occurrences of the node. If we perform the LC operation about node $v_1$ then we will find the word $m(K_n*v_1)=m(S_n)=v_1 v_n v_{n-1} \dots v_2 v_1 \dots v_n$.

Now let us prove analytically the scaling of the LC orbit of $K_n$ and $S_n$. In Appendix~\ref{App:k(G)_of_Kn}, we show that $k(K_n)=k(S_n)=2^{n-1}$. The next step is to find $e(K_n)=e(S_n)$. We know that $S_n$ is LC-equivalent to $K_n$, and it will be easier to analyze the former, as it has $n-1$ leaves. Then, we can use the fact that if $G$ and $G'$ are LC-equivalent, it holds that $e(G)=e(G')$. If we pick a vertex $v$ as any of the leaves of $S_n$ [see Fig.~\ref{fig:ReductionStarRGS}(a)] we find:
\begin{equation}
    e(S_n)=2e(S_n-1)+2^{n-1}.
\end{equation}
The last term arises due to the $n-1$ isolated nodes [see Eq.~(\ref{Eq:e(G)})]. Letting $T(n)=e(S_n)$, we have the recursion:
\begin{equation}
    T(n)=2T(n-1)+2^{n-1},
\end{equation}
and solving this we get [see Appendix~\ref{App:CompleteGraphRECURSION}]:
\begin{equation}
    T(n)=e(S_n)=2^{n-1}(n+1).
\end{equation}
Therefore, we see that the size of the LC orbit of the complete graph (and star graph) scales linearly:
\begin{equation}
    l(K_n)=\frac{e(K_n)}{k(K_n)}=\frac{2^{n-1}(n+1)}{2^{n-1}}=n+1.
\end{equation}

\subsection{Repeater graphs}
To prove that the RGS is a circle graph, it suffices to construct an alternance word. Let the core nodes be labeled as $v_1,\dots,v_n$ and the outer leaves as $w_{1},\dots,w_n$, where we traverse clockwise first the core nodes and then the leaves. The subword of the core nodes can be constructed as $v_1\dots v_n v_1\dots v_n$. For each leaf $w_j$ of core node $v_j$, we place one $w_j$ before and one after the second occurrence of $v_j$ to form $m(K_n^n)=v_1\dots v_n w_1v_1 w_1\dots w_nv_nw_n$ [see also Fig.~\ref{fig:CircleG}(c)]. This word represents any RGS isomorph if we replace $v_j$ and $w_j$ with the actual labels $1,\dots,2n$.

To find the size of the LC orbit of an RGS, we use our above results for complete graphs. We denote the repeater graph of $2n$ nodes by $K_n^n$, which can be thought of as a complete graph of $n$ core nodes to which we attach one leaf node per core node. The superscript indicates the number of leaf nodes, so that $K_n^0= K_n$.
In Fig.~\ref{fig:ReductionStarRGS}(b), we show the first decomposition of $K_n^n$ if we remove one of its leaves. The operation $*vwv$ is an edge pivot, which, in this case, simply exchanges the position of $v$ and $w$; the removal of $v$ leaves $w$ isolated. Thus, if we were to continue the decomposition of this last graph shown in Fig.~\ref{fig:ReductionStarRGS}(b), we would then remove $w$, and we would multiply $e(K_{n-1}^{n-1})$ by 2 [see Eq.~(\ref{Eq:e(G)})]. In Appendix~\ref{App:RGSRECURSION}, we show that $e(K_n^n)$ is given by:
\begin{equation}
    e(K_n^x)=2^{x}\sum_{k=0}^x \begin{pmatrix}
        x \\ k
    \end{pmatrix}e(K_{n-k}^0), n-1 \geq x \geq 1, n\geq 3.
\end{equation}
Based on our analysis of $K_n$, we know that $e(K_{n-k}^0)=2^{n-k-1}(n-k+1)$. Thus, solving the above sum for $x=n-1$, and solving a recursion (see Appendix~\ref{App:RGSRECURSION}), we get:
\begin{equation}
    e(K_n^n)=6^{n-1}(3+2n)+2^{n-1}.
\end{equation}
At this point, we can also verify previous results. For $e(K_2^2)=e(P_4)$, the above expressions gives 44, and for $e(K_1^1)=e(P_2)=6$, which are the correct results for path graphs [see Ref.~\cite{BouchetDiscreteMathematics1993} for $e(P_n)$]. We further tested the expression numerically by recursively evaluating Eq.~(\ref{Eq:e(G)}) and found that the two results agree (see Appendix~\ref{App:RGSOrbit_Enumer}).

In Appendix~\ref{App:k(G)_of_RGS} we further prove that $k(K_n^n)=2^n$, giving rise to the following scaling of the LC orbit:
\begin{equation} \label{Eq:RGS_Orbit}
    l(K_n^n)=\frac{1+3^{n-1}(3+2n)}{2}.
\end{equation}
We see that the LC orbit of the RGS scales exponentially, in contrast to the linear scaling of $K_n$. The reason for this exponential scaling is because the 4-regular multigraphs of the RGS are highly symmetric and lead to an exponential number of Euler Tours (see Appendix~\ref{App:RGSOrbit_Enumer}). 

Although the scaling is exponential, if we generate the orbits of small-sized RGSs, we will notice that they contain a large number of isomorphs. If we can generate the non-isomorphic LC orbit, then it would be easier to analyze the graphs within the orbit and find their preparation cost. By inspection, we find that the pattern of consecutive LCs to generate the unique non-isomorphic graphs of RGSs is:
\begin{equation}
\begin{split}
    \{\emptyset\}\cup\{v_1\}\cup\{v_2\}\cup \{v_{n+1}\}\cup\{v_3\}\cup\{v_4\}\cup\{v_{n+3}\}\cup \dots \\ \cup\{v_{n-2}\}\cup\{v_{n-1}\}\cup\{v_{2n}\},
    \end{split}
\end{equation}
where $v_1,\dots,v_n$ are core nodes, and $v_{n+1},\dots,v_{2n}$ are leaf nodes, and we terminate at a core node for odd $n \geq 3$ or at a leaf node for even $n\geq 4$, without repeating indices. To give an example, let us consider the $K_3^3$ graph, with $(v_1,v_2,v_3)=(1,2,3)$, and $(v_{4},v_5,v_{6})=(4,5,6)$. The LC operations we apply based on the above formula are:
\begin{itemize}
    \item $K_3^3* \emptyset$
    \item $K_3^3* 1$
    \item $K_3^3 *1*2$ 
    \item $K_3^3 * 1*2*4$ 
    \item $K_3^3 *1 * 2 * 4 * 3$.
\end{itemize}
Following this pattern, we generate an orbit of linear size. We provide  Algorithm~\ref{alg:MapOutRGSOrbit} in Appendix~\ref{App:Algorithms}, which prepares the non-isomorphic orbit for any RGS size. We also verified with brute-force generation of the entire orbit (that includes isomorphs) and then removal of the isomorphic graphs that we obtain the same graphs as in the case of using the above pattern. By numerical inspection, we find that the size of the non-isomorphic LC orbit of $K_n^n$ scales as:
\begin{equation}
 l^{\text{non-iso}}(K_n^n)=\frac{(3(2n+1)-(-1)^{n+1})}{4},  
\end{equation}
which is the sequence of numbers non-divisible by 3, known as A001651~\cite{A0001651Seq}. In our case, we shift this sequence because we start from $n=3$. More surprisingly, the size of the RGS orbit when we exclude isomorphs is independent of the number of leaves per core qubit, and a very similar pattern can be used to generate the orbit of an RGS with multiple leaves per core node. In Appendix~\ref{App:Algorithms}, we also provide Algorithm~\ref{alg:MapOutRGSOrbitManyLeaves}, which generates the LC orbit of an RGS with multiple leaves. Of course, if we were to perform the enumeration of the orbit that includes the isomorphs, we would find an even bigger scaling than in Eq.~(\ref{Eq:RGS_Orbit}), because we would have to do more reductions compared to the RGS with a single leaf per core qubit. It is interesting that this structure of adding more leaves per core qubit does not at all affect the pattern we can use to generate the non-isomorphic orbit, and generalizations of this feature could be used as a guide to create more easily the non-isomorphic orbits of graphs with multiple leaves.

\begin{figure*}[!htbp]
    \centering
    \includegraphics[scale=0.74]{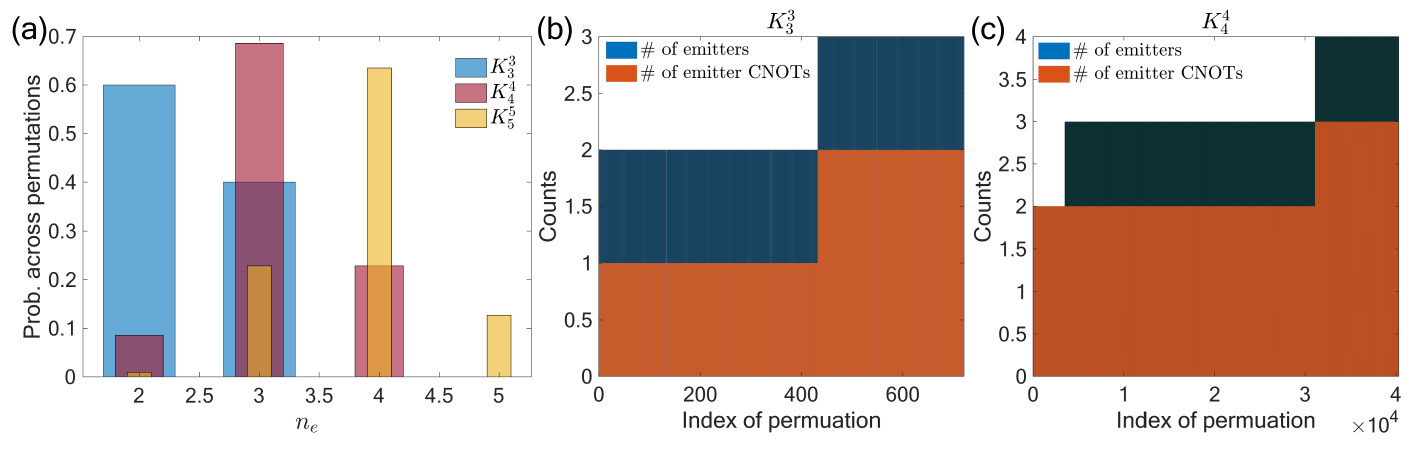}
    \caption{(a) Histogram of emitter resources 
    for preparing $K_3^3$ (blue), $K_4^4$ (pink) and $K_5^5$ (yellow). The probability is calculated across all isomorphs. (b) Emitter and CNOT counts versus emission ordering for $K_3^3$. (c) Same as in (b), for $K_4^4$. The CNOT counts were obtained via the Heuristic $\#1$ optimizer. } 
    \label{fig:RGS_Emitters_vs_CNOTs}
\end{figure*}
\section{Optimal resources for unencoded and 
encoded repeater graphs\label{Sec:OptPrepRGS}}

In this section, we focus on the preparation of repeater states and study the relation between emitter  and CNOT resources. We begin by finding the number of emitters needed to prepare an unencoded RGS of $n$ core qubits. In Fig.~\ref{fig:RGS_Emitters_vs_CNOTs}(a), we show a histogram of emitter resources, for any emission ordering, and up to $K_5^5$. The upper bound of emitters we need to prepare an RGS equals $n$. This is expected because, in the worst-case scenario, we prepare $K_n$ on $n_e=n$ emitters, and then we perform two rounds of emissions before finally measuring the emitters. The first emission round creates the core photons, and the second one creates the leaf photons. For this case of maximal number of emitters used, we obtain easily the optimal number of emitter CNOTs. To create a complete graph on $n_e=n$ emitters, we need a minimum of $n_e-1$ emitter CNOTs. The emission rounds can be performed without intermediate emitter CNOTs (see generation circuit in Appendix~\ref{App:RGS_Parallel_Emissions}). If we view this generation circuit backwards, all emitter CNOTs are performed in the final disentanglement step, which requires at least $\tilde{n}_e-1$ emitter CNOTs, where $\tilde{n}_e$ is the number of emitters that are still entangled with each other right after all photons have been absorbed. In Appendix~\ref{App:Algorithms}, we provide Algorithm~\ref{alg:FinalDisentanglement} which performs the last disentanglement step.

In Figs.~\ref{fig:RGS_Emitters_vs_CNOTs}(b),(c), we plot the  emitter versus CNOT counts across any isomorph of $K_3^3$ and $K_4^4$. For illustration purposes, we sort the data in terms of increasing emitter resources. We obtain the CNOT counts using Heuristics $\# 1$ optimizer, for which we set the following options: i) we do not enable global back-substitution, and ii) we enable the extra test for free photon absorption. We find that for both $K_3^3$ and $K_4^4$ the optimal number of CNOTs is fixed across isomorphs that require the same number of emitters. This is non-trivial, because among the isomorphs that require the same number of emitters we find non-LC equivalent graphs, but they yield the same number of CNOTs. It is also remarkable that our least expensive optimizer can minimize globally the CNOT cost of RGSs for any emission ordering.

\begin{figure*}[!htbp]
    \centering
    \includegraphics[scale=0.67]{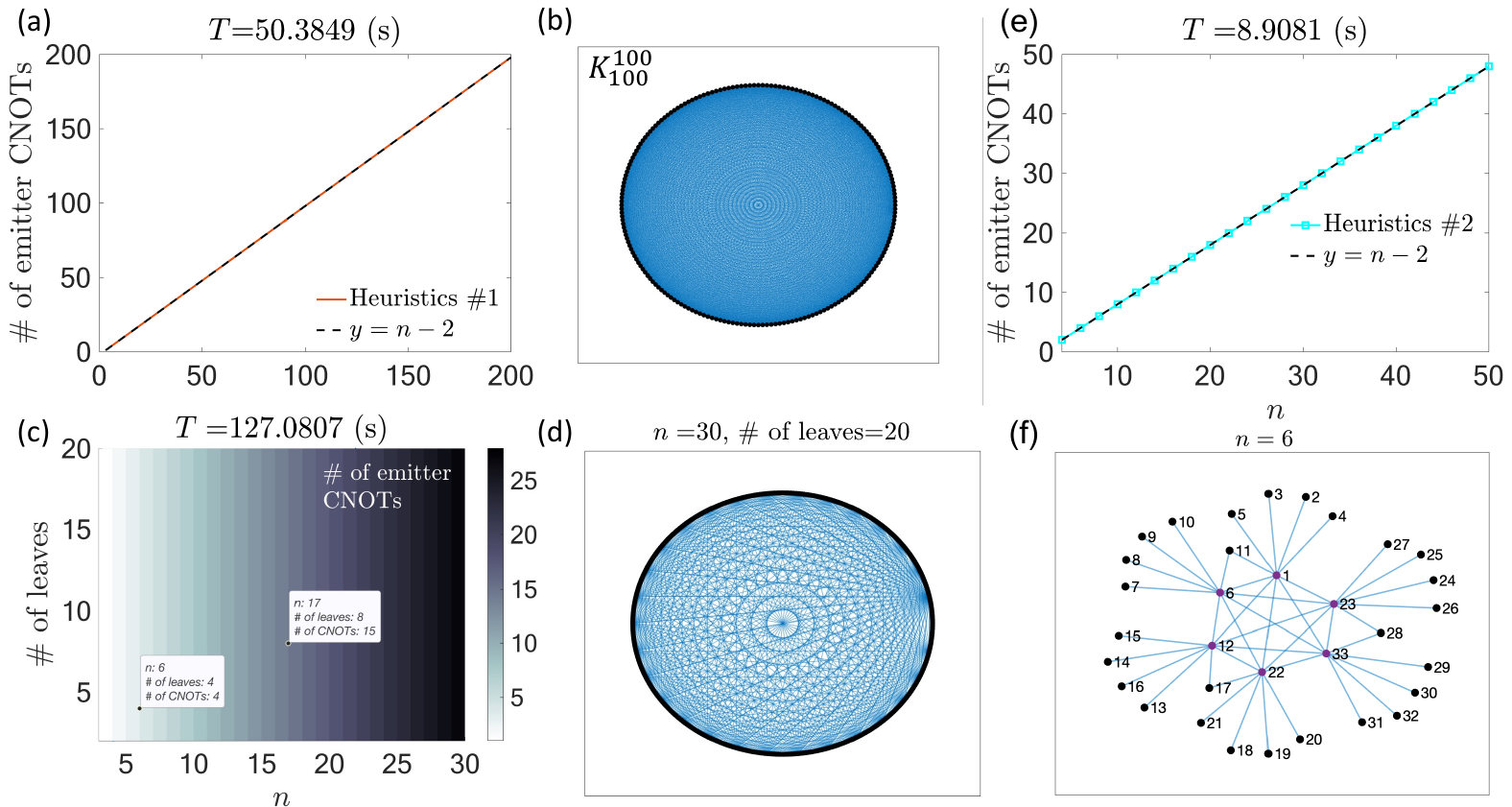}
    \caption{(a) Optimal number of emitter CNOTs for each size of $K_n^n$, given the optimal emission ordering. The orange line is obtained from Heuristics $\# 1$. The dashed line is $y=n-2$. $T\approx 50$~s is the total computational time to process all RGSs up to $n=200$. (b) Repeater graph of 100 core qubits. (c) Emitter CNOTs as a function core qubits and leaves per core qubit of the RGS. (d) Repeater graph with 30 core qubits and 20 leaves per core qubit. (e) Number of emitter CNOTs for encoded RGS with branching parameter $b_0=2$ and $b_1=4$ ($b_1$ is the number of leaves per core qubit). (f) Encoded RGS with $n=6$, $b_0=2$, and $b_1=4$. All nodes correspond to photons. The purple nodes indicate the nodes for the $b_0=2$ branching parameter. $T$ is the total computational time.}
    \label{fig:RGS_Opt_Scaling}
\end{figure*}

Of particular interest is the optimal scaling of emitter CNOTs for the ordering that requires only two emitters, irrespective of the size of the RGS. Ref.~\cite{BikunnpjQI2022} refers to the optimal ordering as the natural emission ordering. In Appendix~\ref{App:Algorithms}, we provide Algorithm~\ref{alg:RGSOrdering}, which generates an alternative optimal emission ordering by exchanging the labeling of two nodes, and again requires two emitters. Figure~\ref{fig:RGS_Opt_Scaling} shows the emitter CNOTs if we use the ordering of Algorithm~\ref{alg:RGSOrdering}, for RGSs of size $n\in[3,200]$. We note that the optimal number of emitter CNOTs scales linearly with the number of core qubits and is given by $n-2$. (We arrive at the same result if we use instead the natural emission ordering.) This linear scaling has also been mentioned in Ref.~\cite{Ghanbari2024arXiv}, with the difference that there the authors claim that particular graph structures in the LC orbit of the RGS achieve this optimal scaling. The authors perform a search over labelings and LC orbits to identify the optimal graph. They propose two different structures from the LC orbit of the optimally labeled RGS, depending on whether the number of core qubits is even or odd. Based on their optimization, they can analyze RGSs of up to 55 photons in total, where for an RGS of 55 photons, their random search algorithm takes 100 seconds, and an informed method based on edge minimization takes about 1 second. In our case, we do not even need to search the labeling of the repeater graphs or their LC orbits because we know the optimal emission ordering. Notably, we can find the optimal emitter CNOT counts for RGSs of up to 400 photons within $\sim$50~s in total (with parallel processing), using our least expensive optimizer, Heuristics $\# 1$. In Fig.~\ref{fig:RGS_Opt_Scaling}(c), we plot the emitter CNOTs as a function of core and leaf nodes of an RGS, where each core node has more than one leaf. In Appendix~\ref{App:Algorithms},
we provide Algorithm~\ref{alg:RGSDepth1Ordering} which generates the optimal emission ordering for any RGS size with multiple leaves. We see that the number of leaves does not affect the CNOT counts. Using our Heuristics $\# 1$ optimizer, we find the same $n-2$ scaling, within 127~s of processing all 532 repeater graphs with $n\in[3,30]$ and number of leaves within $[2,20]$. In Appendix~\ref{App:Circuits_RGS_Many_Leaves}, we provide the circuits to generate an RGS of any number of leaves and core nodes.

Although we showed how to prepare repeater graphs optimally, in order for these states to be useful for quantum communications they need to be encoded. One possible encoding is to replace each core qubit by a tree~\cite{AzumaNatCommun2015}. As an example, we consider a tree with branching parameters $b_0=2$ and $b_1=l$. The $b_1$ nodes are the leaves of the RGS. The task of finding the optimal ordering is once again straightforward for encoded RGSs. In Appendix~\ref{App:Algorithms} we provide Algorithm~\ref{alg:RGSEncodedOrdering}, which generates the adjacency matrix with emission ordering that requires two emitters, for any encoded RGS with an even number $n\geq 4$ of core qubits, to each of which we attach $b_1=l$ leaves. Each pair of core qubits corresponds to the $b_0=2$ nodes, and they connect with the top node of a tree. We display an example for $n=6$, $l=4$ in Fig.~\ref{fig:RGS_Opt_Scaling}(f). We further show the optimal CNOT counts in Fig.~\ref{fig:RGS_Opt_Scaling}(e), which we obtain using Heuristics $\# 2$ algorithm (we set a future cutoff of 2 steps, and do not enable further recursions). We find that similar to the unencoded case, the encoded RGS can be prepared with $n-2$ emitter CNOTs, irrespective of the number of leaf qubits per core qubit.

\section{Study of LC orbits \label{Sec:RandomLCOrbits}}
We now turn our attention to the study of LC orbits of various graphs. We will show how to achieve the same optimal CNOT cost across the orbit using our optimizers.

\subsection{LC orbits of RGSs}

Let us begin with the LC orbits of RGSs. We assume the optimal emission ordering of Algorithm~\ref{alg:RGSOrdering}, which requires two emitters to prepare an RGS of any size. We also make use of  Algorithm~\ref{alg:MapOutRGSOrbit} to generate only the non-isomorphic RGS orbit~(see Appendix~\ref{App:RGSOrbit_Enumer} for examples of the LC orbit). Although we focus only on the non-isomorphs, our following results hold for the entire LC orbit.

\begin{figure}[!hbtp]
    \centering
    \includegraphics[scale=0.38]{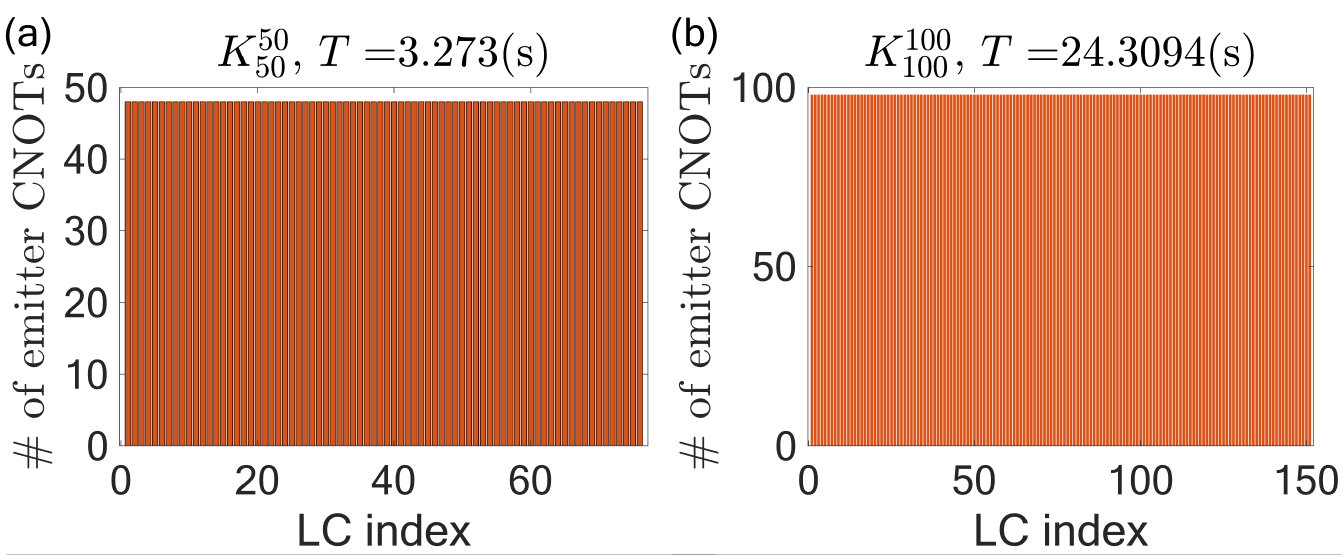}
    \caption{Number of emitter CNOTs for the LC orbit of $K_{50}^{50}$ (a), and $K_{100}^{100}$ (b), using the Heuristics $\#1$ optimizer and the optimal emission ordering. $T$ is the total computational time to extract the number of emitter CNOTs (using parallel processing).}
    \label{fig:RGS_Orbit}
\end{figure}

\begin{figure*}[!htbp]
    \centering
    \includegraphics[scale=0.64]{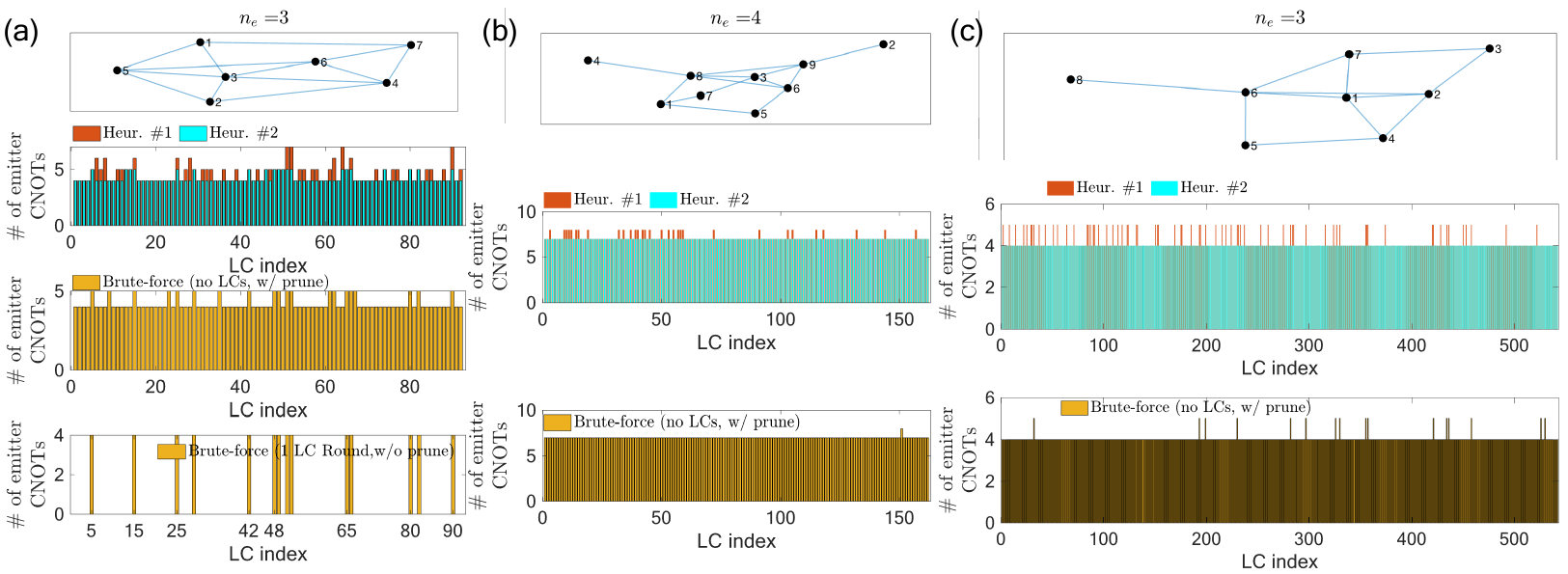}
    \caption{Minimizing the emitter CNOTs across LC orbits of random graphs. (a) Top: Graph representative of LC orbit consisting of $n_p=7$ photons. Middle: Emitter CNOTs across the orbit obtained by the Heuristics $\#1$ (red), Heuristics $\#2$ (cyan), or Brute-force method, w/o allowing LC rounds, (yellow). Bottom: Number of emitter CNOTs obtained by the Brute-force method if we allow 1 LC round, for the graphs for which we did not achieve the minimum CNOT count w/o the LC round for any of the optimizers. (b) Top: Graph representative of LC orbit of $n_p=9$ photons. Middle: Emitter CNOTs across the orbit obtained by the Heuristisc $\#1$ (red), Heuristics $\#2$ (cyan) methods. Bottom: Emitter CNOTs across the orbit obtained by the Brute-force method w/o allowing LCs. (c) Same as in (b), but for an $n_p=8$ random graph that requires only $n_e=3$ emitters to produce.}
    \label{fig:Random_LC_Orbits}
\end{figure*}

In Fig.~\ref{fig:RGS_Orbit}, we show the optimal number of emitter CNOTs for preparing any graph LC equivalent to an RGS consisting of 50 or 100 core qubits. To find the optimal number of emitter CNOTs, we use the Heuristics $\# 1$ algorithm. We see that the cost in emitter CNOTs is the same across the entire non-isomorphic LC orbit and equal to $n-2$. This feature is consistent with the fact that LC graphs have the same entanglement structure (only the connectivity varies), so an algorithm should be capable of finding the same entangling cost. 

\subsection{LC orbits of random graphs}

We now focus on the optimal preparation of LC orbits of random graphs. For generic graphs, there is no efficient algorithm for 
generating the LC orbit. We provide the generic Algorithm~\ref{alg:MapOutOrbit} of preparing the LC orbit of any graph in Appendix~\ref{App:Algorithms}, which has the option to keep or discard isomorphic graphs. In the following analysis, we choose to keep the isomorphs. Since the task of generating the LC orbit becomes difficult as the size of the graph increases, we focus on examples of  graphs with $n_p\leq 9$.

In Fig.~\ref{fig:Random_LC_Orbits}, we display our results for a random graph of 7 photons (a), 9 photons (b), and 8 photons (c). The top panels display one graph representative of the respective orbits. The second panel in (a) shows the number of emitter CNOTs across the entire orbit, if we use the Heuristics $\#1$ or Heuristics $\#2$ methods. For the Heuristics $\#2$ method, we set an emitter cutoff of up to 4, and a future cutoff of up to 4 steps. The third panel in (a) shows the emitter CNOTs we obtained if we do not allow any LCs and if we also prune the total cases we inspect per recursion level. In the bottom panel, we focus on the graphs for which we did not reach the minimum of 4 CNOTs without any LCs for any of the optimizers, and this time we apply the Brute-force method and allow one round of LCs and no pruning of the realizations. We see that we are now able to reach the minimum CNOT count of 4 even for these outliers. Thus, utilizing all our algorithms, we can find the minimum emitter CNOTs across all LC graphs. In Fig.~\ref{fig:Random_LC_Orbits}(b) we repeat the same calculation for the LC orbit of a graph consisting of $n_p=9$ photons, which is prepared by 4 emitters. In this case, we see that the Heuristics $\#2$ method can find the minimum number of emitter CNOTs across the entire orbit. A similar result is shown for another example in Fig.~\ref{fig:Random_LC_Orbits}(c), which shows that Heuristics $\#2$ is again able to find the minimum emitter CNOT count across the entire LC orbit of a graph of $n_p=8$ photons. Thus, we see that with low computational resources, we can verify that any graph in the orbit of a given graph is prepared with the same entanglement cost.

\subsection{Order 6 orbits and conjecture for all LC graphs}
\begin{figure*}[!htbp]
    \centering
    \includegraphics[scale=1.205]{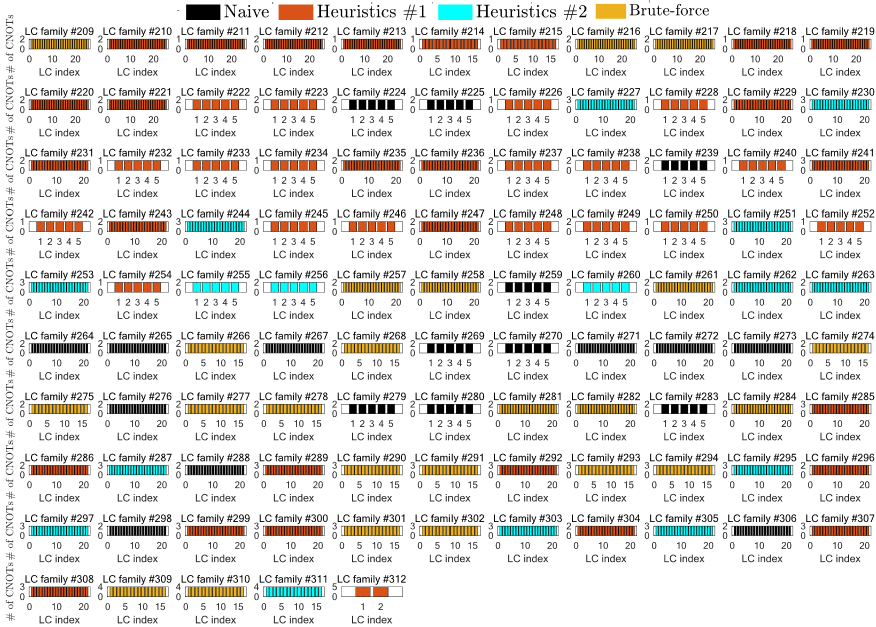}
    \caption{Optimal emitter CNOTs for a subset of the LC families of order 6 graphs. Each panel shows the corresponding LC family index out of 312 families and the number of emitter CNOTs across the non-isomorphic LC graphs of the family. The colors show the optimizer that was used. Black: Naive, red: Heuristics $\# 1$, cyan: Heuristics $\# 2$, yellow: Brute-force.}
    \label{fig:np6_third}
\end{figure*}

Here we study the optimal preparation of graphs of $n_p=6$ photons. To do that, we start by generating all possible adjacency matrices, which amounts to 26,704 connected graphs, including isomorphs. (Note that if we discard isomorphs, we end up with 112 non-isomorphic connected graphs, consistent with Ref.~\cite{hein2006arxiv}.) We then implement Bouchet's LC recognition algorithm~\cite{BouchetDiscreteMathematics1993} to find all the non-LC equivalent graphs that give rise to 312 LC families. Some of these 312 representatives could be isomorphic to each other (but not LC-equivalent), but we do not discard them. The reason why we include isomorphs is because different node labelings lead to different emitter requirements. Once we have a representative for each LC family, we generate the non-isomorphic LC graphs of the representative using Algorithm~\ref{alg:MapOutOrbit}. Note that discarding isomorphs at this step keeps our family-sorting method consistent because the family representative and the collection of graphs LC equivalent to it all require the same number of emitters to be prepared. Thus, we can safely discard isomorphs from within an LC orbit, although we should keep in mind that the circuits to prepare two isomorphic LC graphs will be different.

To find the optimal CNOT cost for every graph within an LC family, we utilize the different optimizers we developed. We test if the Naive, Heuristics, or Brute-force methods achieve the optimal number of emitter CNOTs. If any of those methods does not minimize the emitter CNOT cost across the entire orbit, we add new layers of optimization. This means that we can follow a different optimizer or change the optimization levels it includes. For instance, in the Brute-force optimizer, we can choose to prune or not realizations, and to allow one or more LC rounds. Once we find that the optimizer resolves the ambiguity and gives rise to the same lower bound on emitter CNOTs within the LC family, we terminate the search.

In Fig.~\ref{fig:np6_third}, we display the number of emitter CNOTs as a function of the LC graph index, starting from family $\#$209 and up to family $\#$312. The results for the remaining families and the graph representatives of each orbit are in Appendix~\ref{App:Order6CNOTs}. The different colors correspond to the optimizer that gave the minimum number of CNOT counts. The black bars correspond to the Naive method, the orange (cyan) bars to the Heuristics $\# 1$ ($\#2$) method, and the yellow bars to the Brute-force method.

The optimization for all LC families of order 6 shows that it is possible to achieve globally optimal CNOT counts for any graph within the same LC orbit. Based on our results, we see that no graph from within the same orbit is prepared with fewer CNOT requirements. Since it holds that LC equivalent graphs have the same entanglement structure but different connectivity, it is intuitive that the entangling gate cost will be the same. We conjecture that this holds for any LC family of arbitrary-sized graphs. Although this may seem obvious, we need to clarify the optimization procedure we follow and what it entails. In contrast to Ref.~\cite{Ghanbari2024arXiv}, we do not enable LCs upfront in the backward generation. In other words, we optimize every single graph from within an LC orbit, given its tableau representation as an input. In Ref.~\cite{Ghanbari2024arXiv}, the authors sample the LC orbit of a graph, and then they provide each LC graph as an input in the algorithm of Ref.~\cite{BikunnpjQI2022}, which does not explore all degrees of freedom in the optimization procedure. Consequently, they conclude that some graphs in the orbit can require fewer entangling gates to generate compared to others. This graph is then prepared optimally, and one can apply local gates at the final step of the Clifford circuit to prepare any graph LC equivalent to that, with the same number of entangling gates. Our optimization procedure reveals
an even more surprising result. Even though we do not allow LC operations in the final step of the Clifford circuit, we can still resolve the freedom in the remaining part of the Clifford circuit and achieve the same optimal CNOT counts across an entire LC orbit.

\section{Conclusions}
The efficient generation of photonic graph states is important for future quantum network and quantum communication applications. We developed algorithms that minimize the number of CNOTs required to prepare these entangled states, revealing a reduction of up to 75$\%$ in emitter CNOTs. Our Heuristics algorithms make a strong connection between the graph representation and stabilizer circuits and exploit graph modifications in the construction of the graph that relate to the optimal optimization paths. We also study repeater graph states and find the size of their LC orbit analytically, although this is a $\#$P complete problem for arbitrary graphs. We further find an efficient way to generate the LC orbit of repeater graph states irrespective of their size or photonic emission ordering and show that by discarding isomorphic graphs, the size of the LC orbit scales linearly with the size of the state. We find the optimal photonic emission ordering for unencoded and encoded repeater graph states that gives rise to minimal numbers of emitters and minimal numbers of emitter CNOTs. We also address the problem of preparing LC graphs and show how to minimize the CNOT cost across the entire LC orbit by resolving the freedom over local gates at intermediate steps of the Clifford circuit.

\section{Outlook}
Our work addresses the problem of synethesizing Clifford circuits for emitter-based photonic graph state generation with minimal resources. We show how decisions during the generation impact the total number of emitter CNOTs, and under which cases we can follow nearly optimal optimization paths, especially related to photonic absorption decisions. An aspect we have not explored thoroughly is how to minimize globally the number of emitter CNOTs in the final disentanglement step (i.e., after all photons have been absorbed). If no restriction on the emitters' connectivity is imposed, then one could explore various elimination orderings or identify node separators such that the disentanglement is achieved with fewer steps. This can bring further reduction in the emitter CNOT counts. A different direction would be to implement the parallel elimination algorithm of Ref.~\cite{Patel2008}, which is developed for unrestricted connectivity, and in our case, would achieve a worst-case scaling of $\mathcal{O}((n_e')^2/\log_2(n_e'))$ final emitter CNOTs ($n_e'$ is the number of emitters that are still entangled at the last algorithmic step, where $n_e'\leq n_e$). Another direction would be to sample from the LC orbit of the graph on the $n_e'$ emitters and identify the one with the minimal number of edges, such that an unrestricted elimination with fewer CNOTs is followed subsequently. Lastly, one could initially perform the emitter disentanglement we follow in this work and then apply circuit identities to potentially reduce the number of emitter CNOTs~\cite{Staudacher_2023ZX_Calculus,BriegelPRA2006,Bravyi2021Quantum,SchneiderACM2023}. 

Our algorithms can also serve as a guide for inspecting the preparation cost of resource states in the context of photonic quantum computing (QC) and comparing that to the preparation cost of an all-photonic approach. This could open up the path to understanding the complexity of preparing arbitrary resource states and inspecting the viability of emitter-based architectures for photonic QC.

\section*{Acknowledgments}

S.E.E. acknowledges support by
the Army Research Office (MURI grant no. W911NF2120214). E.B. acknowledges the NSF (award 2137953). All authors acknowledge the Commonwealth Cyber Initiative (CCI), an investment in the advancement of cyber R\&D, innovation, and workforce development.

\nocite{*}

\appendix

\section{Graphical analysis of time-reversed measurement \label{App:Graphical_Proofs}}

In the main text we mentioned that the time-reversed measurement connects the chosen emitter with the neighborhood of the photon to be absorbed. Here we will prove this statement based on the stabilizers. Suppose that we begin with the target photonic graph state and we wish to bring the first emitter back into the graph. This is achieved by starting with an emitter in $|0\rangle$, putting it into $|+\rangle$ and then applying a CNOT$_{\text{em},\text{photon}~(i)}$, which simulates the time-reversed measurement. This means that after the Hadamard on the decoupled emitter, the stabilizers read:
\begin{eqnarray}
    \mathcal{S}&=&\langle K_G^{(a)}\otimes \mathds{1}^{\text{em}}, \mathds{1}^{\otimes n_p}\otimes X^{\text{em}}\rangle, \\K_G^{(a)}&=&X^{(a)}\prod_{j\in N_a}Z^{(j)}.
\end{eqnarray}
It is straightforward to show that the stabilizers after the CNOT$_{\text{em},\text{photon}~(i)}$ are:
\begin{eqnarray}
    \tilde{K}^{\text{em}}&=&X^{\text{em}}X^{(i)} \nonumber \\
    \tilde{K}^{(i)}&=&X^{(i)} \prod_{j\in N(i)}Z^{(j)} \nonumber \\
    \tilde{K}^{b,b\in N(i)}&=&X^{(b)} \prod_{j\in N(b)\cup\{\text{em}\}}Z^{(j)}
   \nonumber \\
    \tilde{K}^{b,b\notin N(i)}&=&X^{(b)}\prod_{j\in N(b)}Z^{(j)}.
\end{eqnarray}
To create the canonical form we can redefine the emitter's stabilizer by performing the substitution $\tilde{K}^{\text{em}}\mapsto \tilde{K}^{\text{em}}\cdot \tilde{K}^{(i)}$, which leads to:
\begin{eqnarray}
    \tilde{K}^{\text{em}}&=&X^{\text{em}}\prod_{j\in N(i)}Z^{(j)} \nonumber \\
    \tilde{K}^{(i)}&=&X^{(i)} \prod_{j\in N(i)}Z^{(j)} \nonumber \\
    \tilde{K}^{b,b\in N(i)}&=&X^{(b)} \prod_{j\in N(b)\cup\{\text{em}\}}Z^{(j)}
   \nonumber \\
    \tilde{K}^{b,b\notin N(i)}&=&X^{(b)}\prod_{j\in N(b)}Z^{(j)}.
\end{eqnarray}
From the first and third stabilizers, we see that the emitter now connects with the neighborhood of the photon.

\section{Optimal circuit of 6-node graph \label{App:Opt_Circuit}}
\begin{figure}[!htbp]
    \centering
    \includegraphics[scale=0.578]{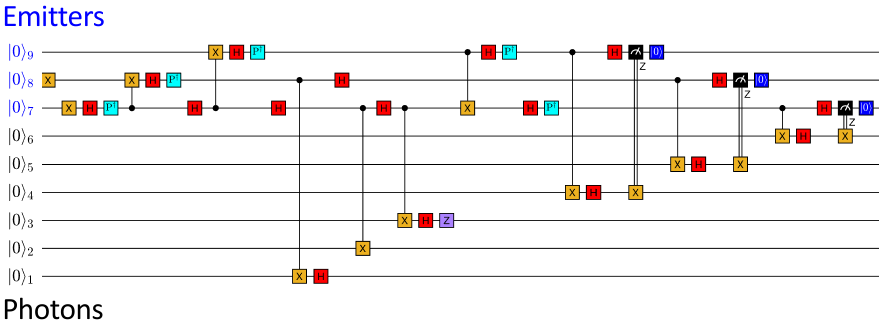}
    \caption{Optimal circuit to prepare the photonic graph state of Fig.~\ref{fig:6NodeGraph}. Red boxes are Hadamard gates, cyan boxes are conjugate Phase gates, black boxes are computational basis measurements, yellow boxes are X gates, and purple boxes are Z gates. The blue boxes correspond to re-initialization of emitters in $|0\rangle$. }
    \label{fig:Opt_Circuit}
\end{figure}

Here we show the optimal circuit of the 6-node graph we presented in Fig.~\ref{fig:6NodeGraph}. In the main text, we showed that generating the graph via the Naive approach requires 5 emitter CNOTs. All our three different optimizers (Brute-force, Heuristics $\# 1$ and Heuristics $\# 2$) verified that the same graph can be prepared with the optimal number of 3 emitter CNOTs. We display the optimal circuit found by Heuristics $\# 1$ method in Fig.~\ref{fig:Opt_Circuit}. To obtain this circuit, we set the Back-substitution option to true and the extra inspection for free PA to false.

\section{Algorithms\label{App:Algorithms}}

Here we provide the algorithms for the various subroutines we mentioned in the main text.

\begin{algorithm}
\caption{Naive Circuit Generator}\label{alg:NaiveCircuitGenerator}
\begin{algorithmic}[1]
\Procedure{NaiveCircuitGenerator}{$G$}
\\
\Comment{Get the Clifford circuit based on the Naive scheme.}\\
\Comment{Input: The target photonic graph, $G$}\\
\State Create canonical stabilizer tableau based on $G$
\State Put stabilizer tableau in RREF 
\State Get height function, $h(x)$
\State $n_e$ $\gets$ $\max(h(x))$ \Comment{Minimal $\#$ of required emitters}
\State Augment stabilizer tableau with emitters in $Z$
\\
\For{$k=n_p+1:-1:2$}
\State photon $\gets$ $k-1$
\If{$k<n_p+1$}
\State Put stabilizer tableau in RREF
\State Get new height function
\EndIf
\\
\State $\delta h$ $\gets$ $h(k)-h(k-1)$
\If{$\delta h<0$} \Comment{Need TRM, cannot absorb}
\State Perform TRM and save the circuit updates
\EndIf
\State Attempt free PA (w/o back-substitution or exhausting all checks) 
\If{free PA did not succeed} 
\State Perform naive PA using emitter CNOTs first 
\State Save the circuit updates
\Else
\State Save the circuit updates of free PA
\EndIf
\If{photon==1}
\State \textbf{break}
\EndIf
\EndFor
\\
\State Put stabilizer tableau in RREF
\State Disentangle emitters (w/o enabling back-substitution)
\State Clean up strings of Zs
\State Fix phases to $+1$ via local gates
\State Verify circuit \\
\Return{The generation circuit}

\EndProcedure
\end{algorithmic}
\end{algorithm}

\begin{algorithm}
\caption{Back-Substitution}\label{alg:BackSubs}
\begin{algorithmic}[1]
\Procedure{BackSubs}{Stab,$n$,rowINDX}
\\
\Comment{Perform back-substitution on input RREF Tableau.}\\
\Comment{Input: Stabilizer Tableau, $\#$ of qubits, row to terminate Back-substitution}\\

\If{isempty(rowINDX)} 
\State exitRow $\gets$ 1
\Else 
\State exitRow $\gets$ rowINDX
\EndIf
\\
\State Rowk $\gets$ $n$

\While{true}

\\

\State SX $\gets$ Stab(:,$1:n$)
\State SZ $\gets$ Stab(:,$n+1:2*n$)
\State SXk $\gets$ SX(Rowk,:)
\State SZk $\gets$ SZ(Rowk,:)
\\
\For{$k$=Rowk-1:-1:1}

\\

\State sx $\gets$ SX($k$,:) 
\State sz $\gets$ SZ($k$,:) 
\State w0 $\gets$ sum(sx)+sum(sz) 
\State w0 $\gets$ w0 $-$ sum(bitand(sx,sz)) \Comment{remove double counting of $Y$'s}
\\
\If{w0$>$1}
\State sx $\gets$ bitxor(sx,SXk)
\State sz $\gets$ bitxor(sz,SZk)
\State wA $\gets$ sum(sx)+sum(sz)
\State wA $\gets$ wA $-$ sum(bitand(sx,sz))
\\
\If{wA$\leq $w0}
\State Stab $\gets$ rowsum(Stab,$n$,$k$,Rowk) 
\EndIf
\\

\EndIf
\\
\EndFor
\\

\State Rowk $\gets$ Rowk$-1$

\If{Rowk $==$ exitRow}
\State \textbf{break}
\EndIf
\\
\EndWhile
\\ 
\State \Return{The updated Stabilizer Tableau `Stab'}

\EndProcedure
\end{algorithmic}
\end{algorithm}

\begin{algorithm}
\caption{Free Photon Absorption}\label{alg:Free_Photon_Absorption}
\begin{algorithmic}[1]
\Procedure{freePA}{inputs}
\\
\Comment{Test if PA w/o emitter CNOTs is possible.}
\\
\Comment{Inputs: Stab, Circuit, extraInspectionStep}
\\
\State Find stabilizer(s) starting from photon to be absorbed.
\State Indx $\gets$ smallest photonic row index
\For{each stabilizer row}
\If{weight on emitter sites is 1} 
\State Absorb the photon
\State successFlag $\gets$ true
\State  \Return{updated inputs and successFlag }
\EndIf
\EndFor
\\
\If{Two photonic rows detected}
\State Multiply the photonic rows
\If{weight on emitter sites is 1 for updated row}
\State Absorb the photon
\State successFlag $\gets$ true
\State  \Return{updated inputs and successFlag}
\EndIf
\EndIf
\\
\State Find stabilizer(s) with support only on emitter sites.
\For{each photonic stabilizer row}
\For{each emitter stabilizer row}
\State Multiply emitter stabilizer on photonic one
\If{weight on emitter sites is 1 for new row}
\State Absorb the photon
\State successFlag $\gets$ true
\State  \Return{updated inputs and successFlag }
\EndIf
\EndFor
\EndFor
\\
\If{extraInspectionStep is true}
\State Back-substitute the tableau till row Indx 
\If{weight on emitter sites is 1 for row Indx}
\State Absorb the photon 
\State successFlag $\gets$ true
\State  \Return{updated inputs and successFlag }
\EndIf
\EndIf
\\
\State successFlag $\gets$ false
\State \Return{same inputs and successFlag}

\EndProcedure
\end{algorithmic}
\end{algorithm}

\begin{algorithm}
\caption{rowsum}\label{alg:rowsum}
\begin{algorithmic}[1]
\Procedure{rowsum}{Stab,$n$,rowH,rowI}
\\
\Comment{Multiply two stabilizer rows (rowH is replaced).}\\
\Comment{Input: Stabilizer Tableau, $\#$ of qubits, row indices}\\

\State cntP $\gets$ 0 \Comment{$\#$ of times +1 power of $i$ appears}
\State SH $\gets$ Stab(rowH,:) \Comment{Stabilizer row `rowH'}
\State SI $\gets$ Stab(rowI,:) \Comment{Stabilizer row `rowI'}
\State rH $\gets$ SH(end) \Comment{Phase of row `rowH'}
\State rI $\gets$ SI(end) \Comment{Phase of row `rowI'}
\\
\State xHzI $\gets$ bitand(SH($1:n$),SI($n+1:2*n$))
\State xIzH $\gets$ bitand(SI($1:n$),SH($n+1:2*n$))

\\
\State antiCom $\gets$ xor(xHzI,xIzH) \Comment{Anticommuting Paulis}

\State indices $\gets$ find(antiCom) \Comment{Get qubit positions}
\State $L$ $\gets$ length(indices)
\\
\For{each $k$ in indices}
\\
\If{SH($k+n$)==0} \Comment{`X'}
\\
\If{SI($k$)==1} 
\State cntP $\gets$ cntP +1 \Comment{Has to be `Y'} 
\EndIf
\\
\ElsIf{bitand(SH($k$),SH($k+n$))==1} \Comment{`Y'}
\\
\If{SI($k$)==0} 
\State cntP $\gets$ cntP+1 \Comment{Has to be `Z'}
\EndIf
\\
\Else  \Comment{`Z'}
\\ 
\If{SI($k+n$)==0} 
\State cntP $\gets$ cntP+1 \Comment{Has to be `X'}
\EndIf
\\
\EndIf
\\
\EndFor
\\
\State cntM $\gets$ $L-$cntP \Comment{$\#$ of times $-$1 power of $i$ appears}
\State temp $\gets$ mod(2$*$rH + 2$*$rI + (cntP $-$cntM),4)
\If {temp==0}
\State rH $\gets$ 0
\Else
\State rH $\gets$ 1
\EndIf
\State Stab(rowH,:) $\gets$ bitxor(SH,SI)
\State Stab(rowH,end) $\gets$ rH
\\
\State \Return{The updated Stabilizer Tableau `Stab'}

\\

\EndProcedure
\end{algorithmic}
\end{algorithm}

\begin{algorithm}
\caption{Map Out RGS Orbit}\label{alg:MapOutRGSOrbit}
    \begin{algorithmic}[1]
\Procedure{MapOutRGSOrbit}{$n$,ordering}
\\
\Comment{Create the LC orbit of unencoded RGSs.}
\\
\Comment{$n$: $\#$ of core nodes, ordering: labels of nodes}
\\
\State A $\gets$ createRGS($n$) \Comment{RGS with CW ordering of core qubits $\in [1,n]$, and CW ordering of leaves $\in[n+1,2n]$}
\\
\State nodes $\gets$ [1,2]
\State allNodes $\gets$ [nodes]
\State $l$ $\gets (3*(2*n+1)-(-1)^{(n+1)})/4$ \Comment{Analytical scaling of RGS orbit w/o isomorphs}
\\
\While{true}
\State w $\gets$ n+nodes(1)
\State nodes $\gets$ nodes+2
\State allNodes $\gets$ [allNodes,w,nodes]
\If{length(allNodes)$\geq$ $l$-1}
\State break
\EndIf
\EndWhile
\\
\State [$\sim$,b] $\gets$ unique(allNodes,`first') \Comment{Remove duplicates}
\State allNodes $\gets$ allNodes(sort(b)) \Comment{Restore ordering to that before sorting}
\If{$(-1)^n==1$} \Comment{If even core, terminate at leaf}
\State allNodes(end) $\gets$ []
\EndIf
\\
\State B $\gets$ A
\State AdjLC $\gets$ $\{\}$ \Comment{Initialize empty stack}
\State Enqueue B in AdjLC
\For{every $v$ in allNodes} \Comment{Generate orbit}
\State B $\gets$ LocalComplement(B,$v$)
\State Enqueue B in AdjLC
\EndFor
\\
\State P $\gets$ eye(2$*$n)
\State P(:,$1:2*n$) $\gets$ P(:,ordering) \Comment{Permutation matrix}
\For{every B in AdjLC} 
\State B $\gets$ $P*$B$*P^T$  \Comment{Requested labeling}
\EndFor \\
\Return{AdjLC}
\EndProcedure
    \end{algorithmic}
\end{algorithm}

\begin{algorithm*}
\caption{Map Out RGS Orbit with multiple leaves}\label{alg:MapOutRGSOrbitManyLeaves}
    \begin{algorithmic}[1]
\Procedure{MapOutRGSOrbitManyLeaves}{$n$,$l$}
\\
\Comment{Create the LC orbit of unencoded RGSs with multiple leaves per core node.}
\\
\Comment{$n$: $\#$ of core nodes, $l$: $\#$ of leaves per core node}
\\
\State A $\gets$ createRGSManyLeaves($n$,$l$) \Comment{RGS with many leaves and optimal ordering}

\State $N$  $\gets$ $n + l*n$ \Comment{Max label of nodes}
\State coreLabels $\gets$ $l+1:l+1:N$ \Comment{Labels of cores}
\State leafLabels $\gets$ setdiff(1:N,coreLabels) \Comment{Leaf labels}
\\
\For{$k$=2:length(coreLabels)} \Comment{Optimal ordering}
\State shift $\gets$ $(k-1)*l$
\State temp $\gets$ coreLabels($k$)
\State coreLabels($k$) $\gets$ leafLabels(1+shift)
\State leafLabels(1+shift)$\gets$ temp
\EndFor
\\
\State nodes $\gets$ [1,2]
\State allNodes $\gets$ [nodes]
\State $l$ $\gets (3*(2*n+1)-(-1)^{(n+1)})/4$ \Comment{Analytical scaling of RGS orbit w/o isomorphs}
\\
\While{true}
\State w $\gets$ nodes(1)-1+nodes(1)$*l$
\State nodes $\gets$ nodes+2
\State allNodes $\gets$ [allNodes,w,nodes]
\If{length(allNodes)$\geq$ $l$-1}
\State break
\EndIf
\EndWhile
\\
\State [$\sim$,b] $\gets$ unique(allNodes,`first') \Comment{Remove duplicates}
\State allNodes $\gets$ allNodes(sort(b)) \Comment{Restore ordering to that before sorting}
\\
\State cntL $\gets$ 0
\State cntC $\gets$ 0
\\
\For{every $v$ in allNodes} \Comment{Rename to optimal ordering}
\If{allNodes($v$)$\leq n$}
\State cntC $\gets$ cntC+1
\State allNodes($v$) $\gets$ coreLabels(cntC)
\\~~~~~~~~~\textbf{else}
\State cntL $\gets$ cntL+1
\State allNodes($v$) $\gets$ leafLabels(cntL+(cntL-1)$*l$)
\EndIf
\EndFor
\\
\If{$(-1)^n==1$} \Comment{If even core, terminate at leaf}
\State allNodes(end) $\gets$ [] 
\EndIf
\\
\State B $\gets$ A
\State AdjLC $\gets$ $\{\}$ \Comment{Initialize empty stack}
\State Enqueue B in AdjLC
\For{every $v$ in allNodes} \Comment{Generate orbit}
\State B $\gets$ LocalComplement(B,$v$)
\State Enqueue B in AdjLC
\EndFor
\\
\Return{AdjLC}
\EndProcedure
    \end{algorithmic}
\end{algorithm*}

\begin{algorithm}
\caption{Final emitter disentanglement}\label{alg:FinalDisentanglement}
\begin{algorithmic}[1]
\Procedure{FinalDisentanglement}{Stab,Circuit}
\\
\Comment{Disentangle the emitters at the end of the protocol.}
\\
\Comment{Inputs: Stabilizer Tableau, photonic generation circuit at this step}
\\
\State disentangledE $\gets$ [] \Comment{Disentangled emitters}

\While{true}
\State Call Subroutine(Stab,Circuit,disentangledE)
\If{flag}
\State \textbf{break}
\EndIf

\EndWhile

\State \Return{The updated inputs (Stab, Circuit)}

\\
\Procedure{subroutine}{Stab,Circuit,disentangledE}
\State allEmitters $\gets$ $n_p+1:n_p+n_e$
\State Remove disentangledE from allEmitters
\State cnt $\gets$ 0  \For{every emitter in disentangledE}
\If{emitter in product state}
\State Enqueue emitter in disentangledE 
\Else
\State cnt $\gets$ cnt+1
\EndIf
\EndFor
\\

\If{cnt==0}
\State flag $\gets$ true
\State \Return{}
\EndIf
\\

\If{BackSubs}
\State Stab $\gets$ Backsubstitution(Stab,n,[])
\EndIf
\\
\State row $\gets$ stabilizer row with smallest weight
\State emZ $\gets$ emitters with Pauli Z in row
\State cntZ $\gets$ length(emZ)
\If{cntZ==2}
\State Apply CNOT on emitters in emZ
\State \Return{Stab, Circuit, Emitters}
\EndIf
\\
\State emX $\gets$ emitters with Pauli X in row
\For{every emitter in emX}
\State Apply H on emitter
\State emZ $\gets$ [emZ,emitter]
\State cntZ $\gets$ cntZ+1
\If{cntZ==2}
\State Apply CNOT on emitters in emZ

\State \Return{Stab, Circuit, Emitters}
\EndIf
\EndFor

\\
\State emY $\gets$ emitters with Pauli Y in row
\For{every emitter in emY}
\State Apply P on emitter
\State Appply H on emitter
\State cntZ $\gets$ cntZ+1
\State emZ $\gets$ [emZ,emitter]
\If{cntZ==2}
\State Apply CNOT on emitters in emZ

\State \Return{Stab, Circuit, Emitters}
\EndIf
\EndFor

\EndProcedure
\\

\EndProcedure
\end{algorithmic}
\end{algorithm}

\begin{algorithm}
\caption{Optimal Ordering for RGS}\label{alg:RGSOrdering}
    \begin{algorithmic}[1]
\Procedure{Adj=RGSOrdering}{$n$}
\\
\Comment{Generate an ordering for unencoded RGS that requires 2 emitters to be prepared.}
\\
\Comment{$n$: $\#$ of core nodes}\\
Create adjacency of RGS with natural emission ordering. \\
Exchange labeling of 2nd to last core qubit with that of last leaf. \\
Return the updated adjacency matrix.
\EndProcedure
    \end{algorithmic}
\end{algorithm}

\begin{algorithm}
\caption{Optimal Ordering for RGS with multiple leaves}\label{alg:RGSDepth1Ordering}
    \begin{algorithmic}[1]
\Procedure{Adj=RGSManyLeavesOrdering}{$n$,$l$}
\\
\Comment{Get emission ordering of RGSs with multiple leaves that need 2 emitters to be prepared.}
\\
\Comment{$n$: $\#$ of core nodes, $l$: $\#$ of leaves per core node.}\\
Generate adjacency of RGS where each core has $l$ leaves and with natural emission ordering. \\
For every core qubit with labels $(v_1,v_2,\dots)$ consider only those with labels $v_2,\dots$. \\
Exchange labeling of those core nodes with one of their leaves. \\
Return updated adjacency.
\EndProcedure
    \end{algorithmic}
\end{algorithm}

\begin{algorithm}
\caption{Optimal Ordering for encoded RGS with $b=[2,2,l]$}\label{alg:RGSEncodedOrdering}
    \begin{algorithmic}[1]
\Procedure{Adj=RGSEncodedOrdering}{$n$,$l$}
\\
\Comment{Get emission ordering of encoded RGSs that need 2 emitters to be prepared.}
\\
\Comment{$n$: $\#$ of core nodes (even), $l$: $\#$ of leaves per core node.}\\
\State triangles $\gets$ $n/2$
\State core $\gets$ 1
\State allCores $\gets$ $\{\}$
\\
\For{l=1:triangles}
\State Connect core with first $l$ leaves
\State Core $\gets$ core+1
\State Connect this core with next $l$ leaves
\State Connect a top node $w$ with the two core nodes
\State Add core nodes at the end of the list allCores
\If{$l\geq 2$}
\State Exchange labeling of last core with top node
\State Remove last entry from allCores 
\State Add top node in allCores
\EndIf
\EndFor
\\
\State Create connections among all nodes stored in allCores
\\
\Return{Adjacency matrix}
\EndProcedure
    \end{algorithmic}
\end{algorithm}

\begin{algorithm}
\caption{Map Out Orbit}\label{alg:MapOutOrbit}
\begin{algorithmic}[1]
\Procedure{MapOutOrbit}{A, keepIsomorphs}
\\
\Comment{Generate the LC orbit of a random graph.}
\\
\Comment{A: adjacency matrix, keepIsomorphs: true or false}
\\
\State n $\gets$ length(A)
\State cnt $\gets $1

\State AdjList(cnt).Adj $\gets$ A \Comment{Enqueue A in 1st level of list}
\\
\State isoCond $\gets$ @(A,B) isisomorphic(graph(A),graph(B))
\State equalCond $\gets$ @(A,B) all(all(A==B))
\\
\State \textbf{switch} keepIsomorphs
\State ~~case true
\State ~~~~cond $\gets$ @(A,B) equalCond(A,B)
\State ~~case false
\State ~~~~cond $\gets$ @(A,B) equalCond(A,B) $||$ isoCond(A,B)
\State \textbf{end switch}
\\
\While{true}  \Comment{Till we generate no new LC graphs}
    \State AdjList(cnt+1).Adj$\gets \{\}$ \Comment{Next level empty stack} 
    \\
    \For{every A in level cnt} 
    \State nodes $\gets$ $1:n$
    \State Detect leaf nodes of A 
    \State Remove leaves from nodes
    \\
    \For{every $v$ in nodes}
    \State flag $\gets$ true
    \State Local complement A about $v$ 
    \State allA $\gets$ all adjacencies generated so far 
    \\
    \For{every B in allA}
    \If{cond(B,A)} \Comment{Store A?}
    \State flag $\gets $ false 
    \State break
    \EndIf
    \EndFor
    \\
    \If{flag} \Comment{Keep A}
    \State Enqueue A to stack of level cnt+1
    \EndIf
    \\
    \EndFor
    \\
    \EndFor
    \\
    \If{no adjacency exists in level cnt+1} 
    \State break \Comment{Saturated the orbit}
    \EndIf
    \\
    \State cnt $\gets$ cnt+1
    \EndWhile
\\
\\
\Return{AdjList}
\EndProcedure
\end{algorithmic}
\end{algorithm}

\clearpage


\section{Runtime\label{App:Runtime}}
\begin{figure}[!htbp]
    \centering
    \includegraphics[scale=0.76]{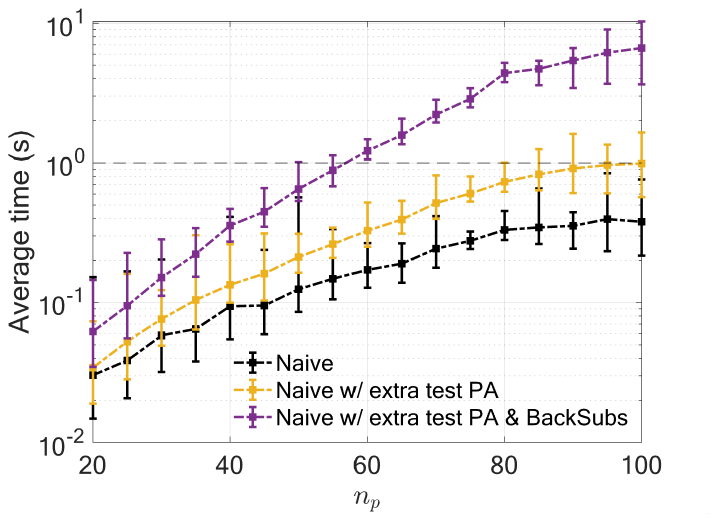}
    \caption{Average runtime for the Naive variants as a function of the size of the graph. We sample 50 random graphs per $n_p$. The error bars mark the minimum and maximum runtime. The dashed line marks the 1~s runtime.  }
    \label{fig:RunTimeNaiveVariants}
\end{figure}
Here we test the runtime of different variants of the Naive method for graph states of up to $n_p=100$ photons. For each $n_p$, we sample 50 graphs and find the average runtime if we allow or not back-substitution. In particular we examine the following optimizers in Fig.~\ref{fig:RunTimeNaiveVariants}:
\begin{itemize}
    \item Naive method without back-substitution on the RREF tableau, and no extra inspection for free photon absorption subroutine. This is shown with a black line.
    \item Naive method with back-substitution on the RREF tableau, and extra inspection for free photon absorption. This is shown with a purple line.
    \item Naive method without back-substitution on the RREF tableau, and extra inspection for free photon absorption. This is shown with a yellow line.
\end{itemize}
The extra inspection step for free photon absorption is mentioned in Algorithm~\ref{alg:Free_Photon_Absorption}, and is performing back-substitution only if the algorithm does not exit in the previous steps. This step is more important than doing globally the back-substitution. If we want to have lower runtime than the one obtained by the purple curve, we can allow the overhead of back-substitution only to verify that free photon absorption is possible or not, which corresponds to the yellow line.

\section{CNOT rules for decoupling two emitters \label{App:CNOT_Rules}}
Here we provide the transformation rules we follow to decouple two emitters whenever we encounter a weight-2 stabilizer with support only on emitter sites. We resort to a look-up table which, depending on the Pauli operators of the two emitters, either:
\begin{itemize}
    \item applies no local gates, and only a CNOT between the emitters,
    \item applies a local gate on one emitter and then a CNOT between the emitters,
    \item applies a local gate per emitter, and then a CNOT between the emitters.
\end{itemize}
The rules are summarized in Table~\ref{tab:CNOT_Rules}. Note that there are only 6 cases of Pauli operators on two qubits on which if we apply local gates and a CNOT we succeed in disentangling one emitter. All other Pauli operators on two qubits do not result in reducing the weight of the stabilizer by one. We also put a condition to throw an error if none of these cases of operators is encountered. In all the tests we did, we never entered this error condition.

\begin{table}[!htbp]
    \centering
    \begin{tabular}{|c|c|c|}
    \hline
         Pauli on emitter 1& Pauli on emitter 2   & Operation \\
         \hline
         $Z_1$ & $Z_2$ & CNOT$_{12}$ \\ 
                \hline
        $X_1$ & $X_2$  & CNOT$_{12}$ \\
        \hline
        {\color{black}$Y_1$} & {\color{black}$Z_2$} & CNOT$_{21}$ \\
        \hline
        $X_1$ & $Z_2$ & $H_2$, then CNOT$_{12}$ \\
        \hline
        {\color{black}$X_1$} & {\color{black}$Y_2$} & CNOT$_{21}$\\ \hline
        $Y_1$ & $Y_2$ &  $P_2$, then CNOT$_{12}$  \\
        \hline
    \end{tabular}
    \caption{Transformation rules we follow to disentangle one of the two emitters when we encounter a stabilizer of weight 2 on emitter sites.}
    \label{tab:CNOT_Rules}
\end{table}

\section{Flowcharts of Heuristics optimizers \label{App:Flowcharts}}

\begin{figure*}[!htbp]
    \centering
    \includegraphics[scale=0.65]{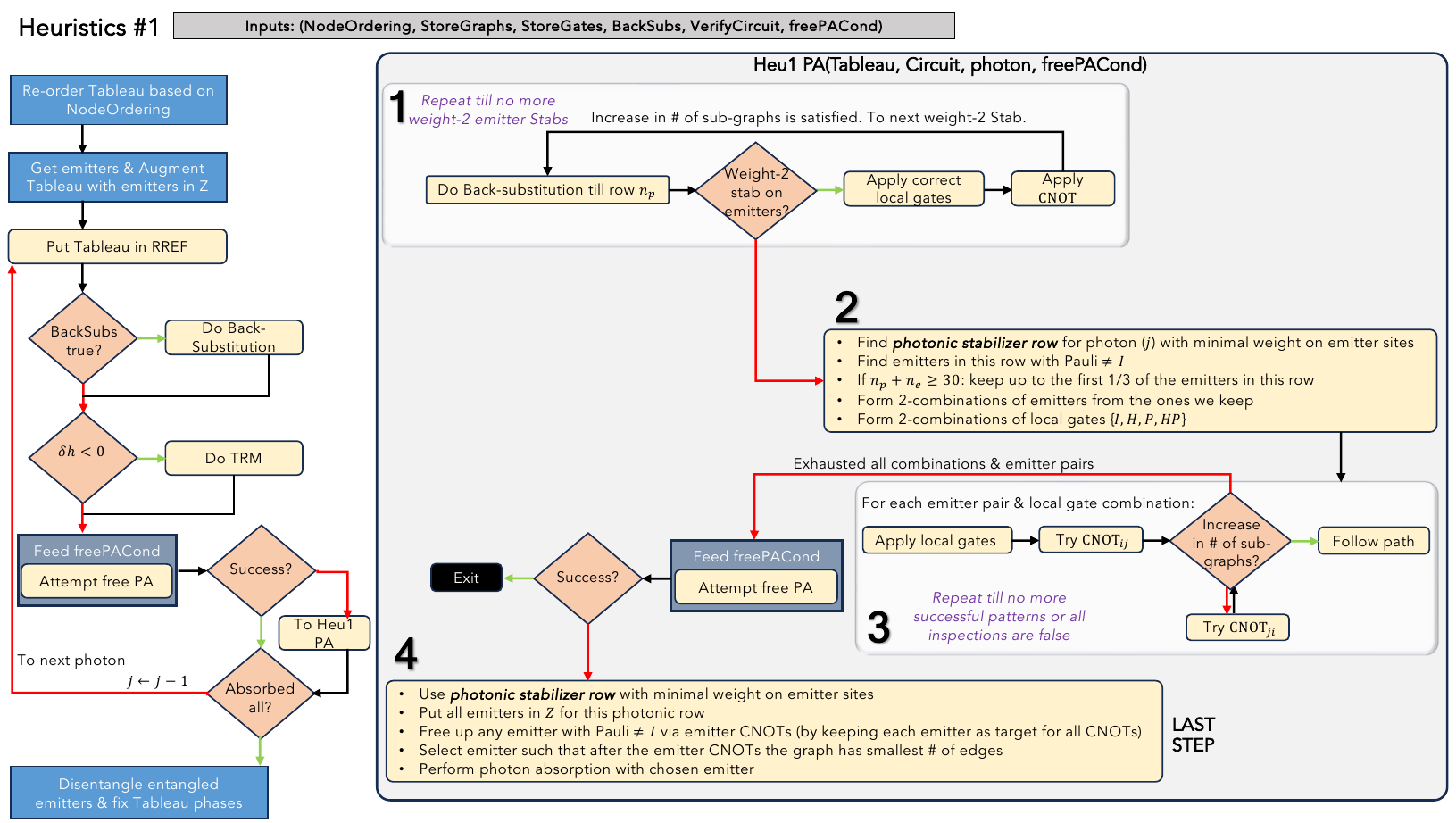}
    \caption{Flowchart that summarizes the optimization procedure we follow according to Heuristics $\#1$. Green (red) arrows answer yes (no) to the question inside pink boxes. TRM is time-reversed measurement and PA is photon absorption. }
    \label{fig:Heuristics1_FlowChart}
\end{figure*}

\begin{figure*}[!htbp]
    \centering
    \includegraphics[scale=0.65]{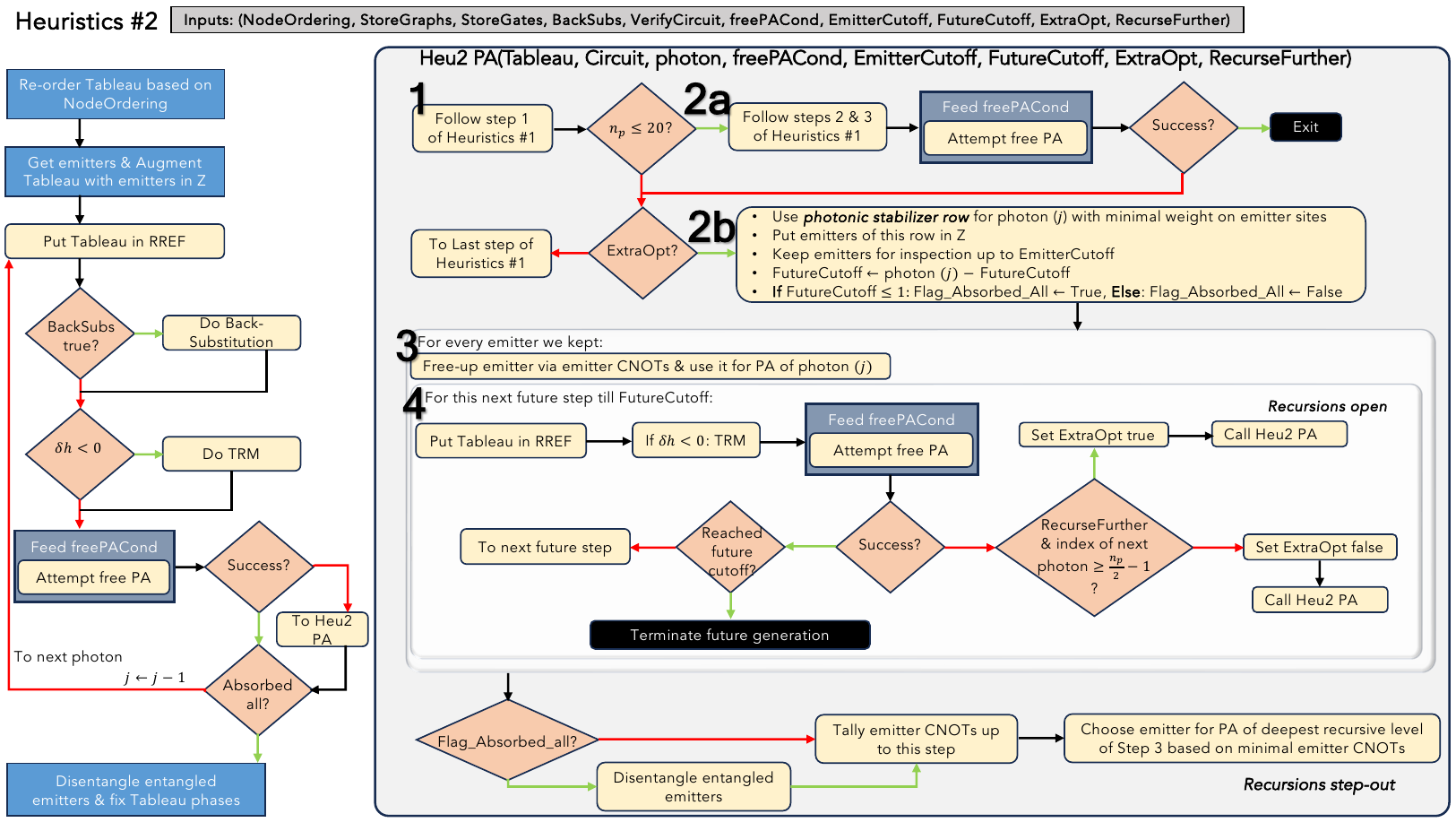}
    \caption{Flowchart that summarizes the optimization procedure we follow according to Heuristics $\#2$. Green (red) arrows answer yes (no) to the question inside pink boxes. TRM is time-reversed measurement and PA is photon absorption. }
    \label{fig:Heuristics2_FlowChart}
\end{figure*}

Here we provide the flowcharts of our Heuristics optimizers.
We begin with Heuristics $\#1$ optimizer, whose flowchart is shown in Fig.~\ref{fig:Heuristics1_FlowChart}. The left part is the main generation routine whose backbone is the algorithm of Ref.~\cite{BikunnpjQI2022}, with various improvements on decisions and runtime (e.g., the free-photon absorption step and final disentanglement of emitters). The main modifications are the options for Back-substitution and to exhaust all possibilities for free photon absorption to happen. Those options are provided as inputs and we can set them to true or false. When photon absorption cannot be performed freely (without emitter CNOTs upfront), we then resort to photon absorption based on the Heuristics $\#1$ patterns we mentioned in the main text. These steps are shown in the right part of Fig.~\ref{fig:Heuristics1_FlowChart}. The search for increased connected components is performed in steps 1 and 3. Immediately after step 3, we re-attempt free photon absorption. If it succeeds, we follow this path and exit the subroutine. If all of these steps fail, then we go to step 4. We pick the photonic stabilizer row whose first non-trivial Pauli starts from the photon to be absorbed. There can be at most two photonic stabilizer rows due to the RREF, and we select the one with minimal weight on emitter sites. We test all cases of free-ing up an emitter, and count the number of edges in the resulting graph. We select to free-up an emitter such that we obtain the minimal edges in the graph.

We also show in Fig.~\ref{fig:Heuristics2_FlowChart} an approximate flowchart of Heuristics $\# 2$. It is approximate, because the procedure is more involved and difficult to depict due to the recursions we open at various stages, but we will try to highlight what this optimizer does. If free photon absorption (PA) fails, we resort to the Heuristics $\# 2$ PA. This follows the whole procedure of Heuristics $\#1$ optimizer (but does not repeat the last step of Fig.~\ref{fig:Heuristics1_FlowChart}). Once we enable the extra optimization level (set ExtraOpt to true), we inspect what happens if we pick any of the emitters (potentially up to a an emitter cutoff we set) to absorb the photon at the first ambiguous level we encounter. We free-up every emitter sequentially, but in every case we proceed with next algorithmic steps of future photon absorptions, potentially time-reversed measurements, and disentanglement of emitters. How many future steps we inspect is set by the future cut-off, which refers to how many next photonic absorptions we test. For each such future step, if we succeed with a free PA we follow this path and go to the next future step. If we fail, then we have the option to open further recursions (set by an input option), as long as it also holds true that we have not consumed yet about half of the graph size. This is because we want to target decisions that happen early on in the consumption of the graph, while at the same time not overshooting the runtime of this optimizer. If further recursions are enabled, we proceed to next future steps, and call again the same subroutine. At some point we will reach a future cut-off. Then, recursions will start to step out, and we will tally emitter CNOTs up to the generation point we are in, for the first potential emitter of step 3. After that, the next emitter of step 3 is inspected, and we enter a new round of recursions. Once we exhaust all emitters, we will have made a decision about the emitter choices of the deepest recursive level, and then we will start to step-out recursions to make emitter choices till the top-most ambiguous level (i.e., the first photon absorption level where we could not make a decision of which emitter is better to use for PA).

\section{Performance of optimizers based on edge probability\label{App:CNOT_cnts_Vs_Edge_Prob}}
\begin{figure}[!htbp]
    \centering
    \includegraphics[scale=0.5]{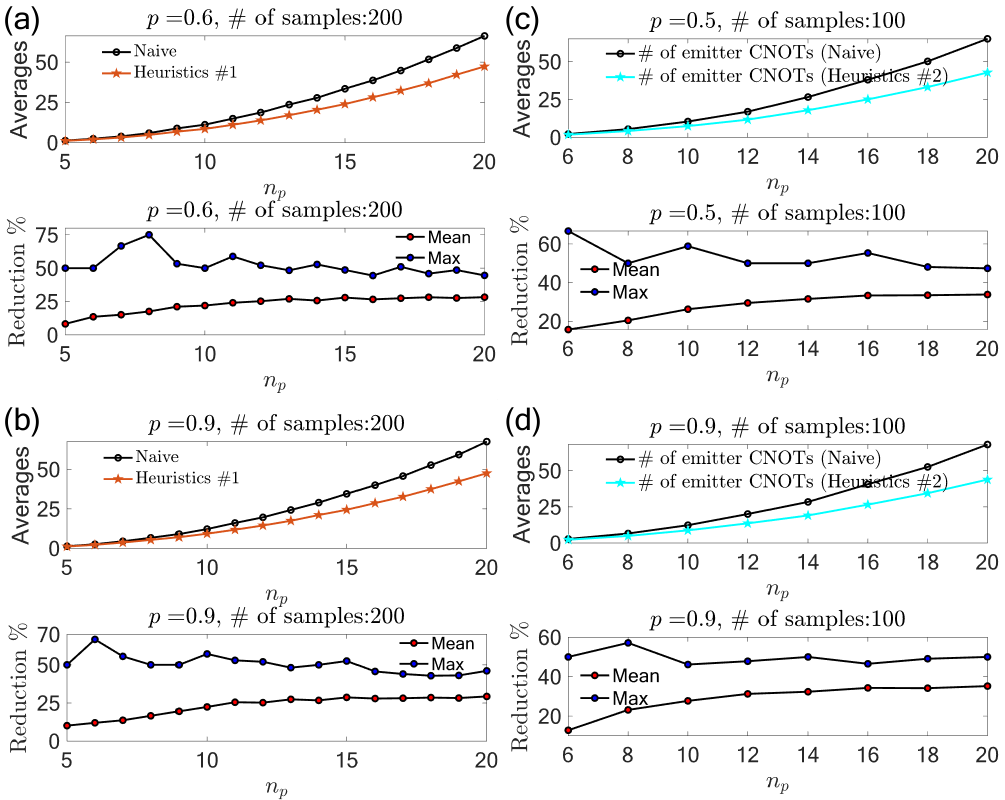}
    \caption{Performance of Heuristics optimizers versus graph size, and for different edge probabilities. (a) Top: average number of emitter CNOTs obtained by the Naive and Heuristics $\#1$ methods for a sampling size of 200 per $n_p$, and edge probability $p=0.6$ Bottom: Mean and maximum reduction in emitter CNOTs across $n_p$. (b) Same as in (a), for $p=0.9$. (c) Same as in (a), but for $p=0.5$, sampling size of 100, and the Heuristics $\#2$ method. (d) Same as in (c) for $p=0.9$.}
    \label{fig:CNOTs_vs_Edge_Prob}
\end{figure}

Here we test if our optimizers can faithfully optimize graphs irrespective of their graph structure, and in particular irrespective of how dense the graphs are. To test this, we generate random adjacency matrices by biasing the probability, $p$, for the non-zero entries to occur. We explore graphs up to $n_p=20$ photons. In Fig.~\ref{fig:CNOTs_vs_Edge_Prob}(a) we show the average CNOTs we obtain based on the Naive method, as well as the mean and max reduction. We consider 100 samples, and set the edge probability to $p=0.6$. Similar to the main text, we see that the mean reduction is about $\sim 25\%$, whereas the max reduction depends on the size of the sample set, and the random generation. The max reduction is again consistent with the main text, since we found for the sample set of the main text a max reduction of $\geq 45\%$ for graphs of size up to $n_p=20$. In Fig.~\ref{fig:CNOTs_vs_Edge_Prob}(a) we generate another sample set of 200 random graphs per $n_p$, and till $n_p=20$, but now with an edge probability of $p=0.9$. Once again, we see a very similar picture as in Fig.~\ref{fig:CNOTs_vs_Edge_Prob}(a), and also similar to the results we presented in the main text. In Figs.~\ref{fig:CNOTs_vs_Edge_Prob}(c) and (d) we repeat the calculation assuming the Heuristics $\# 2$ optimizer and probabilities of $p=0.5$, and $p=0.9$ respectively. For the CNOT counts obtained by the Heuristics $\#2$ optimizer, we set a future cutoff of 2 photon absorptions, a cutoff of up to 4 emitters, and we do not enable further recursions [we set RecurseFurther to false, see also Fig.~\ref{fig:Heuristics2_FlowChart}]. We verify, again, that the results stand in good agreement with each other, and with the results of the main text, irrespective of the edge probability.

\section{Circle graphs\label{App:Circle_Graphs}}
In this section we will explain how circle graphs can be recognized based on Bouchet's algorithm. We will also explain how the size of their LC orbit can be enumerated. Before doing so, we will introduce some notations.

\subsection{Notations}

The graphs we studied in the main text are simple, undirected graphs, denoted as $G$, with vertex set $V(G)$ and edge set $E(G)$. The order of the graph is $n=|V(G)|$, where $|\cdot|$ denotes the size (number of elements) of the set. The neighborhood of a node $v$, $N_v$, is defined as $N(v)=\{w|w\in V(G), (w,v)\in E(G)\}$. We will also denote an edge $(x,y)\in E(G)$ as $e_{xy}$. The addition of two sets $X$ and $Y$ is defined as the symmetric difference of the sets, i.e., $X+Y=X\cup Y - X\cap Y$. We also define the function $\nu_G(e_{xy})=N(x)\cap N(y)$ as the ``edge-neighbor'' function, where $x$ and $y$ are two node labels of the graph $G$. For a set $Q\subseteq E(G)$ with $Q=\{e_{xy},\dots,e_{(Q-1)Q}\}$, we define $\nu_G(Q)=\sum_{q\in Q}\nu_G(q)=\nu_G(e_{xy})+\dots+ \nu_G(e_{(Q-1)Q})$.

\subsection{Bouchet's circle graph recognition algorithm}
\begin{figure}[!htbp]
    \centering
    \includegraphics[scale=0.7]{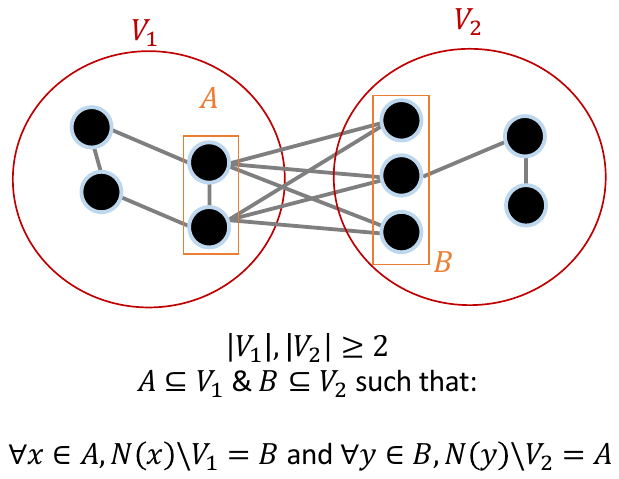}
    \caption{Example of a split in a graph. The split is the parition of nodes into the sets $V_1$ and $V_2$, such that they satisfy the condition that $|V_1|,|V_2|\geq 2$ and $A\subseteq V_1$ and $B\subseteq V_2$ form the complete bipartite graph.}
    \label{fig:Splits}
\end{figure}

The recognition algorithm of circle graphs as presented by Bouchet~\cite{BouchetCombinatorica1987} first inspects for splits of the graph. A graph which has no splits is called prime. We show an example of a graph with split $V_1|V_2$ in Fig.~\ref{fig:Splits}. Let us first explain how we can inspect for splits of a graph in a brute-force way. By definition of the split we care about the neighborhood outside of $V_1$ and the neighborhood outside of $V_2$. For those two neighborhoods, we need to test if the edges across $V_1$ and $V_2$ form a complete bipartite graph. Thus, we can follow the procedure below:
\begin{enumerate}
    \item Form all possible $V_1 \subset V(G)$, with $n-2\geq |V_1|\geq 2$. For each one of these subsets, define $V_2=V(G)\backslash V_1$, for which we also ensure that $|V_2|\geq 2$.
    \item For each possible $V_1|V_2$ partition of step 1,  find $N(V_1)\backslash V_1$. $N(V_1)$ is defined as $N(V_1)\backslash V_1=\{y\in V_2: (x,y)\in E(G),\forall x\in V_1\}:=B$. Also, find $N(V_2)\backslash V_2$ which is defined as $N(V_2)\backslash V_2 =\{x\in V_1: (x,y)\in E(G), \forall y\in V_2\}:=A$. 
    \item Given $A$ and $B$ from the above step, check if $\forall x \in A$ and $\forall y\in B$, there is an edge $(x,y)\in E(G)$. If for any $x\in A$ and $y\in B$, $(x,y)\notin E(G)$, then the partition $V_1|V_2$ is not a split. If $\forall x \in A$ and $\forall y\in B$, there is an edge $(x,y)\in E(G)$, then exit and store the split $V_1|V_2$, as well as the sets $A$ and $B$.
    \item If the routine has not terminated yet and there are still remaining $V_1|V_2$ partitions to be inspected, continue the above steps for checking for splits. If all partitions $V_1|V_2$ have been checked, exit and announce that the graph is prime.
\end{enumerate}

The above subroutine is needed for checking for splits during the circle graph recognition. Note, that there are more efficient implementations in the literature for checking for primality and identifying a split~\cite{SpinradSplitDecomp,CharbitSplitDecomp,Cunningham1982DecompositionOD}.  Let us now explain how to test if a graph is circle, when the input graph is prime (split-free). If the graph is split-free and has order $n\geq 6$, then one starts removing nodes following Eq.~(\ref{Eq:VertexMinors}). Because the input graph is prime, there has to be some vertex $v$ so that one of the three subgraphs of $G$ of Eq.~(\ref{Eq:VertexMinors}) is prime. For each $v\in \{1,\dots,n\}$, we find the three minors of Eq.~(\ref{Eq:VertexMinors}) and test if one of them is still prime. Once we find a prime subgraph, we terminate the search and exit the loop. We store the node $v$ we removed together with the LC operation which is either $\emptyset$, $*v$, or $*vwv$. We terminate this reduction when we reach a graph that consists of 5 nodes. Suppose this graph is called $G_5$. At this step, according to Bouchet, $G_5$ is necessarily LC equivalent to $C_5$ (the circle graph of 5 nodes), because $G_5$ is prime. Thus, we can test to which LC graph of $C_5$ our graph matches with, and we construct the relevant alternance word, $m(G_5)$. Then, using the last removed node and the information of which LC operation we applied, we insert the node back into the word by respecting the previous alternances (i.e., those of $m(G_5)$), and additionally we try to match our new word to $m(G_6)$. Similarly, we proceed to reach the top-most level until we have exhausted the list of nodes we removed (and the respective list of LC operations). If at any step in the construction of the word we cannot satisfy the alternances, it is because the graph is not a circle graph.

Now, let us turn to the case when the input graph is not prime. We first separate the input graph into prime ones, using a split decomposition, and then we repeat the procedure of constructing the word for each prime graph. As an example, if we have a graph of order $n=6$ and it happens to have a split $V_1 =\{1,3\}$ and $V_2=\{2,4,5,6\}$, with $A=\{1\}$ and $B=\{2,4\}$, then we take the subgraph spanned by $V_1$, and connect all nodes from within $A$ with a new dummy node. We also take the subgraph spanned by $V_2$ and connect all nodes from within $B$ to another dummy node. Thus, we have two separate graphs and if those separate graphs are prime, we repeat the subroutine of circle graph recognition for each piece.
If the new graphs are not prime, we keep on identifying the splits and introducing new dummy nodes. If we fail to respect alternances for any of the prime subgraphs then we know the input graph is not prime. 
Suppose that we succeed in constructing a word for each prime subgraph. Those words for each prime graph include dummy nodes, which are entirely artificial and do not correspond to nodes of the original input graph. Suppose we have a word $m_1(G_1)=vA_1vB_1$ and a word $m_2(G_2) = vA_2vB_2$ (we can always do cyclic permutation of the words so that one occurrence of the dummy node, $v$, is at the beginning). Suppose that $v$ was used to split a bigger non-prime graph composed of both words. To remove the artificial dummy node, we simply construct the new word $m_{12}=A_1A_2B_1B_2$, and this represents the non-prime graph before splitting it into prime ones. The reason why we can join the two words $m_1$ and $m_2$, is because if the corresponding graphs $G_1$ and $G_2$ are circle graphs, and $G_1$ and $G_2$ are composable, then the composition of $G_1$ and $G_2$, denoted as $G_1 c G_2$ will also be a circle graph. The two graphs $G_1$ and $G_2$ are composable if $|V(G_1)|,|V(G_2)|\geq 3$, $|V(G_1)\cap V(G_2)|=1$. Then, the composition of $G_1$ and $G_2$ is defined on the vertex set $V(G_1)\backslash \{v\} \cup V(G_2)\backslash \{v\}$, and it's edge-set is given by:
\begin{equation}
\begin{split}
    E(G_1cG_2)&=E(G_1\backslash\{v\})\cup E(G_2\backslash\{v\})\cup \\& 
    \{(v_1,v_2):(v_1,v) \in E(G_1) \text{~and~} (v_2,v)\in E(G_2)\}.
    \end{split}
\end{equation}
This implies that the composition of $G_1$ and $G_2$ recovers the graph before performing the split operation, and we remove the dummy node $v$.
This is the main idea of how we can then start gluing the words to construct the final alternance word. We also provide the code for this circle graph recognition in Ref.~\cite{ETgithubCode2024}.

\subsection{Bouchet's LC orbit enumeration algorithm \label{App:index_k(G)}}

In the main text we mentioned that the size of the LC orbit is defined as the ratio of the index $e(G)$ and $k(G)$. Here we will explain how $k(G)$ is calculated using information from Ref.~\cite{AxelJofMathPhysics2020}, and later on we will provide examples of how this index is calculated for particular graph families.

Let us begin with the definition of $k(G)$. It is defined as:
\begin{equation}\label{Eq:k(G)}
    k(G)=\begin{cases}
        2^{\dim(\mathcal{L}(G)^\perp)}, \text{if } G\notin \mu \\
        2^{\dim(\mathcal{L}(G)^\perp)}+2,\text{if } G\in \mu.
    \end{cases}
\end{equation}
To calculate it, we need to know if the graph belongs in class $\mu$. In Ref.~\cite{AxelJofMathPhysics2020}, the authors mention that for a graph to belong in $\mu$ it has to satisfy all the following properties:
\begin{enumerate}
    \item All nodes of $G$ have odd degree.
    \item $\nu_G(e_{xy})$ has even size, $\forall (x,y)\in E(\bar{G})$. In other words, for every edge $e_{xy}=(x,y)\in E(\bar{G})$ we test if in the original graph, $G$, $\nu_G(e_{xy})=N_G(x)\cap N_G(y)$ has even size. Note that here $N_G(x)$ means that we find the neighborhood of $x$ in the original graph, $G$.
    \item $\forall C\in \mathfrak{C}$, if the cycle $C$ has an even (odd) number of edges in its set, then $\nu_G(C)$ should have even (odd) size. Here $\mathfrak{C}$ is the space composed of all possible cycles of the graph. In other words, if $C=\{e_{v_j,v_i},\dots,e_{v_k,v_l}\}$, we test if $\nu_G(C)=\sum_{e\in C}\nu_G(e)$ has even (odd) size when the set $C$ has even (odd) size. The summation here is taken as the symmetric difference. This condition can be tested only on the cycle basis of $G$~\cite{AxelJofMathPhysics2020}, which can be found by promoting the sets into binary vectors and performing Gaussian elimination.
\end{enumerate}
After we find if the graph belongs in class $\mu$, we need to find the dimension of $\mathcal{L}(G)^\perp$.
The space $\mathcal{L}(G)^\perp$ is the orthogonal complement of the bineighborhood space, $\mathcal{L}(G)$. $\mathcal{L}(G)$ is formed from the neighborhood cycle space, $\nu_G(\mathfrak{C})$, and the neighborhood complement edge space, $\nu_G(\mathfrak{B})$. Note that $\mathfrak{C}$ is the regular cycle space of the graph, but we need to apply the neighborhood function $\nu_G$ on every cycle (or the basis of cycles) to obtain the space $\nu_G(\mathfrak{C})$ (and similarly we need to apply the function $\nu_G$ on the complement edge space to obtain $\nu_G(\mathfrak{B})$). We can calculate the dimension of $\mathcal{L}(G)^\perp$ as:
\begin{equation}
\begin{split}
 \dim(\mathcal{L}(G)^\perp)&=n-\dim(\mathcal{L}(G))
 \\&=
n-\dim(\nu_G(\mathfrak{B})\cup \nu_G(\mathfrak{C})).
 \end{split}
\end{equation}

To find the cycle space it suffices to find a cycle basis, which for example, numerically, can be done with the help of a package (in Matlab there is a built-in function, which outputs the nodes contained in each cycle of the cycle basis). Once we have the information of the nodes, we construct the edges that are contained in each cycle. For example, if we get the nodes $\{v_1,v_2,v_3\}=\{1,4,5\}$, then the elements of the cycle, $C$, are $\{\{1,4\},\{4,5\},\{5,1\}\}$. Suppose that $l$ is the last node contained in the cycle. In other words, we construct each cycle by adding the element $\{v_i,v_{i+1}\}$ into the set $C$, where $i$ runs from 1 and till the last node that forms the cycle, with the condition that $v_{1+l}=v_1$. At this stage, we have a cycle $C=\{e_{xy},\dots,e_{zx}\}$, and then for each edge we calculate the function $\nu_G(e_{lm})=N(l)\cap N(m)$. In the end, we add all of them (symmetric difference) as $\nu_G(C)=N(x)\cap N(y)+\dots + N(z)\cap N(x)$.
 
We can then promote the set $\nu_G(C)$ into a binary vector. If $\nu_G(C)$ is the empty set, we define a vector of all zeros of size $n$ ($n$ is the order of $G$). Otherwise, we set the value $1$ to the $i$-th entries which correspond to the nodes $i$ that are within the set $\nu_G(C)$.  For example, if $n=5$ and $\nu_G(C)=\{3,4,5\}$ we then set the binary vector $\tilde{\nu}_G(C)=[0,0,1,1,1]$. 

We also need to find the complement edge space, $\mathfrak{B}$. We first create the complement of the graph $\bar{G}$. For every edge $(x,y)\in E(\bar{G})$ we calculate in the original graph $G$, the function $\nu_G(x,y)=N_G(x)\cap N_G(y)$, and again promote these sets to binary vectors.

Once we have all the binary vectors for the neighborhood complement edge space and the neighborhood cycle space, we put them in a matrix and perform Gaussian elimination. The number of non-zero rows gives us the value of  $\dim(\mathcal{L}(G))$. Then, we know that $\dim(\mathcal{L}(G)^\perp)=n-\dim(\mathcal{L}(G))$, and we can then find $k(G)$.

\subsection{Index \texorpdfstring{$k(G)$}{k(G)} of \texorpdfstring{$K_n$}{Kn} \label{App:k(G)_of_Kn}}
We will begin by finding the index $k(G)$ for a complete graph: $G=K_n$, for $n\geq 3$. It is easy to prove that the total number of independent cycles is $(n-1)(n-2)/2$. This is because every cycle for $K_n$ (triangle with all-to-all connectivity) is defined based on the $ijk$ indices. First, we fix the index $i=1$, and we consider $k>j$ to avoid double-counting. Since $i,j,k$ all run up to $n$, where $n=|V(K_n)|$, then we know that the number of $ijk$ triplets with fixed $i$ is $(n-1)(n-2)/2$. Next we know that for any cycle $C=\{(x,y),(y,z),(z,x)\}$  of a $K_n$ graph, the following holds:
\begin{equation}
\begin{split}
    \nu_G(C)&=N_G(x)\cap N_G(y)+N_G(y)\cap N_G(z)\\&\qquad+N_G(z)\cap N_G(x)
    \\&=V\backslash\{x,y\}+V\backslash\{y,z\}+V\backslash\{z,x\}
    \\&=V\backslash\{x,y\}\cup V\backslash\{y,z\}-V\backslash\{x,y\}\cap V\backslash\{y,z\}\\&\qquad+V\backslash\{z,x\}
    \\&=V\backslash\{y\}-V\backslash\{x,y,z\}+V\backslash\{z,x\}
    \\&=\{x,z\}+V\backslash\{x,z\}=V=\{1,\dots,n\}.
    \end{split}
\end{equation}
The above equality holds because every node $x\in V(K_n)$ is connected to all remaining nodes, and hence the intersection of the neighborhoods $N_G(x)$ and $N_G(y)$ is the entire vertex set except for $x$ and $y$. The disjoint union of such intersections over a cycle is then the entire vertex set. Therefore, if we take any given cycle of $K_n$, we find that there is only one basis vector for $\nu_G(\mathfrak{C})$ and hence $\dim(\nu_G(\mathfrak{C}))=1$. We also know that $\bar{K}_n$ is the fully disconnected graph and thus, $\dim(\nu_G(\mathfrak{B}))=0$. This is true, because $\mathfrak{B}$ contains no edges, i.e., it is the empty set. This implies that $\dim(\mathcal{L}(G))=\dim(\nu_G(\mathfrak{C}))=1$ and thus we have:
\begin{equation}
    \dim(\mathcal{L}(K_n)^\perp)=|V|-1=n-1.
\end{equation}
Therefore, $k(K_n)=2^{n-1}$. Let us also prove that $K_n$ is not in class $\mu$. For odd $n$, each vertex $v$ has $n-1$ neighbors, so their degree is even. Hence, $K_n$ fails to belong in class $\mu$ by condition 1. If $n$ is even,  any cycle $C=\{e_{xy},e_{yz},e_{zx}\}$ has odd size, but it holds that $|\nu_G(C)|=|\{1,\dots,n\}|$ has even size. In this case, $K_n$ fails to belong in class $\mu$ by condition 3. Thus, $K_n\notin \mu$.

\subsection{Index \texorpdfstring{$k(G)$}{k(G)} of \texorpdfstring{$K_n^n$}{Knn}\label{App:k(G)_of_RGS}}
The next step is to calculate $k(K_n^n)$, where $G=K_n^n$ is a repeater graph. We know that the RGS has the same neighborhood cycle space as the complete graph, because the cycles of $K_n^n$ only involve core qubits, and because the leaf nodes do not lie in any of the intersections in $N(x)\cap N(y)+N(y)\cap N(z)+N(z)\cap N(x)$, where $x,y,z$ are all core qubits. Therefore we have that $\nu_G(C_j)=\{\text{core nodes}\}, \forall C_j \in \mathfrak{C}$. The complement $\bar{K}_n^n$ is not fully disconnected, but for any $e_{xy}\in E(\bar{K}_n^n)$ we will find that $N_G(x)\cap N_G(y)$ is either the empty set, or contains a single core qubit. Thus:
\begin{equation}
    \nu_G(e_{xy})=\begin{cases}
        \{\text{a single core node}\}, \\
        \{\emptyset\}
    \end{cases}.
\end{equation}
Therefore,  $\dim(\nu_G(\mathfrak{B}))=n$ since there are $n$ distinct elements (i.e., $n$ distinct core qubits). The symmetric addition of all core nodes gives rise to the entire set of core nodes, which overlaps with the only element of $\nu_G(\mathfrak{C})$. Thus, the dimension of $\mathcal{L}(G)$ is $\dim(\mathcal{L}(G))=\dim(\nu_G(\mathfrak{C}))+\dim(\nu_G(\mathfrak{B}))-\dim(\nu_G(\mathfrak{C})\cap \nu_G(\mathfrak{B}))=1+n-1=n$. Therefore, we find that $\dim(\mathcal{L}(K_n^n)^\perp)=2n-n=n$, giving rise to $k(K_n^n)=2^n$. We can prove that the RGS is not in class $\mu$, due to the fact that the only element in its neighborhood cycle basis, $\nu_G(\mathfrak{C})$, fails to satisfy condition 3.

\subsection{Recursion formula for complete/star graph \label{App:CompleteGraphRECURSION}}
In the main text we mentioned that the recursion formula for evaluating $e(K_n)=e(S_n)$ is:
\begin{equation}
    T(n)=2T(n-1)+2^{n-1}.
\end{equation}
We can solve this recursion by considering the following generating function:
\begin{equation}
    f(X):=\sum_{n=1}^\infty T(n)X^n.
\end{equation}
We know that $T(1)=e(P_1)=2=a$ which holds true for $|V|=1$ [see also Bouchet's paper~\cite{BouchetDiscreteMathematics1993}]. Thus, we have:
\begin{equation}
    \begin{split}
        f(X)&=\sum_{n=1}^\infty T(n)X^n  = T(1)X+\sum_{n=2}^\infty T(n)X^n \\
        &=aX+2\sum_{n=2}^\infty [T(n-1)+2^{n-2}]X^n \\
        &=aX+2\sum_{n=2}^\infty T(n-1)X^n +\sum_{n=2}^\infty 2^{n-1}X^n \\
        &=aX+2\sum_{k=1}^\infty T(k)X^{k+1}+\frac{2X^2}{1-2X}\\
        &=aX+2Xf(X)+\frac{2X^2}{1-2X},
    \end{split}
\end{equation}
in the second-to-last line we have replaced the geometric sum with its value $2X^2/(1-2X)$. Solving for $f(X)$ we find:
\begin{equation}
    f(X)=\frac{aX}{1-2X}+2\frac{X^2}{(1-2X)^2}.
\end{equation}
We can rewrite $f(X)$ as:
\begin{equation}
\begin{split}
    f(X)&=a \sum_{n=1}^{\infty}2^{n-1}X^n +\sum_{n=1}^\infty (n-1)2^{n-1}X^n \\
    &=\sum_{n=1}^\infty X^n (2^{n-1}(a+n-1))=\sum_{n=1}^\infty T(n) X^n.
\end{split}
\end{equation}
Therefore, we conclude that $T(n)=2^{n-1}(a+n-1)=2^{n-1}(n+1)=e(K_n)$.

\subsection{Recursion formula for RGS \label{App:RGSRECURSION}}
In the main text, we found the recursion formula:
\begin{equation}\label{Eq:RecRGS}
    e(K_n^x)=2e(K_n^{x-1})+2e(K_{n-1}^{x-1}).
\end{equation}
Let us assume a single leaf qubit, attached to only one core qubit, so $x=1$. In this case we have:
\begin{equation}
    e(K_n^1)=2[e(K_n^0)+e(K_{n-1}^0)].
\end{equation}
For $x=2$ we have:
\begin{equation}
\begin{split}
    e(K_n^2)&=2[e(K_n^1)+e(K_{n-1}^1)]
    \\&=4[e(K_n^0)+2e(K_{n-1}^0)+e(K_{n-2}^0)].
    \end{split}
\end{equation}
For $x=3$ we have:
\begin{equation}
    e(K_n^3)=8[e(K_n^0)+3e(K_{n-1}^0)+3e(K_{n-2}^0)+e(K_{n-3}^0)].
\end{equation}
Thus, we see that the binomial coefficients appear in this sequence and hence we can write the formula:
\begin{equation}
    e(K_n^x)=2^x\sum_{k=0}^x \begin{pmatrix}
        x \\
        k
    \end{pmatrix}e(K_{n-k}^0), n-1\geq x \geq 1, n\geq 3.
\end{equation}
We also know that $e(K_n^0)=2^{n-1}(n+1)$, which implies that $e(K_{n-k}^0)=2^{n-k-1}(n-k+1)$. Let us consider the case of $x=n$, which corresponds to the un-encoded RGS. Solving the above sum for $x=n-1$ we find:
\begin{equation}
    e(K_n^{n-1})=2^n 3^{n-2}(n+2).
\end{equation}
Plugging this into Eq.~(\ref{Eq:RecRGS}) we obtain:
\begin{equation}
    e(K_n^n)=2^{n+1} 3^{n-2}(n+2)+2e(K_{n-1}^{n-1}), ~n>1.
\end{equation}
We can now follow a similar procedure as before to solve this recursion. We let $e(K_n^n):=T(n)$ and thus we have:
\begin{equation}
    T(n)=2T(n-1)+2^{n+1} 3^{n-2}(n+2).
\end{equation}
We further know that $T(1)=e(K_1^1)=e(P_2)=e(K_2)=6$. We consider again the generating function:
\begin{equation}
    f(X):=\sum_{n=1}^\infty T(n)X^n,
\end{equation}
for which we have:
\begin{equation}
    \begin{split}
        f(X)&=\sum_{n=1}^\infty T(n)X^n 
        =e(K_2)X +\sum_{n=2}^\infty T(n)X^n 
        \\&
        =6X + \sum_{n=2}[2T(n-1)+2^{n+1} 3^{n-2}(n+2)]X^n 
        \\&
        =6X + 2\sum_{k=1}^\infty T(k)X^{k+1}+\sum_{n=2}^\infty 2^{n+1} 3^{n-2}(n+2)X^n
        \\&
        =6X + 2X f(X) - \frac{16X^2 (9X-2)}{(6X-1)^2}.
    \end{split}
\end{equation}
Solving for $f(X)$ we find:
\begin{equation}
    f(X) = 6\frac{X}{1-2X}- 16 \frac{X^2(9X-2)}{(6X-1)^2(1-2X)}.
\end{equation}
We now need to work with the second term and re-write it as:
\begin{widetext}
\begin{equation}
    \begin{split}
&-16\frac{X^2(9X-2)}{(6X-1)^2(1-2X)}=\frac{AX+B}{1-2X}+\frac{C}{(6X-1)}+\frac{D}{(6X-1)^2}
\\&=
\frac{(AX+B)(36X^2-12X+1)}{(1-2X)(6X-1)^2}+    \frac{C(6X-1)(1-2X)}{(6X-1)^2(1-2X)}+  \frac{D(1-2X)}{(6X-1)^2(1-2X)}  
\\&=
\frac{36AX^3 - 12AX^2 +AX +36BX^2 - 12BX+B}{(1-2X)(6X-1)^2} +    \frac{C(-12X^2+8X -1)}{(6X-1)^2(1-2X)}+\frac{D-2DX}{(6X-1)^2(1-2X)}  
\\&=
\frac{36AX^3+(-12A+36B-12C)X^2+(A-12B+8C-2D)X+B-C+D}{(6X-1)^2(1-2X)}
    \end{split}
\end{equation}
From the above decomposition we find that $A=-4$, and we have the following conditions:
\begin{eqnarray}
    -12(-4)+36B-12C&=&32 \\
    -4-12B+8C-2D &=& 0 \\
    B-C+D &=& 0.
\end{eqnarray}
The first condition can be re-written as $9B-3C=-4$, and the second condition as $-2-6B+4C-D=0$. Plugging in the last condition into the second one we get $-5B+3C=2$. Using the first and second condition we find $B = -1/2$, and hence $C=-1/6$, $D=1/3$. Thus, we can now re-write $f(X)$ as:

\begin{equation}
\begin{split}
    f(X) &= 6 \frac{X}{1-2X} - \frac{4X+1/2}{1-2X}-\frac{1}{6}\frac{1}{6X-1}+\frac{1}{3}\frac{1}{(6X-1)^2}
    \\&=
    -\frac{1}{2}\frac{1}{1-2X}-\frac{1}{6}\frac{1}{6X-1}+2 \frac{X}{1-2X} +\frac{1}{3}\frac{1}{(6X-1)^2}
    \\& = 
    -\frac{1}{2}\sum_{n=0}^\infty 2^n X^n
    +\frac{1}{6}\sum_{n=0}^\infty 6^n X^n
    +2 \sum_{n=1}^\infty 2^{n-1}X^n 
    + \frac{1}{3}\sum_{n=0}^\infty 6^n (1+n)X^n
    \\&=
    -1/2 + 1/6 +1/3 + \sum_{n=1}^\infty \Big(-2^{n-1}+6^{n-1}+2^n + 6^n\frac{1+n}{3}\Big)X^n
    \\&= \sum_{n=1}^\infty \Big(   
    6^{n-1}(2(1+n)+1)+2^{n-1}(-1+2)
    \Big) X^n
    \\& = \sum_{n=1}^\infty \Big(6^{n-1}(3+2n)+2^{n-1}\Big)X^n.
    \end{split}
\end{equation}

\end{widetext}
Thus, we conclude that:
\begin{equation}
    e(K_n^n)=6^{n-1}(3+2n)+2^{n-1}.
\end{equation}


\section{Scaling of the RGS LC orbit\label{App:RGSOrbit_Enumer}}

\begin{figure}[!htbp]
    \centering
    \includegraphics[scale=0.53]{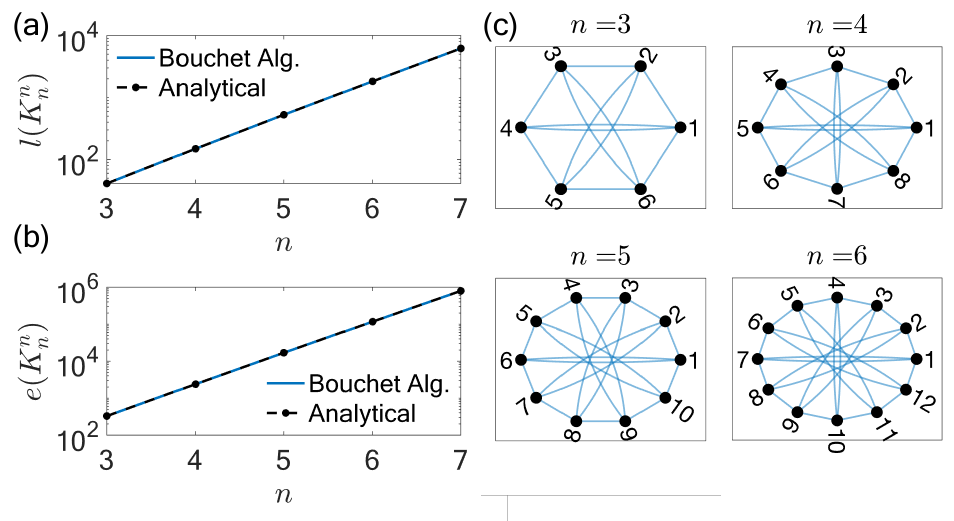}
    \caption{(a) Size of the LC orbit $l$ of the repeater graph $K_n^n$, which includes isomorphs, as a function of the core qubits $n$. (b) Index $e(K_n^n)$ of the RGS as a function of the core qubits. The solid lines were obtained by numerical evaluation using Bouchet's algorithm and the dashed lines correspond to our analytical expressions. (c) Multigraphs of RGSs for $n=3$ up to $n=6$ core qubits.}
    \label{fig:RGSOrbit_Enumer}
\end{figure}

\begin{figure}[!htbp]
    \centering
    \includegraphics[scale=1]{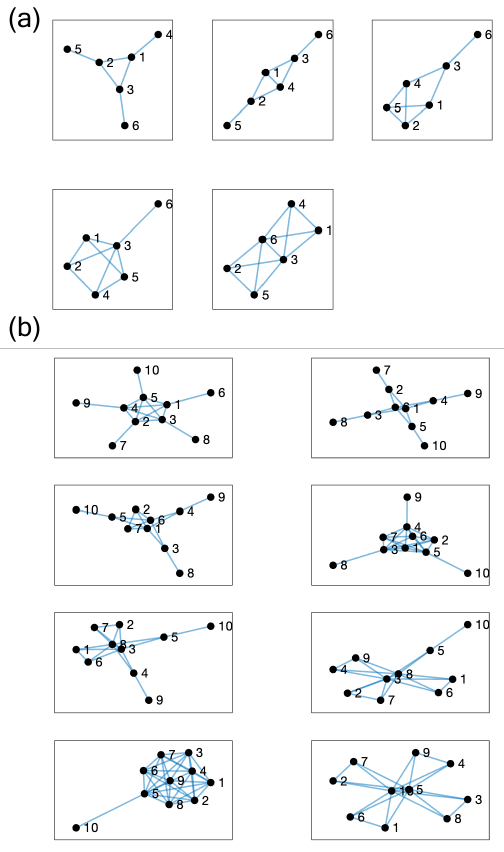}
    \caption{Non-isomorphic LC orbit of RGS for $n=3$ core qubits (a), and $n=5$ core qubits (b).}
    \label{fig:nonIsoRGSOrbit}
\end{figure}

Here we plot the scaling of the orbit of repeater graph states. Figure~\ref{fig:RGSOrbit_Enumer}(a) shows the size of the LC orbit as a function of the number of core qubits of the RGS. The solid line shows the result we obtain from Bouchet's orbit enumeration algorithm and the dashed line shows the analytical expression we showed in the main text. The two lines agree exactly with each other. In Fig.~\ref{fig:RGSOrbit_Enumer}(b) we also show the index $e(K_n^n)$. The solid line shows again the value we obtain by running Bouchet's algorithm and the dashed line shows the analytical result we extracted. The reason why the scaling of the RGS LC orbit (including isomorphs) is exponential is because there are a lot of symmetries in the corresponding multigraphs (where enumeration of the Euler tours gives rise to the index $e(K_n^n)$. The multigraphs for $K_3^3$ up to $K_6^6$ are shown in Fig.~\ref{fig:RGSOrbit_Enumer}(c). When we discard the isomorphs the orbit follows the linear scaling we mentioned in the main text. As an example, we show the non-isomorphic LC orbit of the RGS for $n=3$ and $n=5$ core qubits in Fig.~\ref{fig:nonIsoRGSOrbit}.

\section{Preparing RGSs with parallel emissions\label{App:RGS_Parallel_Emissions}}
\begin{figure*}[!htbp]
    \centering
    \includegraphics[scale=0.67]{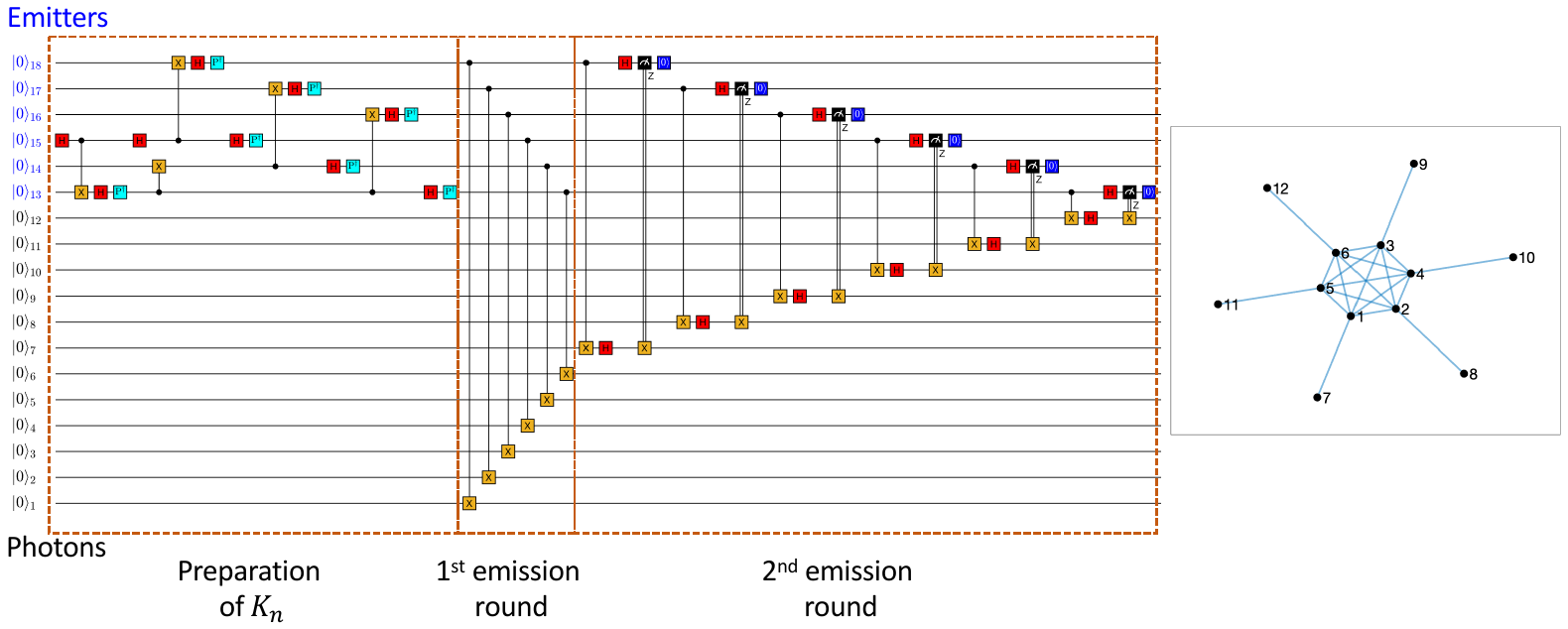}
    \caption{Circuit to prepare the RGS shown on the right, with clockwise emission ordering. The first step  prepares $K_n$. The next step performs all core-emissions in parallel, and the last step performs all leaf-emissions in parallel. The emitters are measured in the end. Red boxes are Hadamard gates, cyan boxes are conjugate Phase gates, yellow boxes are X gates, and black boxes are computational basis measurement.}
    \label{fig:RGS_Parallel_Emissions}
\end{figure*}

Here we show how to prepare repeater graph states where we label first clockwise the core qubits, and follow again clockwise labeling of leaf qubits. The labeling of leaf qubits starts from the first node we picked to begin the labeling of core qubits. An example for the $K_6^6$ graph is shown in the right panel of Fig.~\ref{fig:RGS_Parallel_Emissions}. The left panel of Fig.~\ref{fig:RGS_Parallel_Emissions} shows the circuit to prepare it, which consists of three main steps: i) preparation of a complete graph, ii) first round of parallel emissions, iii) second round of parallel emissions and measurements of the emitters. This circuit was obtained from the Heuristics $\# 1$ method, by setting the Back-substitution option to true, and setting the extra inspection tests for free PA to false.

\section{Optimal Circuits to prepare RGSs with many leaves \label{App:Circuits_RGS_Many_Leaves}}
\begin{figure*}[!htbp]
    \centering
    \includegraphics[scale=0.82]{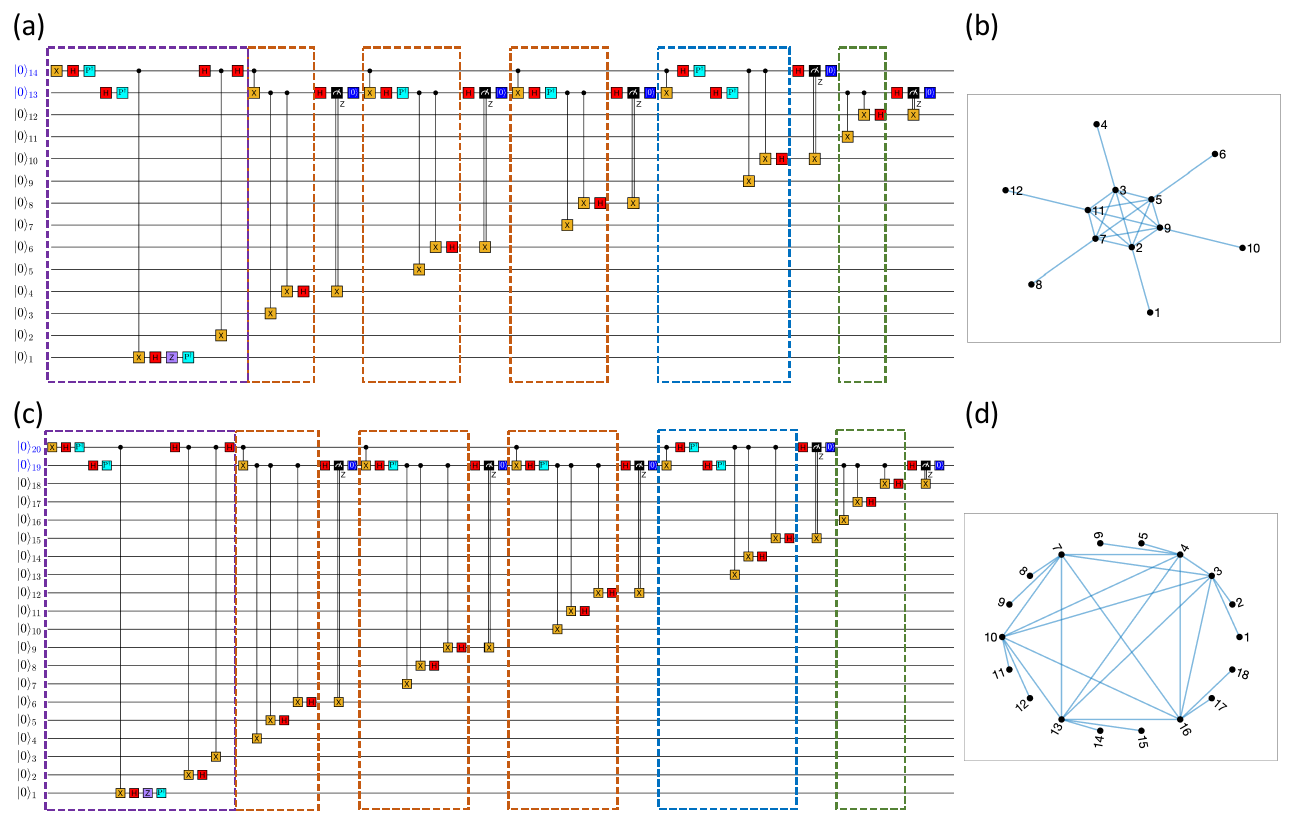}
    \caption{(a) Circuit with optimal number of CNOTs to prepare the RGS shown in (b). (c) Circuit with optimal number of CNOTs to prepare the RGS with 2 leaves per core qubit shown in (d). The purple boxes correspond to $Z$ gates. }
    \label{fig:OptCircuitsRGSManyLeaves}
\end{figure*}

In Fig.~\ref{fig:OptCircuitsRGSManyLeaves} we show the optimal circuits to prepare an RGS with 6 core qubits. We generate an optimal emission ordering using Algorithm~\ref{alg:RGSDepth1Ordering}, which also allows us to select the  number of leaves per core node. The circuits are obtained by using our Heuristics $\# 1$ optimizer by setting the Back-substitution option to true, as well as the extra inspections for free PA to true. (Similar circuits are obtained using the Naive method, by setting the same conditions to true.) Figure~\ref{fig:OptCircuitsRGSManyLeaves}(a) shows the optimal circuit if we have one set of leaves. The orange boxes (besides the first one) indicate the emission part that gets repeated (the first orange box has two less gates compared to the rest). The purple box corresponds to the first 2 emissions and contains additional local gates. The blue box is similar to the orange boxes, but with additional $H$ and $P^\dagger$ gates on the control emitter after the emitter-emitter CNOT. The green box corresponds to the last two emissions by the emitter qubit that was the target in the previous blue box. The orange boxes, as well as the orange and blue boxes, are interleaved by measurements of the same emitter. The circuit in Fig.~\ref{fig:OptCircuitsRGSManyLeaves}(c) follows a very similar pattern as in Fig.~\ref{fig:OptCircuitsRGSManyLeaves}(a), since it prepares the RGS shown in Fig.~\ref{fig:OptCircuitsRGSManyLeaves}(d), which now has one extra leaf qubit per core node. Every indicated box has now an extra emission. For the purple box, we apply Hadamards on the first photons besides the last photon included in this box. For the orange boxes we apply Hadamards on all but the first photon included in this box. For the blue box and green boxes the Hadamards on the photons follow the same pattern as in the orange box.

\section{Optimal emitter CNOTs for order 6 graphs\label{App:Order6CNOTs}}
\begin{figure*}[!htbp]
    \centering
    \includegraphics[scale=0.73]{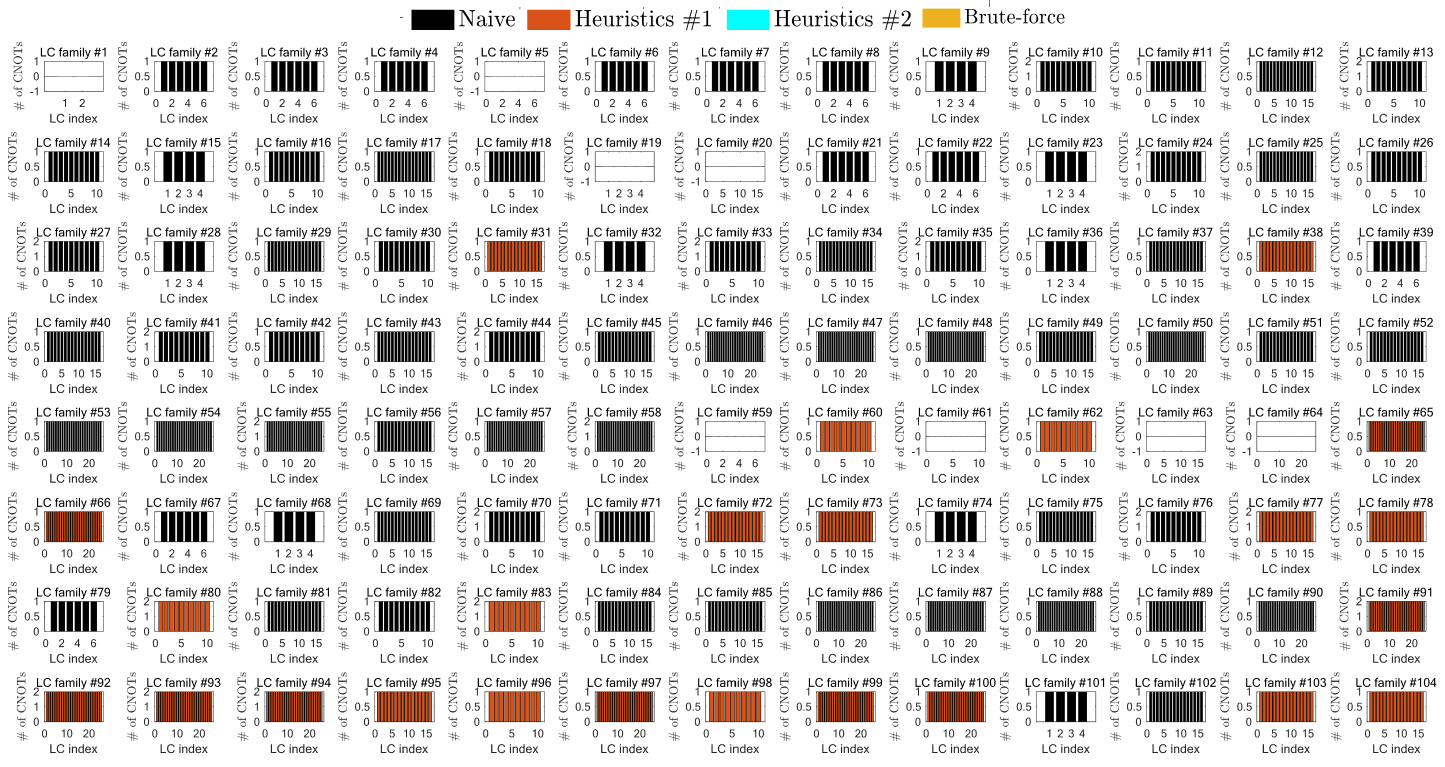}
    \caption{Optimal emitter CNOTs for the first 104 LC families of $n_p=6$ photonic graphs. Each panel shows the corresponding LC family index out of 312 families, and the number of emitter CNOTs across the non-isomorphic LC graphs of the family. The colors show the optimizer that was used. Black: Naive, red: Heuristics $\# 1$, cyan: Heuristics $\# 2$, yellow: Brute-force.}
    \label{fig:np6_first}
\end{figure*}

\begin{figure*}[!htbp]
    \centering
    \includegraphics[scale=0.72]{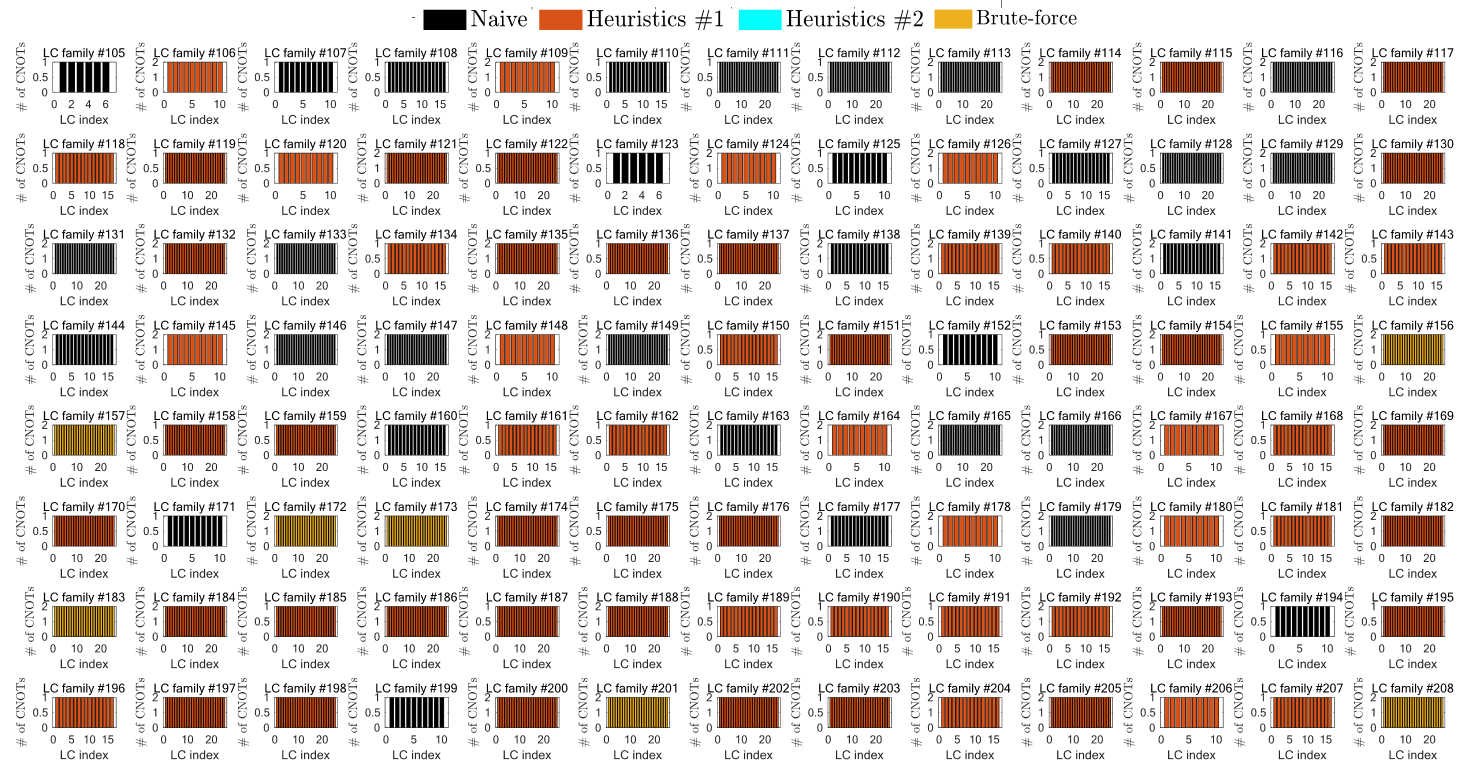}
    \caption{Optimal emitter CNOTs for the second set of 104 LC families of $n_p=6$ photonic graphs. Each panel shows the corresponding LC family index out of 312 families, and the number of emitter CNOTs across the non-isomorphic LC graphs of the family. The colors show the optimizer that was used. Black: Naive, red: Heuristics $\# 1$, cyan: Heuristics $\# 2$, yellow: Brute-force.}
    \label{fig:np6_second}
\end{figure*}

\begin{figure*}[!htbp]
    \centering
    \includegraphics[scale=0.73]{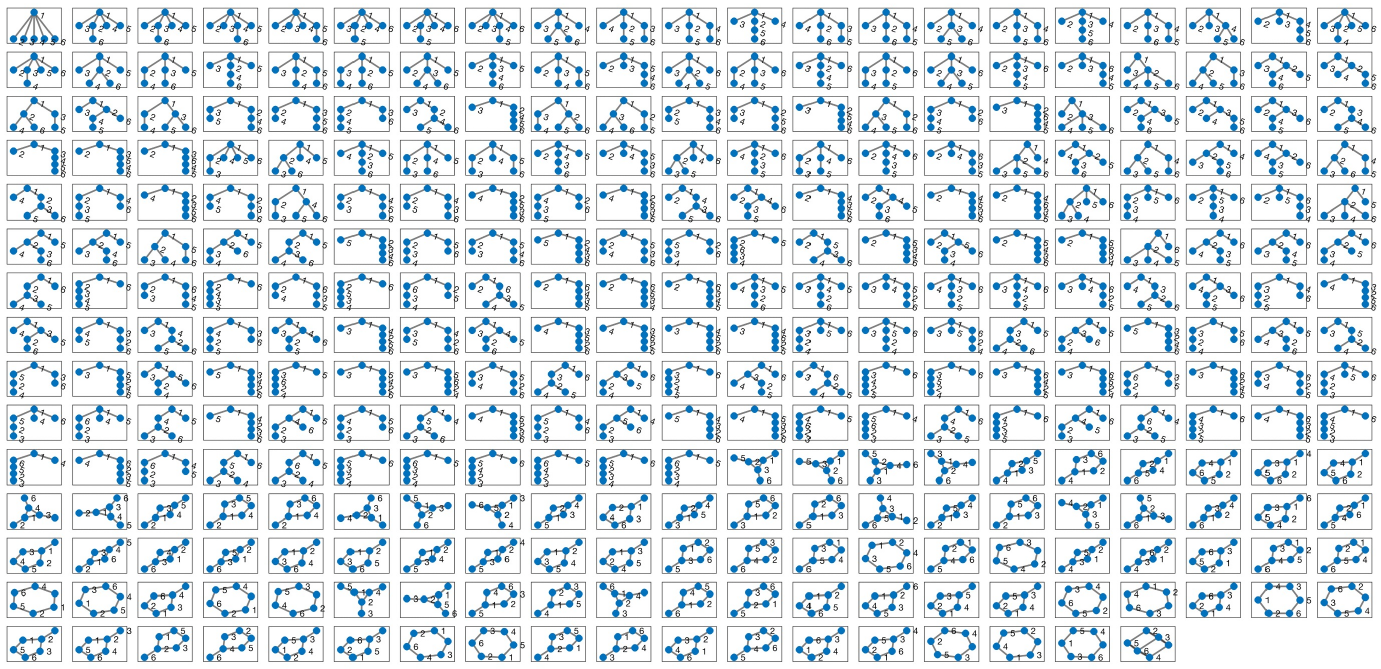}
    \caption{Graph representatives of all 312 LC orbits (including isomorphs) of order 6.}
    \label{fig:n6Orbits}
\end{figure*}

Here we provide the results for the optimization of all LC families of $n_p=6$ photons. As we mentioned in the main text, there are 312 non-LC equivalent families. In Fig.~\ref{fig:np6_first} and Fig.~\ref{fig:np6_second} we show the emitter CNOT counts across all LC families up to LC family $\# 208$. The different colors correspond to the optimizer that managed to minimize the CNOT counts across all graphs within the particular orbit. In Fig.~\ref{fig:n6Orbits}, we further show all 312 graph representatives.

\section{Statistics of improvement for Heuristic methods}
\begin{figure}[!htbp]
    \centering
    \includegraphics[scale=0.5]{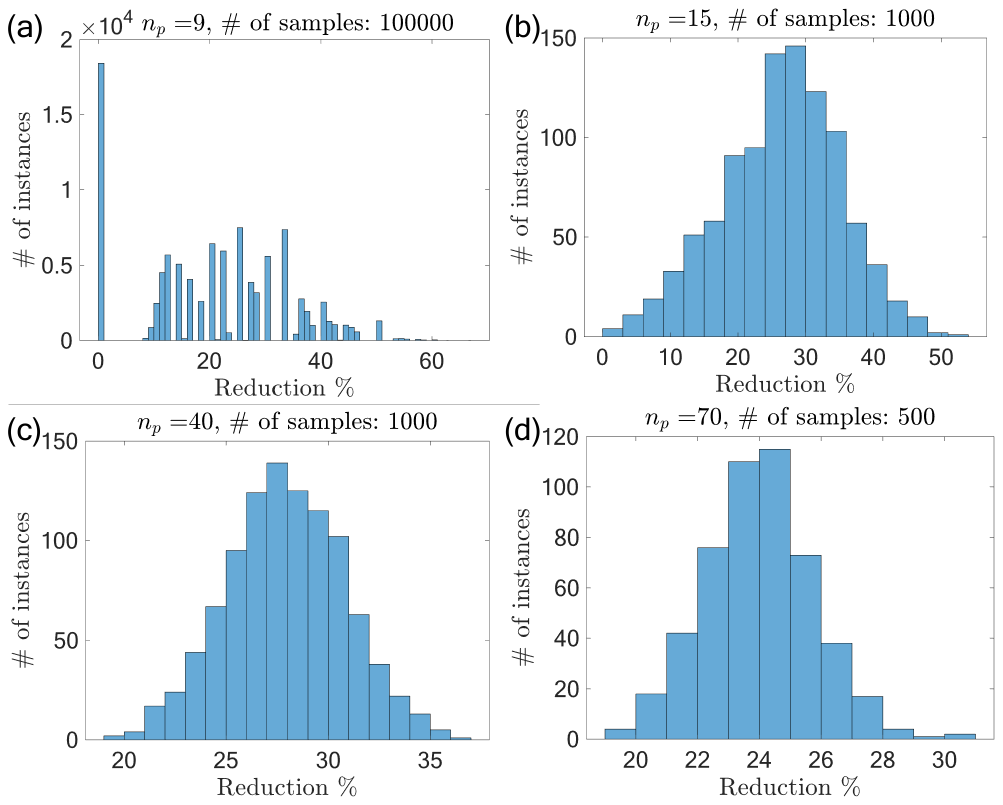}
    \caption{Reduction of emitter CNOTs across different sampling sizes and number of photons in the graph. We use the Heuristics $\#1$ optimizer. We set $n_p=9$ in (a), $n_p=15$ in (b), $n_p=40$ in (c) and $n_p=70$ in (d).}
    \label{fig:StatHeu1}
\end{figure}
Here we provide more data on the improvement we can get for random graphs if we use our Heuristics optimizers.
In Fig.~\ref{fig:StatHeu1} we show the percentage reduction in emitter CNOTs, if we use the Heuristics $\#1$ optimizer. We test two variants of Heuristics $\#1$ optimization, where in the first one we set the Back-substitution option to false, and the extra test for free photon absorption to false, and in the other one we enable both options. From the two methods, we pick the one that gave the best reduction in emitter CNOTs. In Fig.~\ref{fig:StatHeu1}(a) we show our results for $n_p=9$, for $10^5$ samples. The maximum reduction we find is $66.67 \%$. We repeat the same calculation for $n_p=15$ and 1000 samples in (b), for $n_p=40$ and 1000 samples in (c), and for $n_p=70$ and 500 samples in (d). We see that the distributions saturate to about $25-30\%$ reduction for the larger-sized graphs we consider.

\begin{figure}[!htbp]
    \centering
    \includegraphics[scale=0.49]{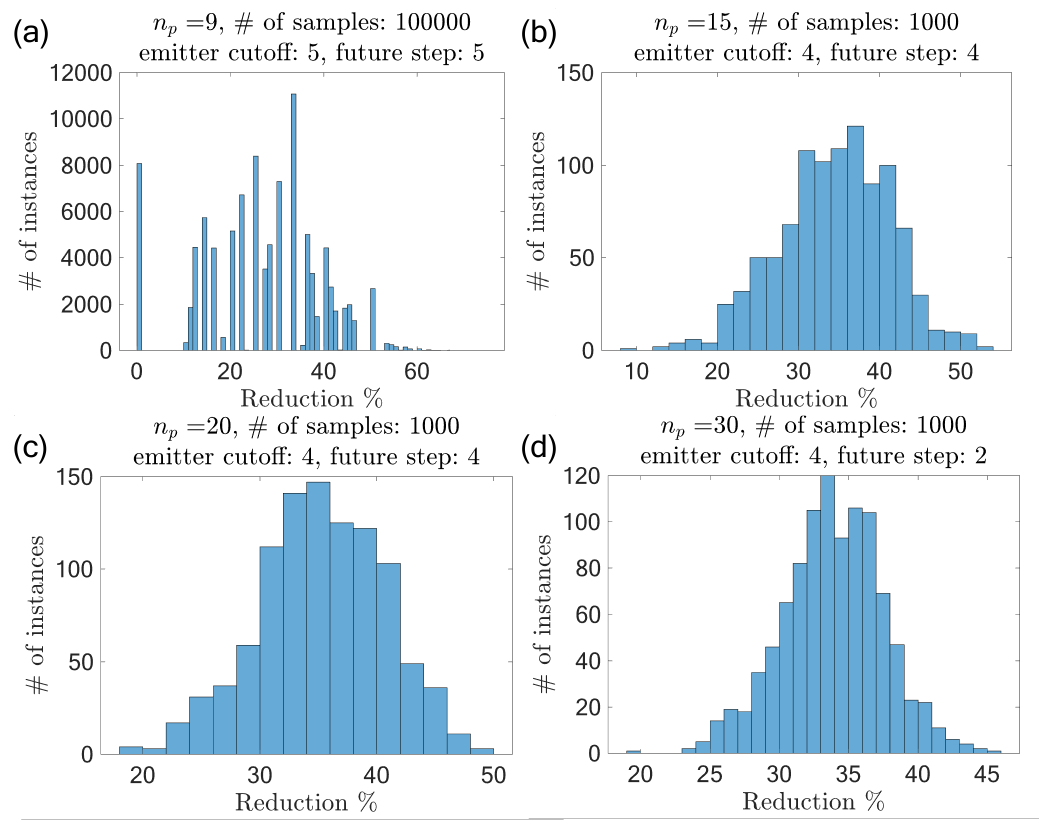}
    \caption{Reduction of emitter CNOTs across different sampling sizes and number of photons in the graph. We use the Heuristics $\#2$ optimizer. We set $n_p=9$ in (a), $n_p=15$ in (b), $n_p=20$ in (c) and $n_p=30$ in (d).}
    \label{fig:StatHeu2}
\end{figure}

In Fig.~\ref{fig:StatHeu2}, we use the Heuristics $\#2$ optimizer and test three variants involving the Back-substitution and extra inspection for free PA options, namely:
\begin{enumerate}
    \item both options set to false
    \item first option set to true, and second option set to false 
    \item first option set to false, and second option set to true.
\end{enumerate}

From these variants, we select the one that gave the minimum CNOT counts. Those are the same variants we explored for the Heuristics $\#2$ method in Fig.~\ref{fig:Comp_2_heuristics_w_Naive}. In Fig.~\ref{fig:StatHeu2}(a) we consider $n_p=9$ for $10^5$ samples. The maximum reduction we find is $75\%$. In this plot, we allow the further recursions, and we set emitter and future cutoffs to 5. In Fig.~\ref{fig:StatHeu2}(b) we consider $n_p=15$, 1000 samples and set the emitter and future cutoffs to 4. We see a maximum reduction of $52.94\%$, and a mean reduction of $\sim 34.35\%$. We see similar results in Fig.~\ref{fig:StatHeu2}(c) where we set $n_p=20$, and in (d), where we set $n_p=30$. In Fig.~\ref{fig:StatHeu2}(d) we change the future cutoff to 2.



%


\end{document}